\documentclass[12pt,preprint]{aastex}

\newcommand{\kmpers}{{{\rm \; km\;s}^{-1}}}
\newcommand{\logOH}{{\log{{\rm(O/H)}}}}
\newcommand{\Htot}{{N{\rm(H_{tot})}}}
\newcommand{\OH}{{\rm O/H}}
\newcommand{\NHI}{{N\rm(OI)}}
\newcommand{\Hmol}{{\rm H_{2}}}
\newcommand{\nH}{{\langle \rm{n_{H}} \rangle}}
\newcommand{\NHmol}{{N\rm(H_{2})}}
\newcommand{\fHmol}{{f\rm(H_{2})}}
\newcommand{\logHI}{{\log{\NHI}}}
\newcommand{\logHmol}{{\log{\NHmol}}}
\newcommand{\logHtot}{{\log{\Htot}}}
\newcommand{\logOI}{{\log{N\rm(OI)}}}
\newcommand{\ppm}{{\rm \; ppm}}
\newcommand{\percc}{{\rm \; cm}^{-3}}
\newcommand{\OHsolar}{{\rm(O/H)}_{\odot}}
\newcommand{\OHdust}{{\rm O_{dust}/H}}
\newcommand{\dex}{{\rm \; dex}}

\shortauthors{Jensen, Rachford, and Snow}

\begin{document}

\title{Abundances and Depletions of Interstellar Oxygen}

\author{Adam G. Jensen, Brian L. Rachford, and Theodore P. Snow}

\affil{Center for Astrophysics and Space Astronomy}
\affil{Department of Astrophysical and Planetary Sciences, University of Colorado at Boulder}
\affil{Campus Box 389}
\affil{Boulder, CO 80309-0389}

\email{Adam.Jensen@colorado.edu, rachford@casa.colorado.edu, tsnow@casa.colorado.edu}

\begin{abstract}

We report on the abundance of interstellar neutral oxygen (OI) for 26 sightlines, using data from the {\it Far Ultraviolet Spectroscopic Explorer} ({\it FUSE}), the {\it International Spectroscopic Explorer} ({\it IUE}), and the {\it Hubble Space Telescope} ({\it HST}).  OI column densities are derived by measuring the equivalent widths of several ultraviolet absorption lines, and subsequently fitting those to a curve of growth.  We consider both our general sample of 26 sightlines and a more restrictive sample of 10 sightlines that utilize {\it HST} data for a measurement of the weak 1355 \AA{} line of oxygen, and are thus better constrained due to our sampling of all three sections of the curve of growth.  The column densities of our {\it HST} sample show ratios of O/H that agree with the current best solar value if dust is considered, with the possible exception of one sightline (HD 37903).  We note some very limited evidence in the {\it HST} sample for trends of increasing depletion with respect to $R_V$ and $\fHmol$, but the trends are not conclusive.  Unlike a recent result from \citet{Cartledge2}, we do not see evidence for increasing depletion with respect to $\nH$, but our {\it HST} sample contains only two points more dense than the critical density determined in that paper.  The column densities of our more general sample show some scatter in O/H, but most agree with the solar value to within errors.  We discuss these results in the context of establishing the best method for determining interstellar abundances, the unresolved question of the best value for O/H in the interstellar medium (ISM), the O/H ratios observed in Galactic stars, and the depletion of gas-phase oxygen onto dust grains.

\end{abstract}


\section{INTRODUCTION AND BACKGROUND}
\label{s:intro}
As the third most abundant element in the Galaxy after hydrogen and helium, oxygen is an important component in interstellar chemistry and dust grain models.  The exact abundance of gas-phase oxygen relative to hydrogen has been in question for quite some time.  The question of whether oxygen accumulates onto dust grains to varying degrees in different environments has also been investigated.  While recent research seems to be moving towards a picture of constant oxygen depletion from the gas, questions remain, especially with regard to the possible dependence of oxygen depletion on cloud density.  There is much work to be done in examining sightlines with more extreme physical properties---sightlines that have been previously ignored due to observational considerations.

Interstellar absorption lines in the ultraviolet were first detected with sounding rockets, e.g. \citet{MortonSpitzer}.  Detailed studies of neutral oxygen in the interstellar medium became possible with the introduction of ultraviolet satellites such as {\it Copernicus} and {\it IUE} in the 1970s.  Studies such as those by \citet{Morton74}, \citet{Morton75}, \citet{Zeippen}, \citet{deBoer79}, and \citet{deBoer81} each focused on the study of an individual sightline, finding that gas-phase oxygen was depleted relative to the solar value.  \citet{MortonEtAl} also studied the column densities of several different atomic species, including neutral oxygen, for five sightlines.

The first extensive survey to determine the general interstellar value of O/H was reported by \citet{York}.  Their study included 53 sightlines.  Of these, oxygen column densities were determined in 14 cases, while column density upper limits were determined in 16 cases and column density lower limits were determined in 20 cases (they did not report on OI for the final three sightlines).  Their analysis used the weak 1355 \AA{} line of OI, relying on the assumption that NI and OI have a similar $b$-value.  They determined column densities by using this $b$-value to apply an appropriate saturation correction to the 1355 \AA{} line when necessary.  In a few cases, the 988 \AA{} triplet of OI was used, but further details were not given.  The conclusion of this study was that interstellar O/H is between 40\% and 70\% of the assumed solar value, from \citet{Withbroe}, of approximately 690 oxygen atoms per million hydrogen atoms (later studies have assumed different solar values of the solar O/H ratio, and the correct value is still very much in dispute; see \S\ref{ss:dust} for a brief discussion).  Additionally, \citet{York} concluded that there is no evidence of systematically enhanced depletion as reddening (i.e. total hydrogen column density) increases.  Rather, they found some limited evidence that sightlines with $\logHtot<20.5$ are more depleted (with the mean O/H ratio approximately 40\% solar) than sightlines with $\logHtot>20.5$ (with the mean O/H ratio approximately 70\% solar).

\citet{Keenan} explored 26 sightlines---including one sightline from \citet{York}---and found an average O/H ratio that is 50\% of the solar value by \citet{Withbroe}.  In contrast to the results of \citet{York}, \citet{Keenan} did not find a significant difference between sightlines with $\logHtot<20.5$ and those with $\logHtot>20.5$.

An important recent survey of interstellar oxygen was performed by \citet{Meyer}.  Their study covered 13 sightlines and involved detailed analysis of the 1355 \AA{} line of OI using the Goddard High Resolution Spectrograph (GHRS) on the {\it Hubble Space Telescope}.  These 13 sightlines included four for which the column density had been previously determined, and three for which a saturation correction (estimated from MgII and NI) was applied.  The study found the mean O/H to be $319\pm14 \ppm$, with the largest deviation from the mean less than 18\%.  Assuming as many as 180 oxygen atoms in the dust phase per million hydrogen atoms, a value consistent with many theoretical models (see \S\ref{ss:dust}), this corresponds to a total O/H ratio approximately two-thirds of the \citet{Grevesse} solar value, $\OHsolar=741\pm130 \ppm$.  Additionally, \citet{Meyer} did not find any systematic, statistically significant variation of O/H with respect to the total column density of hydrogen ($\Htot$), the average volume density of hydrogen along the sightline ($\nH$), or the molecular fraction of hydrogen ($\fHmol=2\NHmol/[2\NHmol+\NHI]$).  We note that the \citet{Meyer} study was confined to stars with $A_V \lesssim 1.0$ mag.

Studies since \citet{Meyer} may exhibit systematic differences from previous studies, in both total oxygen column density and O/H ratio relative to solar, for the following reasons:
\begin{itemize}
\item Revision of the $f$-value for the weak 1355 \AA{} transistion (see \S\ref{ss:1355line}) from $1.25\times10^{-6}$ to $1.16\times10^{-6}$, as suggested by \citet{Welty}
\item Revision of the solar O/H ratio from approximately $700 \ppm$ or more \citep{Withbroe, Grevesse} to approximately $550\ppm$ or less \citep{Allende, Holweger, Asplund}
\end{itemize}
In comparing results with \citet{Meyer}, more recent authors \citep{Cartledge, Andre, Cartledge2} have made a linear correction to the reported O/H ratios of \citet{Meyer} to reflect the change in $f$-value.  With the change, the mean \citet{Meyer} value becomes $\OH=343\pm15\ppm$.  Additionally, assuming $\OHdust$ is as large as $180\ppm$, the total O/H ratio can be reconciled with the revised solar value.

\citet{Cartledge} examined 11 sightlines using the Space Telescope Imaging Spectrograph (STIS) onboard {\it HST} to observe the 1355 \AA{} line.  This study used two methods: profile fitting, where velocity information from other atomic species is incorporated into the fit of the oxygen line; and the apparent optical depth (AOD) method, where it is assumed that the line does not contain unresolved saturated structure and the measurement represents a lower limit on column density if the assumption fails (profile fitting can suffer from the same problem if unresolved, saturated structure is present and none of the velocity information from other species reveals such structure).  \citet{Cartledge} analyzed a subset of seven sightlines, and compared them with the 13 sightlines measured with GHRS from \citet{Meyer}.  They found that five of the seven sightlines measured with STIS lie below the O/H mean (adjusted for the new $f$-value of the 1355 \AA{} line) determined by \citet{Meyer}.  These five sightlines have five of the six highest hydrogen column densities in the combined sample of 20 sightlines.  This result is similar to O/Kr ratios reported in the same work, where some sightlines with higher total oxygen content exhibit oxygen depletion relative to krypton (a noble gas that is likely to be found almost exclusively in the gas phase).  Despite this, the oxygen column densities for only two sightlines have 1-$\sigma$ uncertainties that exclude the possibility of consistency with the \citet{Meyer} mean for O/H.  The evidence that there is oxygen depletion with respect to krypton, however, is somewhat more convincing.

\citet{Moos} were among the first to use {\it FUSE} to measure the equivalent widths of several different transistions for neutral oxygen, and to subsequently measure oxygen column densities by using curves of growth.  They did so for seven sightlines less than $180 {\rm \: pc}$ in length.  Despite large uncertainties and potential systematic errors from inhomogeneous determinations of atomic hydrogen, the seven sightlines were found to have O/H ratios from 234 to $479\ppm$---consistent within a factor of 2.1.  Excluding the sightline BD+28$^\circ$2411, the range is 363 to $479\ppm$---consistent within a factor of 1.3.  Within the uncertainties, these results are consistent with \citet{Meyer}.

The next major survey of interstellar oxygen was by \citet{Andre}.  This survey measured the oxygen content of 19 sightlines using STIS.  Similar to \citet{Cartledge}, \citet{Andre} used both profile fitting and the AOD method to determine column densities.  Their analysis produced a mean $\OH=408\pm13\ppm$.  The standard deviation in their sample was $59\ppm$.  \citet{Andre} included the previous results of \citet{Cartledge} and \citet{Meyer} to find a mean $\OH=362\ppm$, with a standard deviation of approximately 20\%.  \citet{Andre} took the average of values by \citet{Allende} and \citet{Holweger}, $517\pm58\ppm$, as the solar value of O/H.  Both the O/H mean of the 19 sightlines of the \citet{Andre} sample and the O/H mean of the combined sample can be reconciled with this solar value if dust is considered.

The most recent interstellar oxygen survey was by \citet{Cartledge2}, probing 36 sightlines using the same methods as \citet{Cartledge}, and adding previous results for 20 sightlines, discussed results for a cumulative sample of 56 sightlines.  The results of \citet{Cartledge2} reinforce the conclusions of \citet{Cartledge} with many more data points.  Specifically, a 4-parameter Boltzmann function is fit to the ensemble of O/H ratios when plotted against $\nH$, with an apparent transition centered at $\nH\approx1.5\percc$.  This phenomenon is somewhat surprising in that it is not clear why $\nH$ should accurately trace the environment of a sightline; rather it is more commonly expected that reddening parameters, e.g. $A_V$ or $E_{B-V}$, should be a better trace of the individual clouds that dominate a given sightline.  However, the trend seen by \citet{Cartledge2} is not new; results by \citet{SnowRach}, \citet{Jenkinsetal}, \citet{Spitzer}, and many other references cited therein have shown that the sightline characteristic with the most consistent correlation for enhanced depletion of many different elements is $\nH$.  \citet{Spitzer} discussed the possibility that $\nH$ traces the superposition of three phases of the ISM: a warm, low-density phase, a cold, moderate-density phase, and a cold, high-density phase.  Regardless of the theoretical difficulties of determining a comprehensive picture for why $\nH$ traces the interstellar environment so well, this most recent result of \citet{Cartledge2} is the first time that this has been detected for oxygen.  The fit found by \citet{Cartledge2} suggests that $\OHdust=192\pm51 \ppm$ in the higher density sightlines, compared to $\OHdust=86\pm51 \ppm$ in lower density sightlines.  As discussed in that paper, the increase of over $100\ppm$ is difficult to account for with most dust models.  Of note, \citet{Cartledge2} adopt a slightly different value for the solar oxygen abundance than \citet{Andre}.  \citet{Cartledge2} adopt $\OHsolar=476\pm50 \ppm$, the weighted average of \citet{Holweger} and \citet{Asplund}, as opposed to the straight average of \citet{Allende} and \citet{Holweger} adopted by \citet{Andre}.

Taking all these results into account, the emerging picture is that interstellar O/H is relatively constant over a wide range of environments.  If ${\rm O_{dust}/H}\lesssim180\ppm$, then the interstellar O/H ratio can be reconciled with the solar value.  There are two exceptions to the observed constancy.  The first exception is the \citet{Cartledge} study that hints at the evidence of, but does not conclusively prove, enhanced depletions in sightlines with higher total oxygen and hydrogen column densities.  The second exception is the \citet{Cartledge2} study that shows a small but clear depletion effect as $\nH$ becomes larger than $1.5\percc$.  This is in strong contrast to the results for many other atomic species such as iron, silicon, and aluminum, species that show highly enhanced depletion in sightlines with higher reddening.

One assumption concerning these studies that should be noted is that the ionization energies of OI and HI are very similar, at $13.618 {\rm \; eV}$ and $13.598 {\rm \; eV}$, respectively.  This means that the fractional contribution to the total oxygen abundance from ionized oxygen (OII, OIII, etc.) will be mirrored by the fractional contribution to the total hydrogen abundance from HII.  A charge exchange reaction between oxygen and hydrogen also constrains this relationship \citep{Field}.  Since the dissociation energy of molecular hydrogen is only 4.476 eV, the assumption that O/H is independent of ionization fraction holds across all molecular fractions as well.  In particular, the fact that a cloud has a high molecular fraction would tend to reveal that the ionization fraction is very small, since the majority of molecular hydrogen would be destroyed before significant ionization of atomic hydrogen would occur.

The goal of this study is to add to the current body of knowledge concerning interstellar oxygen by determining the O/H ratio for several sightlines observed by {\it FUSE}.  We supplement the {\it FUSE} data with {\it HST} and {\it IUE} data where possible.  Here we report on our measurements of the equivalent widths of several neutral oxygen absorption lines for each sightline, and our fits of these to curves of growth, from which we determined the $b$-values and column densities of oxygen.  The curve of growth method was used in the hope of avoiding assumptions about $b$-values and subsequent saturation corrections.  We then looked for systematic differences in O/H with respect to total hydrogen and various reddening parameters.

As a side goal, we aimed to determine the utility of {\it FUSE} for observing oxygen absorption lines shortward of $\sim1100$ \AA{}, a wavelength region that is not covered by {\it HST} or {\it IUE} and has been observed only to a limited extent with {\it Copernicus}, but where a number of weak-to-medium strength OI lines are found.  With some exceptions, e.g. \citet{Zeippen}, studies of interstellar oxygen have been limited to analysis of the 1355 \AA{} line of oxygen which, though weaker than the {\it FUSE} lines, is sometimes strong enough to be saturated.  While the more recent of these studies have taken into account the profiles of other species to formulate a comprehensive picture of the cloud structure along a given sightline, there remains the possibility of saturation effects for individual velocity components.  It is not yet known whether studies such as this paper and \citet{Moos} can equal or perhaps improve upon the results of previous studies by taking into account all the information that is now available to us through {\it FUSE} in the form of many far-UV OI lines.

\section{OBSERVATIONS AND DATA REDUCTION}
\label{s:obsdata}
Data were taken from {\it FUSE}, {\it HST}, and {\it IUE}.  The sightlines were chosen from a sample observed by {\it FUSE} as part of a molecular hydrogen survey conducted by \citet{Rachford} and continued in \citet{Rachford2}.  The original {\it FUSE} sample included 34 sightlines, and {\it HST} or {\it IUE} data were available for 27 of these sightlines.  For four sightlines (the sightlines towards HD 43384, HD 166734, HD 281159, and NGC 2264 67, a.k.a. Walker 67) we were unable to measure the equivalent widths of any absorption lines.  We do not report further on these sightlines.  Basic stellar data for the stars of the other 30 sightlines is given in Table \ref{stellardata}.  Values for hydrogen column densities and reddening parameters along the lines of sight are given in Tables \ref{hydrogentable} and \ref{reddeningtable}, respectively.

\subsection{{\it FUSE} Data}
\label{ss:FUSEdata}
\begin{figure}[t!]
\begin{center}
\epsscale{1.00}
\plottwo{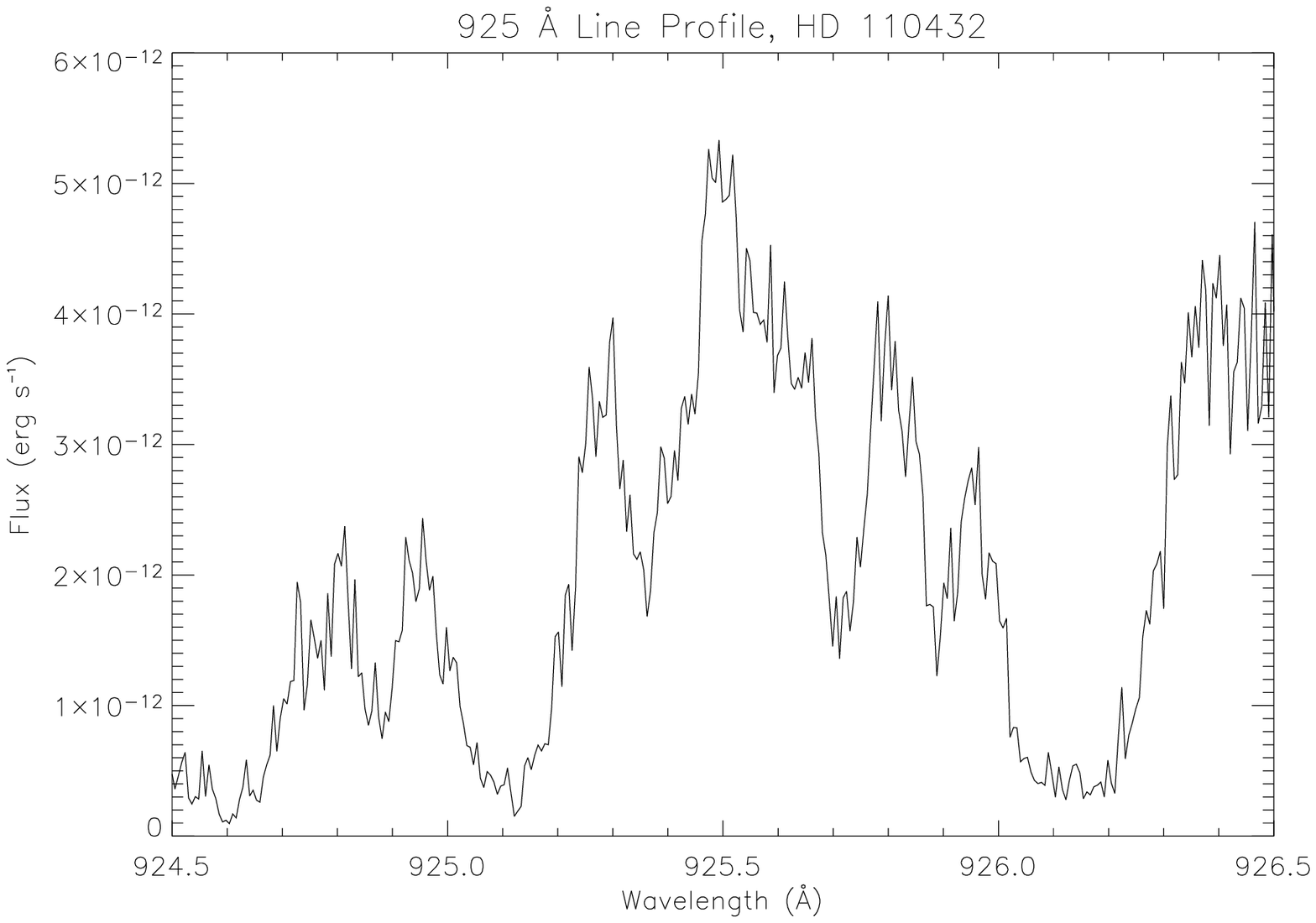}{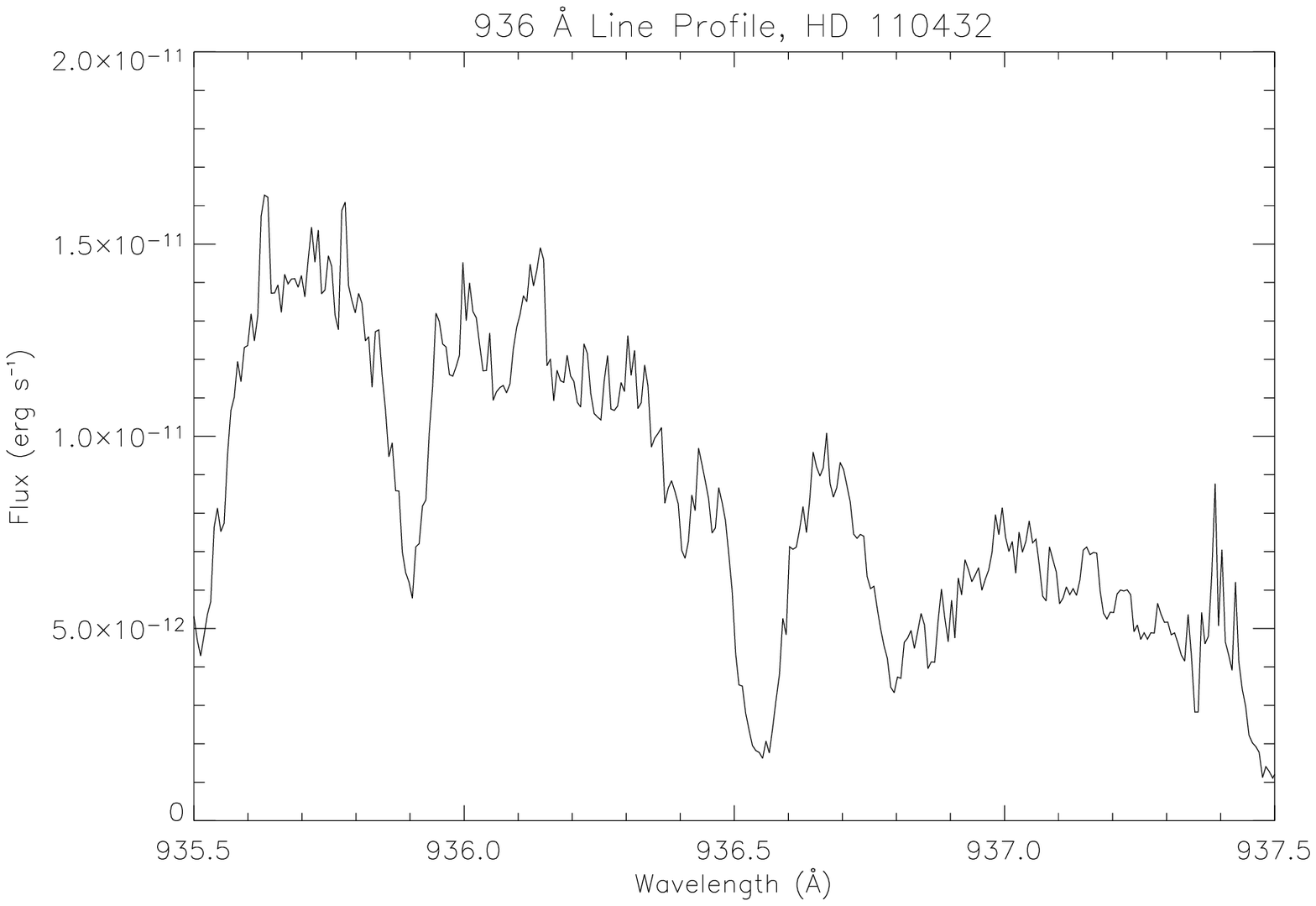}
\end{center}
\begin{center}
\epsscale{1.00}
\plottwo{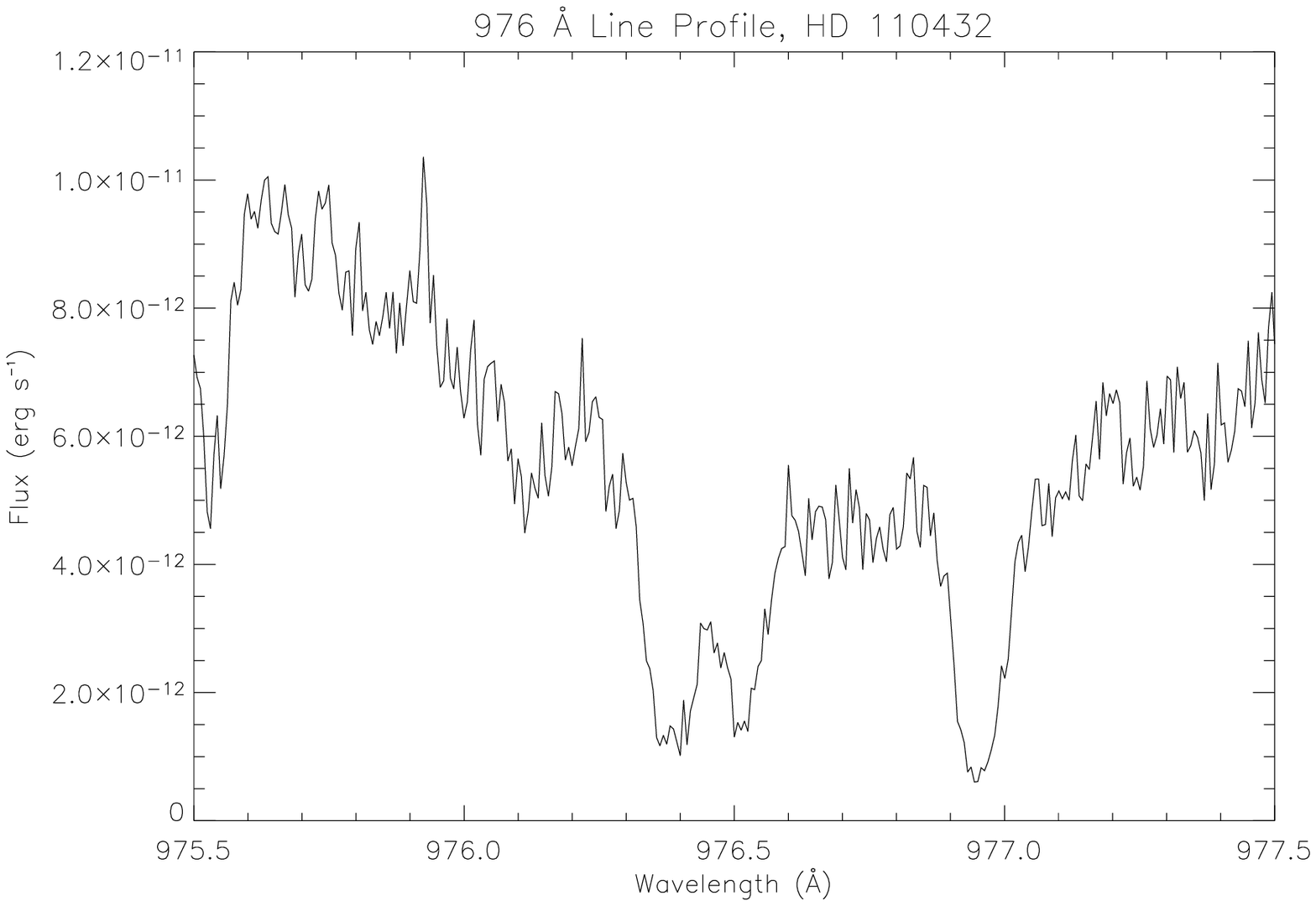}{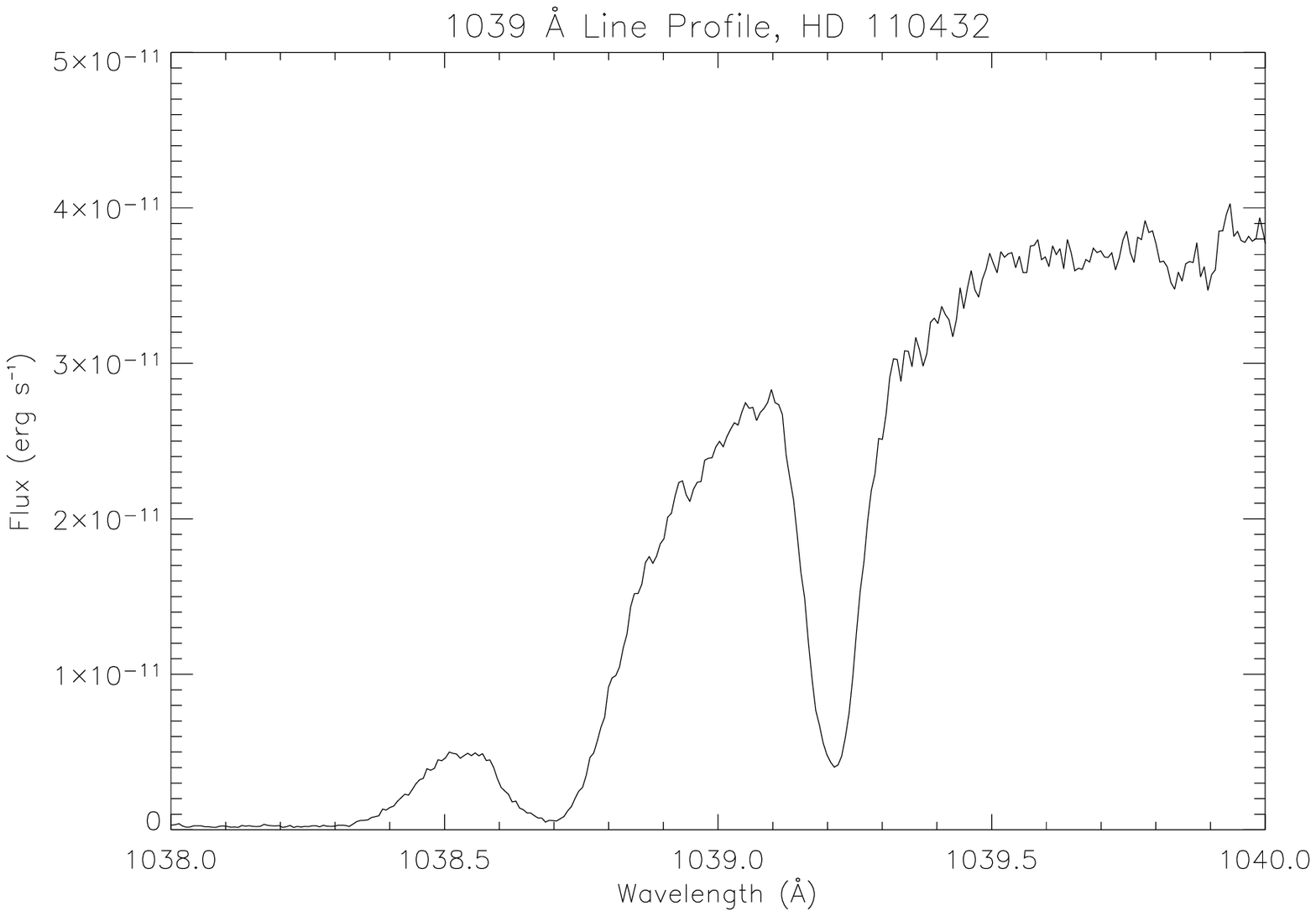}
\end{center}
\caption{Selected portions of the {\it FUSE} spectrum for HD 110432 showing some of the major absorption lines used in this study.  The spectrum are shown unnormalized and without velocity correction.  {\bf Top left}---925.4461 \AA{} absorption line.  {\bf Top right}---936.6295 \AA{} absorption line.  {\bf Bottom Left}---976.4481 \AA{} absorption line.  {\bf Bottom Right}---1039.2304 \AA{} absorption line.}
\label{fig:FUSEspecs}
\end{figure}

Data are collected with the {\it FUSE} satellite by four channels, two with a lithium fluoride detector (LiF1 and LiF2) and two with a silicon carbide detector (SiC1 and SiC2).  Each channel contains two data segments, designated as A and B, covering adjacent wavelength regions.  The lithium fluoride channels cover the wavelength region from 989-1188 \AA{}, while the silicon carbide channels cover the region from 905-1104 \AA{}.  Samples of {\it FUSE} spectra are shown in Figure \ref{fig:FUSEspecs}.

The pixel scale of {\it FUSE} is $\sim0.007$ \AA{}.  Knowledge of the FWHM of weak lines is necessary because the theoretical line must be convolved with instrumental resolution when we make theoretical profile fits.  We measured weak lines in the {\it FUSE} spectra and found the full-width at half-max (FWHM) of these lines to be 0.074 \AA{}.  This is the value for the instrumental resolution element used in our model that iteratively fits a Voigt profile to an absorption line.  This model was used in certain sightlines to fit moderately saturated lines of oxygen (see \S\ref{ss:1000lines} and \S\ref{ss:1039line}).

The data were reduced with the CALFUSE pipeline, versions 2.0.5 or 2.1.6.  Data from different data segments are not coadded; rather, the data segment with the best signal-to-noise (S/N) is selected, with other data segments providing a check for consistency.

\subsection{{\it HST} Data}
\label{ss:HSTdata}
\begin{figure}[t!]
\begin{center}
\epsscale{1.00}
\plottwo{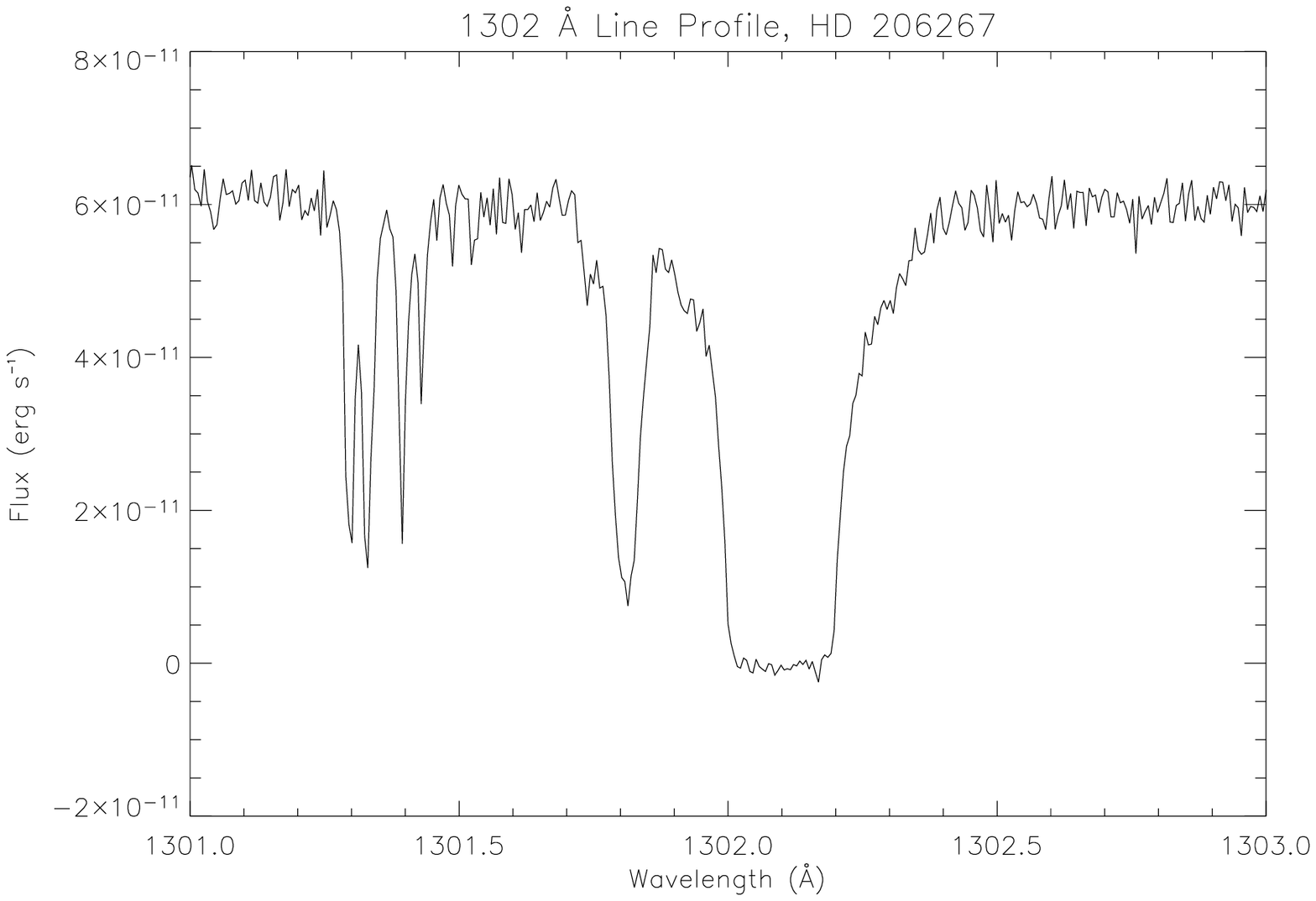}{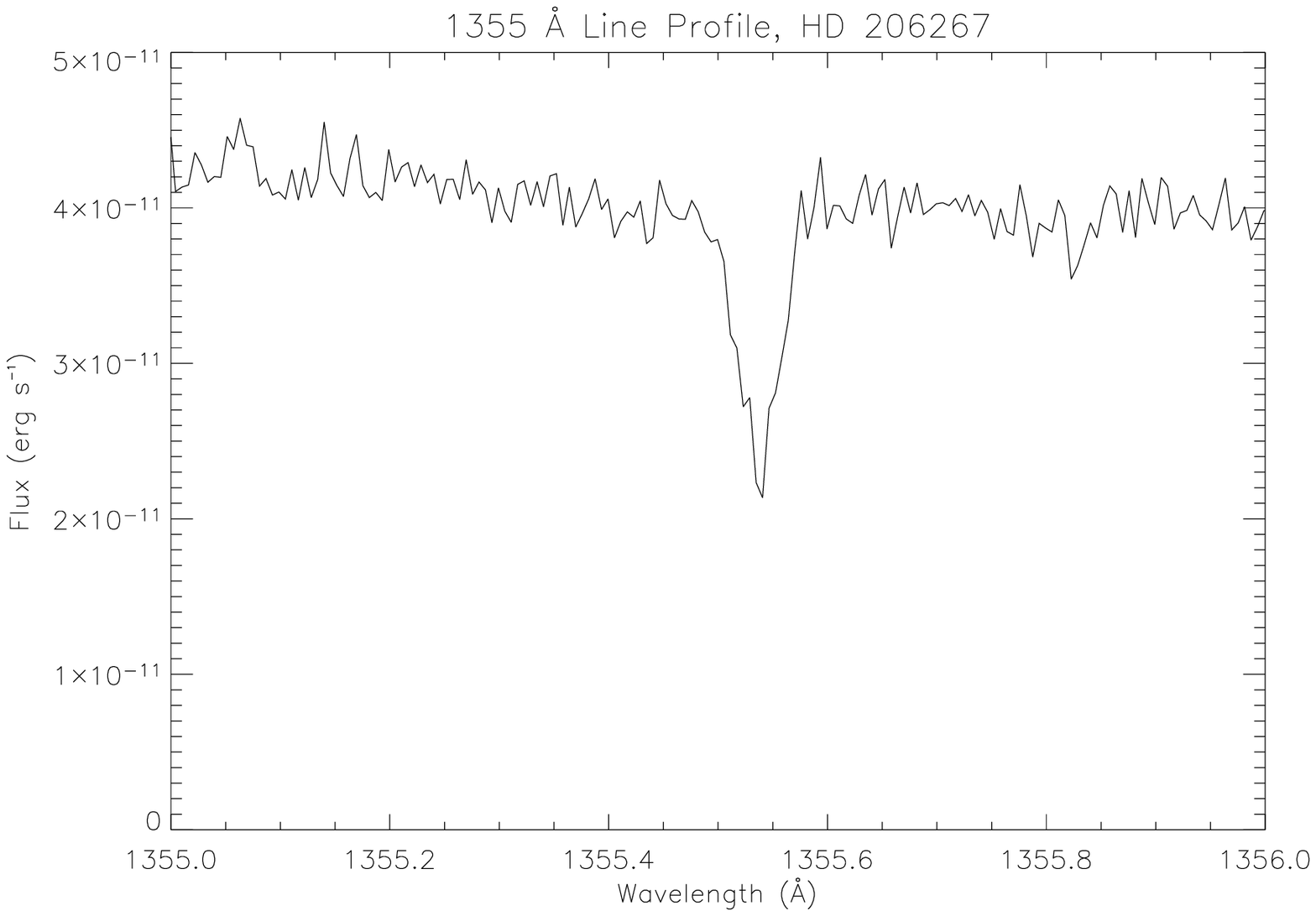}
\end{center}
\begin{center}
\epsscale{1.00}
\plottwo{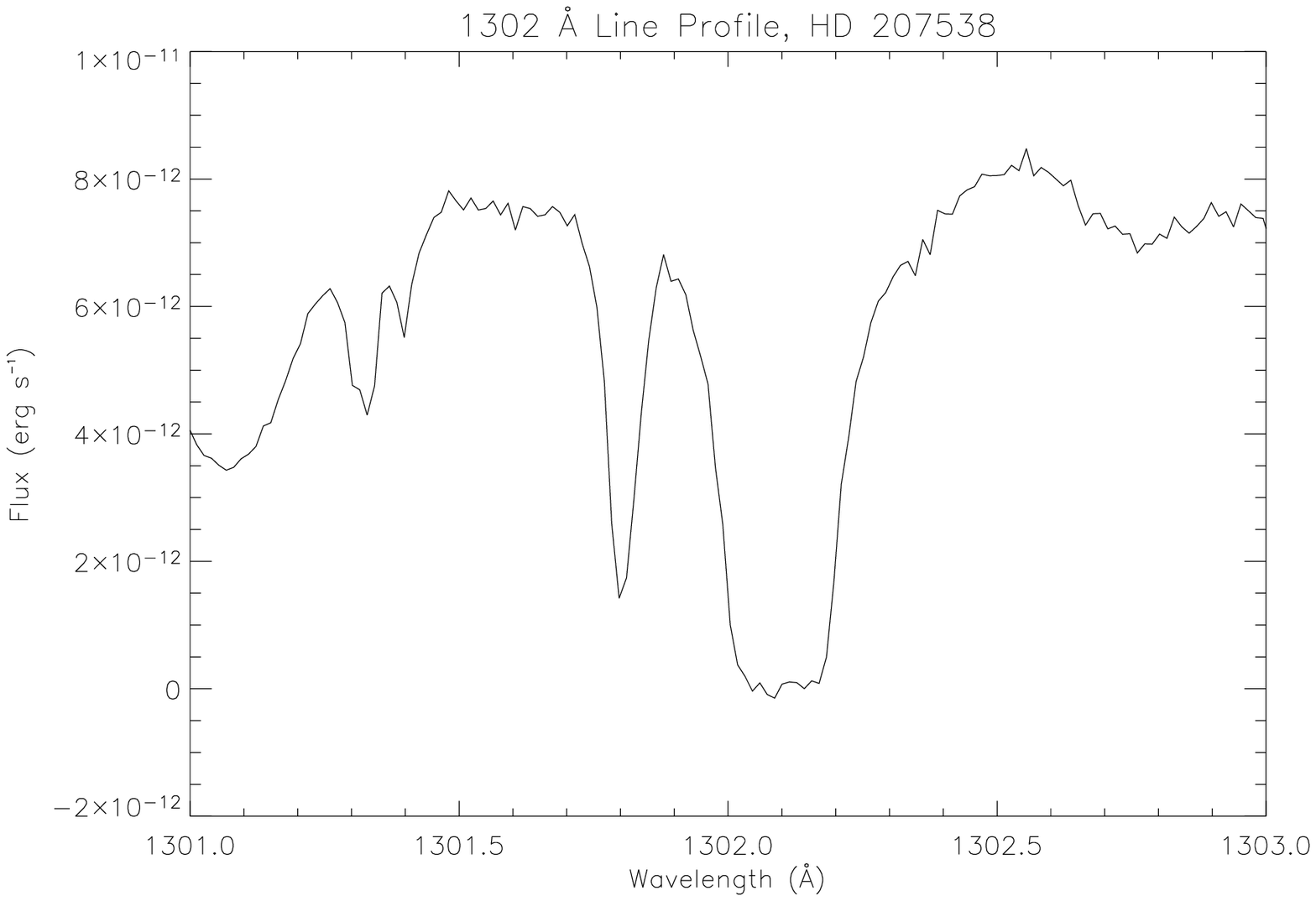}{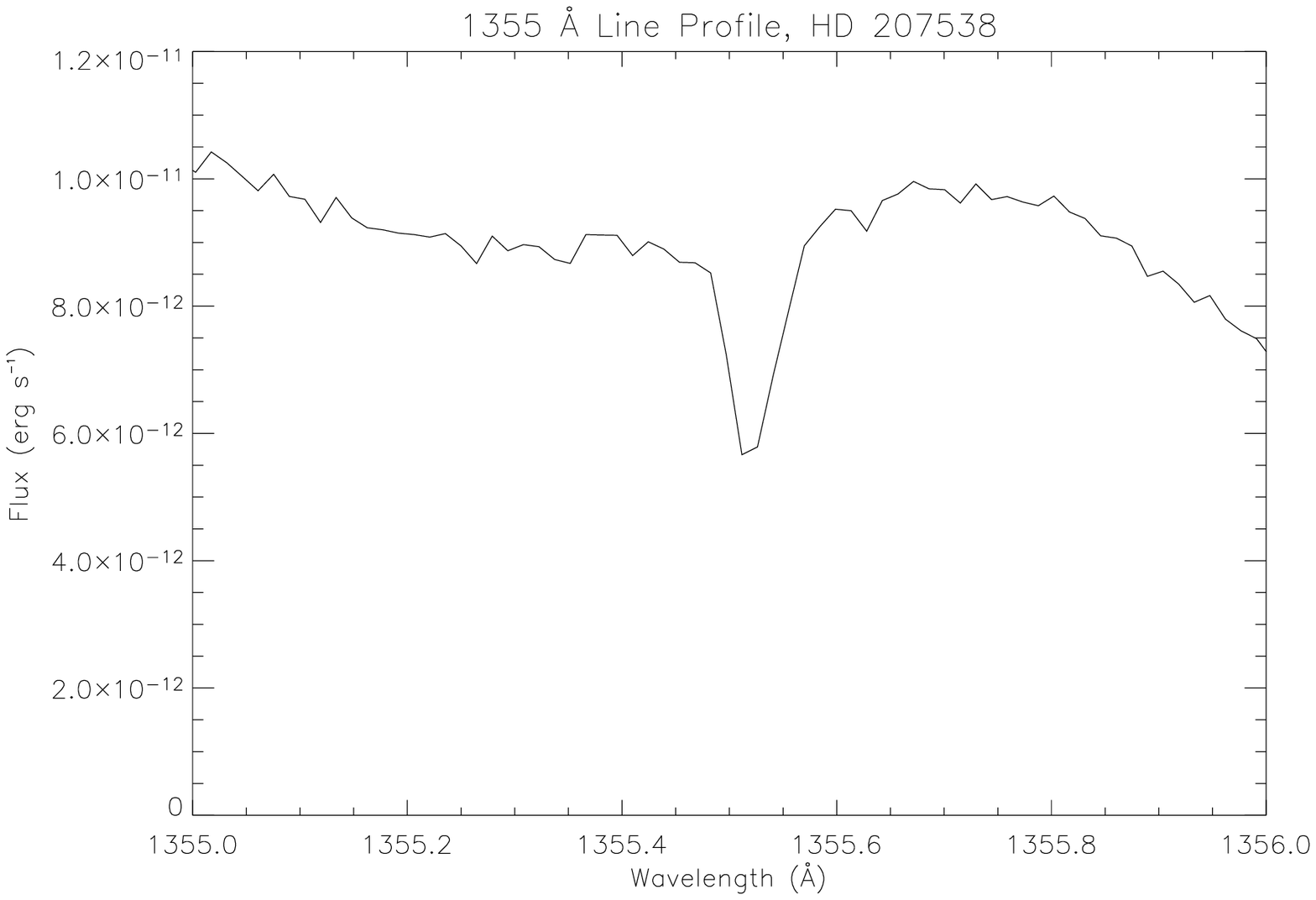}
\end{center}
\caption{Previously unpublished {\it HST} spectra for HD 206267 and HD 207538 showing the weakest and strongest absorption lines used in this study.  Spectra are shown unnormalized and without velocity correction.  {\bf Top left}---1302.1685 \AA{} absorption line in the sightline towards HD 206267.  {\bf Top right}---1355.5977 \AA{} absorption line in the sightline towards HD 206267.  {\bf Bottom left}---1302.1685 \AA{} absorption line in the sightline towards HD 207538.  {\bf Bottom right}---1355.5977 \AA{} absorption line in the sightline towards HD 207538.}
\label{fig:HSTspecs}
\end{figure}

Archived {\it HST} data taken by STIS were available for nine sightlines:  HD 24534, HD 27778, HD 37903, HD 185418, HD 192639, HD 206267, HD 207198, HD 207538, and HD 210839.  The oxygen content of many of these sightlines has been analyzed previously with either STIS or GHRS (see Table \ref{comparison}).  We also use the equivalent width for the 1355 \AA{} line of the HD 154368 sightline that was measured by \citet{Snow154368} using GHRS data.  Samples of {\it HST} spectra are shown in Figure \ref{fig:HSTspecs}.

The pixel scale of the STIS observations is $\sim0.006$ \AA{}.  Measuring weak lines, we find FWHM values as small as 0.017 \AA{}.  In fitting the 1302 \AA{} line of OI (see \S\ref{ss:1302line}), we used this value for the instrumental resolution in our model that fits a Voigt profile.

We used on-the-fly calibrated STIS data.  In many cases we had multiple observations.  We coadded observations of similar echelle order, adding errors in quadrature.  When the targeted line was observed in multiple echelle orders, we analyzed separate echelle orders independently, then took an average of the results in determining our final equivalent widths.

\subsection{{\it IUE} Data}
\label{ss:IUEdata}
\begin{figure}[t!]
\begin{center}
\epsscale{1.00}
\plottwo{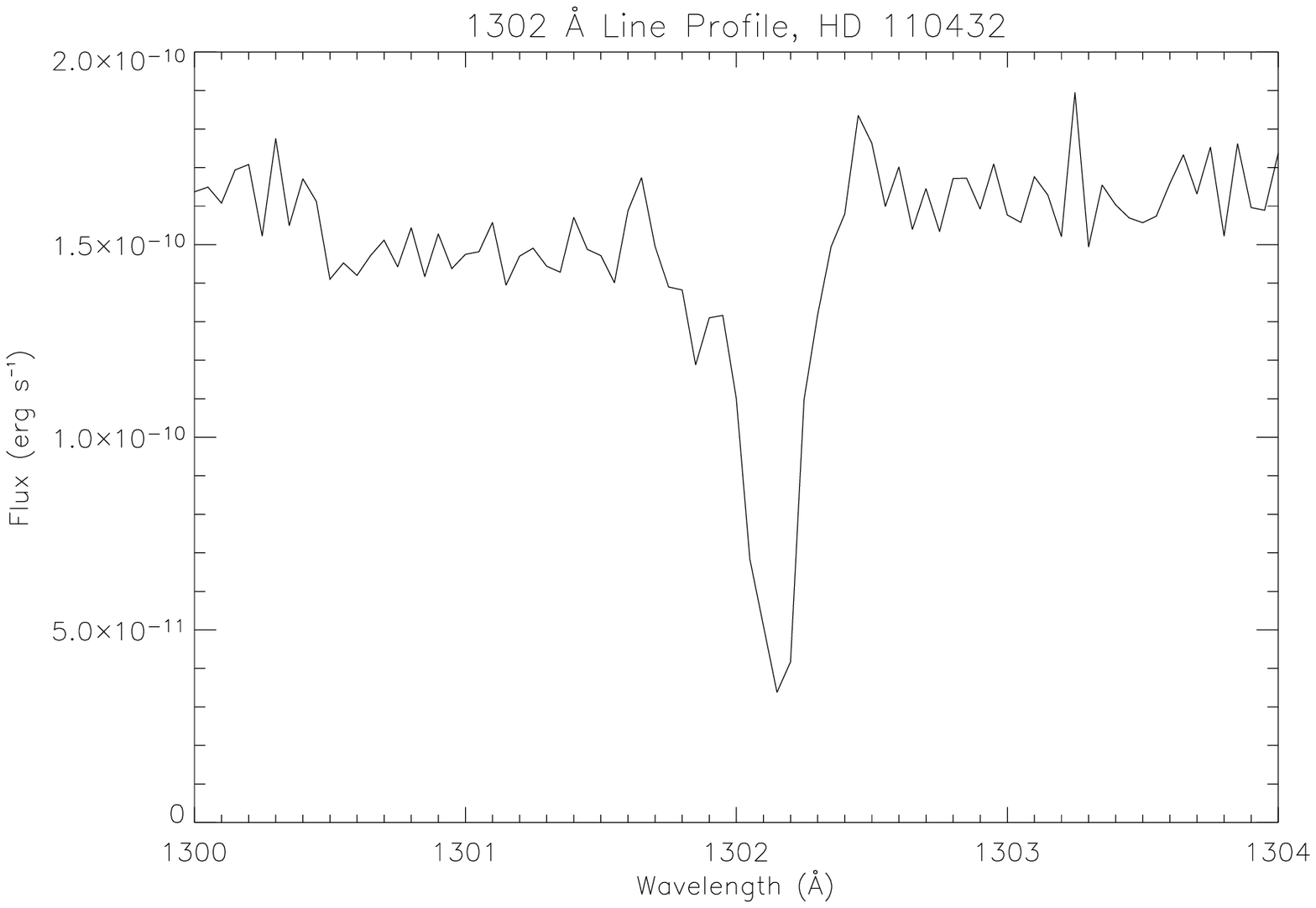}{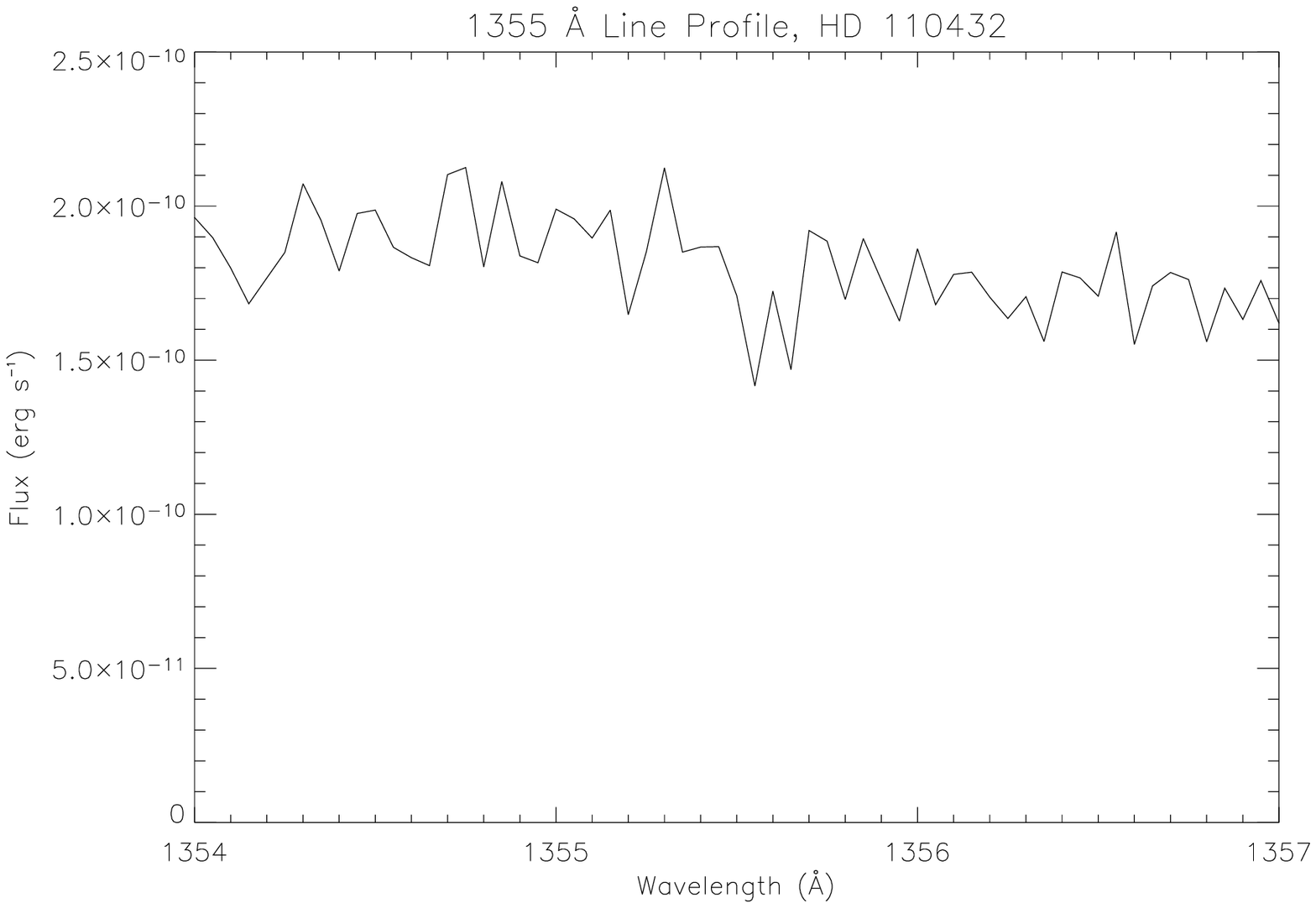}
\end{center}
\caption{Selected portions of the {\it IUE} spectrum for HD 110432 showing the weakest and strongest absorption lines used in this study.  The spectrum is shown unnormalized and without velocity correction.  {\bf Left}---1302.1685 \AA{} absorption line.  The low resolution of {\it IUE} causes the line to appear Gaussian even though it is heavily saturated.  The spectrum is a high-resolution, large-aperture observation from the SWP camera.  {\bf Right}---the 1355.5977 \AA{} absorption line is too weak to be seen in most {\it IUE} spectra.  The spectrum here hints at the presence of the line, but the surrounding continuum noise dictates that the error is larger than the equivalent width in our marginal fit of this line, and must be considered a non-detection.}
\label{fig:IUEspecs}
\end{figure}

High-dispersion data taken with the Short Wavelength Prime (SWP) camera onboard the {\it IUE} satellite were available on the MAST archive for 26 of the original 34 sightlines, and 25 of the 30 sightlines for which we were able to measure at least one equivalent width.  All of the sightlines for which {\it HST} data were available also had available {\it IUE} data, with the exception of HD 185418.  Low-dispersion data were available, but do not clearly resolve either the 1302 \AA{} or 1355 \AA{} line of OI and were not used.  Samples of {\it IUE} spectra are shown in Figure \ref{fig:IUEspecs}.

The pixel scale of {\it IUE} is 0.05 \AA{}.  We measured weak lines in the {\it IUE} spectra and found the FWHM of these lines to be 0.156 \AA{} or more.  This is the resolution element value used in calculating an upper limit on equivalent widths of undetected lines (see \S\ref{ss:limits}).

Where multiple observations were available, we coadded observations of similar aperture.  Bad pixels are corrected by using an interpolation between neighboring points.  Where multiple bad pixels are adjacent, we examine the spectra.  We find multiple adjacent bad pixels to be correlated with saturated absorption lines, and correct for this by setting the flux to zero at these points, if all spectra used in coadding show the same pattern of bad pixels.

When available, we analyzed data from both large and small apertures independently, and adopted a straight average of the resulting equivalent widths.  Results for the two equivalent widths are usually consistent within the errors.  We conclude that there is not a clear systematic difference between the two apertures that affects the two absorption lines that are under examination in this data.  All measured equivalent widths are either from large-aperture observations or an average---small aperture data are never used exclusively.

For eight of the sightlines where both {\it HST} and {\it IUE} data were available, we preferentially used our results from {\it HST} due to the higher resolution and S/N.  We examined the equivalent widths of the 1302 \AA{} line of OI in both {\it HST} and {\it IUE} in the cases where both data sets are available.  In two cases (HD 207198 and HD 207538) the equivalent widths were consistent.  In the other six cases, the IUE results seem to underestimate the equivalent width as compared to the {\it HST} data at the 1- or 2-$\sigma$ level.  There is a ninth case, HD 154368, where there are GHRS results found in the literature \citep{Snow154368}.  As mentioned in \S\ref{ss:compare}, the {\it IUE} equivalent width we measure in this case is larger than the equivalent width determined by \citet{Snow154368}, though that equivalent width is based on a theoretical profile derived from another line and appears to be too narrow when plotted against the data.  It is thus unclear whether or not there is a systematic difference between the {\it HST} and {\it IUE} data, though the possibility exists that {\it IUE} data underestimates the equivalent width of the strong line.

\citet{Massa} examined the scientific quality of the {\it IUE} archived data, and noted several aspects of the spectra that create uncertainty, including time-dependent effects on strong absorption lines and certain special cases where the cores of saturated lines that should be zero are instead a significant fraction (positive or negative) of the continuum.  Potential effects on the 1302 \AA{} and 1355 \AA{} lines and the surrounding spectral region are not specifically discussed.  The 1355 \AA{} is too weak to be observed in most {\it IUE} spectra, so the greatest concern would be with regard to the 1302 \AA{} line and whether or not it exhibits time-dependent effects similar to those \citet{Massa} found in Lyman-$\alpha$, a CIII multiplet at 1175 \AA{}, and NV lines near 1240 \AA{}.  In spite of these problems, additional calibration techniques for the {\it IUE} data are not apparent and we do not correct these spectra further.  Instead, we note here the general effect of uncertainty in a strong line like the 1302 \AA{} line.  If we have underestimated the true equivalent width of the 1302 \AA{} line, then our curve of growth analysis will underestimate the column density and/or overestimate the $b$-value.  On the other hand, if we overestimate the equivalent width of the 1302 \AA{} line, then our curve of growth analysis will overestimate the column density and/or underestimate the $b$-value.  These effects will be the most significant when there are fewer other lines included in the analysis.

\section{OBSERVED OI ABSORPTION LINES}
\label{s:lines}
The first step in determining column densities by a curve of growth is to measure the equivalent widths of multiple absorption lines.  Since there are two degrees of freedom in fitting the equivalent widths to a curve of growth (column density $N$ and velocity parameter $b$), whether or not we have a unique solution and/or a well-determined constraint depends on the number of equivalent widths fit to the curve.  One equivalent width provides a family of solutions, i.e., a unique column density can only be determined if a $b$-value is assumed, or vice-versa.  Two equivalent widths provide a unique solution, but with two free parameters the best $\chi^2$ cannot be converted into meaningful confidence intervals unless either $N$ or $b$ is assumed.  Three or more equivalent widths provide a unique solution and the ability to determine confidence intervals.  Consequently, the goal for each sightline was to measure the equivalent widths of at least three different absorption lines of OI.

The following is a description of the OI absorption lines that were used in our analysis.  Wavelengths, $f$-values, damping constants, and the instrument that was used for the observations of each of these lines are summarized in Table \S\ref{linetable}.  All atomic data in this section is taken from \citet{Morton03}.

\subsection{Lines Below 1000 \AA{}}
\label{ss:1000lines}
We identify six different OI absorption lines below 1000 \AA{} that are useful in our study.  These lines range in oscillator strength from $1.77\times10^{-4}$ to $3.65\times10^{-3}$.  None of these lines have damping constants that are reported in \citet{Morton03}.  These lines can be observed in the SiC1B and SiC2A data segments of {\it FUSE} in the spectra with the best S/N.  In the spectra of most sightlines, these lines are not identified.

Four of the lines (at 921.857 \AA{}, 922.200 \AA{}, 936.6295 \AA{}, and 976.4481 \AA{}) suffer from mild blending with $\Hmol$ lines in some spectra.  In these cases, we attempt simultaneous fitting of the OI line and the blending line if there is a separation between the cores that is resolved.  If there is not separation in the cores, then we do not attempt fitting.

We use Gaussians fits for these lines where possible, but a few cases require a Voigt profile.  We use a fitting routine that generates a Voigt profile from the parameters of column density and $b$-value.  Since we do not know the value of the damping constant $\gamma$ for these lines, we must supply the routine with a placeholding value for $\gamma$, and cannot use data from our fits to contrain column density or $b$-value from a single line.  However, the equivalent widths are still useful in the curve of growth analysis.

We should note that the utility of these lines is ultimately limited because of the large uncertainties associated with them.  The two major sources of uncertainty are blends with other lines and an uncertain continuum.  The latter source of uncertainty is very important for the four lines below 930 \AA{}.  However, the inclusion of these lines where possible still represents an important aspect of our study.

\subsection{1039 \AA{} Line}
\label{ss:1039line}
The 1039.2304 \AA{} line is the second strongest oxygen line seen in the {\it FUSE} spectra, after the strongest component of the 988 \AA{} triplet (which was not used; see \S\ref{ss:omittedlines}).  Its strength and isolation from other lines make it a staple of this study.  The line is visible in four data segments of {\it FUSE}: LiF1A, LiF2B, SiC1A, and SiC2B.  The LiF1A data segment provided the best data and those data were used in most cases.

The line is found in the redward damping wing of an $\Hmol$ bandhead centered at $\sim1037$ \AA{}.  The line has an oscillator strength of $f=9.07\times10^{-3}$.  The line was usually fit with a Gaussian, but in some cases was broad enough that a Voigt profile was required.  Unlike the lines below 1000 \AA{} we know the damping constant of the line, $\gamma=1.87\times10^{8}$, and therefore Voigt profile fits to this line can potentially be used to extract column density and $b$-value.  However, the values obtained may be poorly constrained, limited by the coupling of column density and $b$-value, the resolution of {\it FUSE}, and the data quality of each individual sightline.  However, measuring the equivalent width remains useful for the curve of growth analysis.

\subsection{1302 \AA{} Line}
\label{ss:1302line}
The 1302.1685 \AA{} line is the strongest line used in this analysis, with an oscillator strength of $f=4.80\times10^{-2}$.  The large damping constant, $\gamma=5.65\times10^{8}$, indicates that this line is on or near the ``square-root region'' of the curve of growth.  In our analysis of {\it IUE} spectra, the 1302 \AA{} line was fit with a Gaussian, as the damping wings are not clearly resolved, and the spectrum is crowded with many other nearby lines (also fit with Gaussians) that makes identification of the continuum more difficult.  A Gaussian fit still provides a reasonable approximation for the equivalent width in cases where saturation is marginally detected in the profile of the line.

For cases in which {\it HST} data were available, the damping wings of the line were evident, and the line was fit with a Voigt profile.  As with the 1039 \AA{} line, Voigt profile fits may provide direct information about the column density and $b$-value, but the strength of the constraints vary.  Additionally, while the 1039 \AA{} line is almost always on the ``flat region'' of the curve of growth, where column density and $b$-value are highly coupled, the 1302 \AA{} line always shows damping wings in the {\it HST} data, and is found in a transition region in the curve of growth where the profile depends strongly on both column density and $b$-value.  As the damping wings become stronger, however, the profile begins to become insensitive to $b$-value.  When available, equivalent widths measured from {\it HST} superceded those from {\it IUE} (except for the case of HD 154368; see \S\ref{ss:IUEdata} and \S\ref{ss:compare}).

In the {\it HST} data, either one or two absorption features are seen in the blue wing of the 1302 \AA{} line.  One of these features appeared consistently and is a singly-ionized phosphorous line; the other line was not seen in all spectra and remains unidentified.  These lines were fit independently with Gaussians and divided out of the spectra before the 1302 \AA{} line was fit.

As with many of the lines in this study, the 1302 \AA{} line is one component of a multiplet, in this case a triplet, but the 1304.8576 \AA{} and 1306.0286 \AA{} components of the triplet are absorption lines from two different excited states.  Thus, the column densities corresponding to the equivalent widths of those lines are the column denisities of the excited states, and would not have been useful in our analysis without an assumption about the excitation mechanism.  These excited lines were observed in the {\it HST} spectra, and may be worth further study that is beyond the scope of this paper.

\subsection{1355 \AA{} Line}
\label{ss:1355line}
The 1355.5977 \AA{} line is the weakest line used in this analysis, with an oscillator strength of $f=1.16\times10^{-6}$.  This line has been used extensively to determine oxygen column densities by itself, either by assuming it contains no unresolved saturated structure; by ignoring the possibility of such structure and merely obtaining a lower limit on column density; or by using information from other elements to aid in determining the component structure of the line.  In our analysis, 1355 \AA{} lines were fit with Gaussians, either in {\it HST} or {\it IUE} data.  In cases where a single Gaussian was a poor fit, multiple Gaussians were fit under the assumption that we were dealing with resolved component structure.  This resolved component structure was only observed in the {\it HST} data.  Resolved component structure in the 1355 \AA{} line may suggest the need for a curve of growth analysis that accounts for more complex velocity structure.  Such analysis is for the most part beyond the scope of this paper, but is discussed briefly in \S\ref{ss:bvalues}.  {\it HST} data superceded {\it IUE} data when available.  For one sightline (HD 154368), we used an equivalent width from \citet{Snow154368}, measured from GHRS spectra.

\subsection{Omitted Lines}
\label{ss:omittedlines}
There are many other neutral oxygen lines that fall in the wavelength range covered by {\it FUSE}.  For example, the triplet at 988 \AA{} has been used in at least least one other study \citep{York}, but in our data, there are a few problems.  The first is that the spectral region surrounding the line contains some of the strongest telluric lines in the wavelength region covered by {\it FUSE} \citep{Feldman}, including OI emission from the 988 \AA{} triplet.  In the spectra with the lowest S/N, this airglow dominates.  In other spectra, the strength of the effect is unclear.

Secondly, the resolution {\it FUSE} ($\sim 15$-$20\kmpers$) dictates that the components of the triplet, with the greatest separation between two components $\sim 36\kmpers$, should be resolved if they are weak.  However, the strongest component of the triplet is comparable in strength to the 1302 \AA{} line, with oscillator strengths similar to within 3.5\%.  Therefore, at typical OI column densities, the line is broadened well beyond a single {\it FUSE} resolution element, and the components of the triplet are blended.  This is in addition to some mild blends with weak $\Hmol$ lines, and a very uncertain continuum due to strong $\Hmol$ bands in the surrounding region.

A third problem is that the damping constant of the components of the triplet are unknown.  Using the Eistein A coefficients found in \citet{Morton03}, some lower limits on $\gamma$ for each component might be determined, but the true value would still be somewhat uncertain.  Given its strength, the equivalent width of the strongest component of the 988 \AA{} triplet should be strongly dependent in many cases on its damping constant (though the spectrum will not necessarily reveal a clearly damped profile).  Thus, a direct profile fit of the blended components of the triplet would be very uncertain.  Equally uncertain would be the point of reference for this line on our curve of growth analysis.  We have therefore elected not to include it in our results.

Lines at 924.950 \AA{} and 937.8405 \AA{} were also detected, but strong $\Hmol$ lines make the surrounding continuum impossible to measure.  The latter line may also be suject to contamination from an FeII line, but the FeII line should be much weaker given typical relative column densities of FeII and OI.

In addition to the lines just discussed, Table \ref{nolinetable} lists the remaining OI lines with wavelengths in the {\it FUSE} region and the reasons they are not used in our study.  The reasons for omitting the remaining lines is that they are all either too weak to be detected, are too heavily blended with other atomic or molecular lines, or are completely wiped out by very strong atomic or molecular lines.  There are no other UV absorption lines of oxygen in the range covered by either {\it HST} or {\it IUE}.  Since we do not expect to see strong absorption lines from excited states (such as the ``forbidden'' 6300 \AA{} transition) and were they detectable we would need an assumption on relative populations to use excited states in constraining total column density, we do not consider lines outside of the ultraviolet region.

\subsection{Errors}
\label{ss:errors}
Errors on equivalent width measurements take two forms.  When absorption lines are fit with a Gaussian, errors are taken from standard error propagation of the curve-fitting routine and the functional form of a Gaussian.  When absorption lines are fit to a Voigt profile, errors are not clear due to the lack of a functional form (the Voigt profile is fit iteratively and involves convolution with the instrumental resolution element).  Therefore, we estimate 1-$\sigma$ errors by first taking the difference between the fit and the data (both normalized) at each point.  The uncertainty is then estimated to be twice the standard deviation of this difference, multiplied by the wavelength range of the fit.  This is a very conservative estimate of the error, but is adopted for lack of a more rigorous method.

\subsection{Upper Limits}
\label{ss:limits}
We attempted to obtain an upper limit on the equivalent width of these lines for cases where the line was not detected.  This is done by the relationship $W_{\lambda, {\rm max}}=\Delta \lambda N_{\sigma} / (S/N)$, where $\Delta\lambda$ is the FWHM of a weak, unresolved line (i.e., the resolution element of the instrument), $N_{\sigma}$ is the $N$-$\sigma$ confidence level desired, and S/N is the signal-to-noise (the average of the continuum divided by its standard deviation).  Upper limits calculated in this manner assume that the line is unsaturated (i.e. on the linear part of the curve of growth), which may not necessarily be the case.

\begin{figure}[t!]
\begin{center}
\epsscale{1.00}
\plottwo{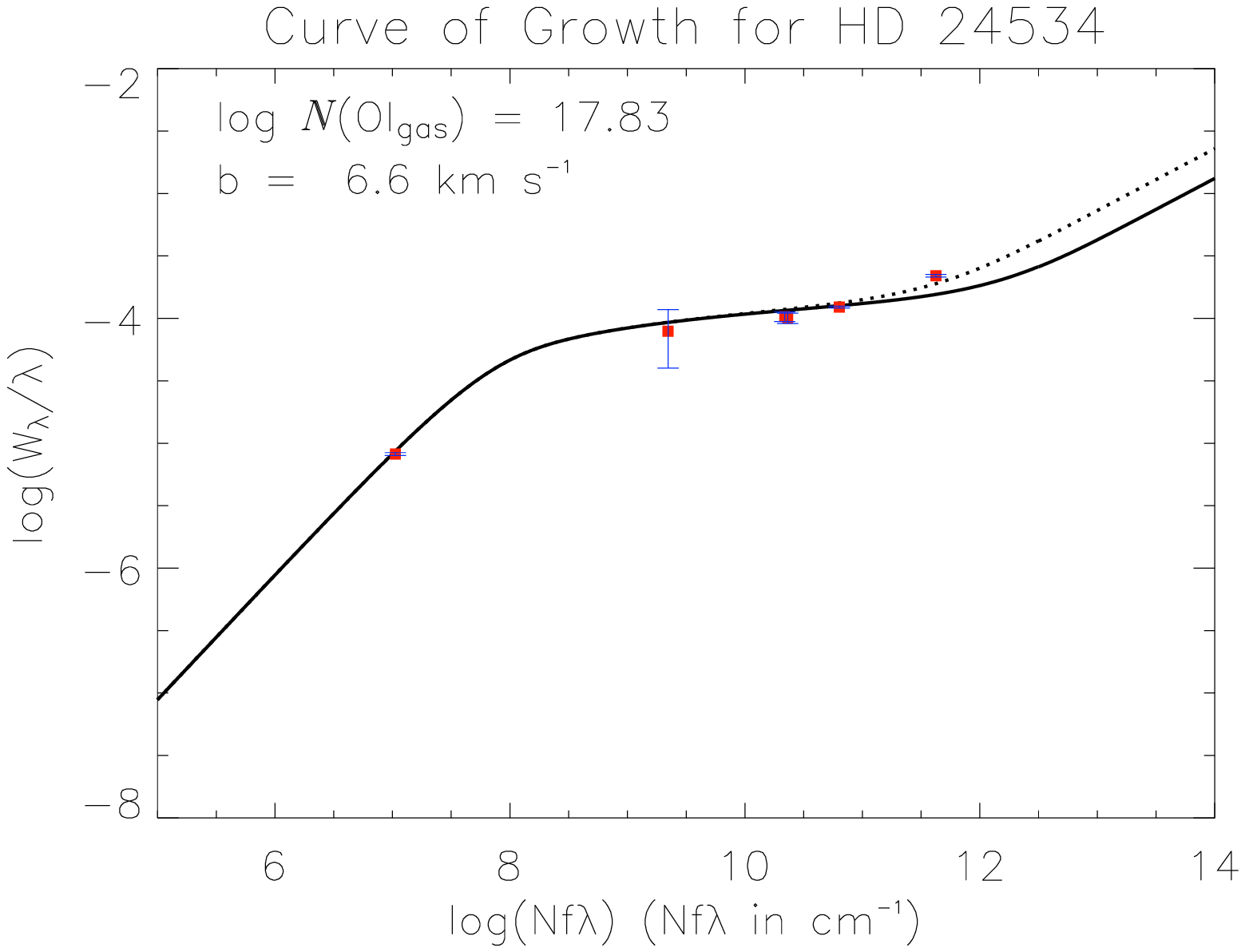}{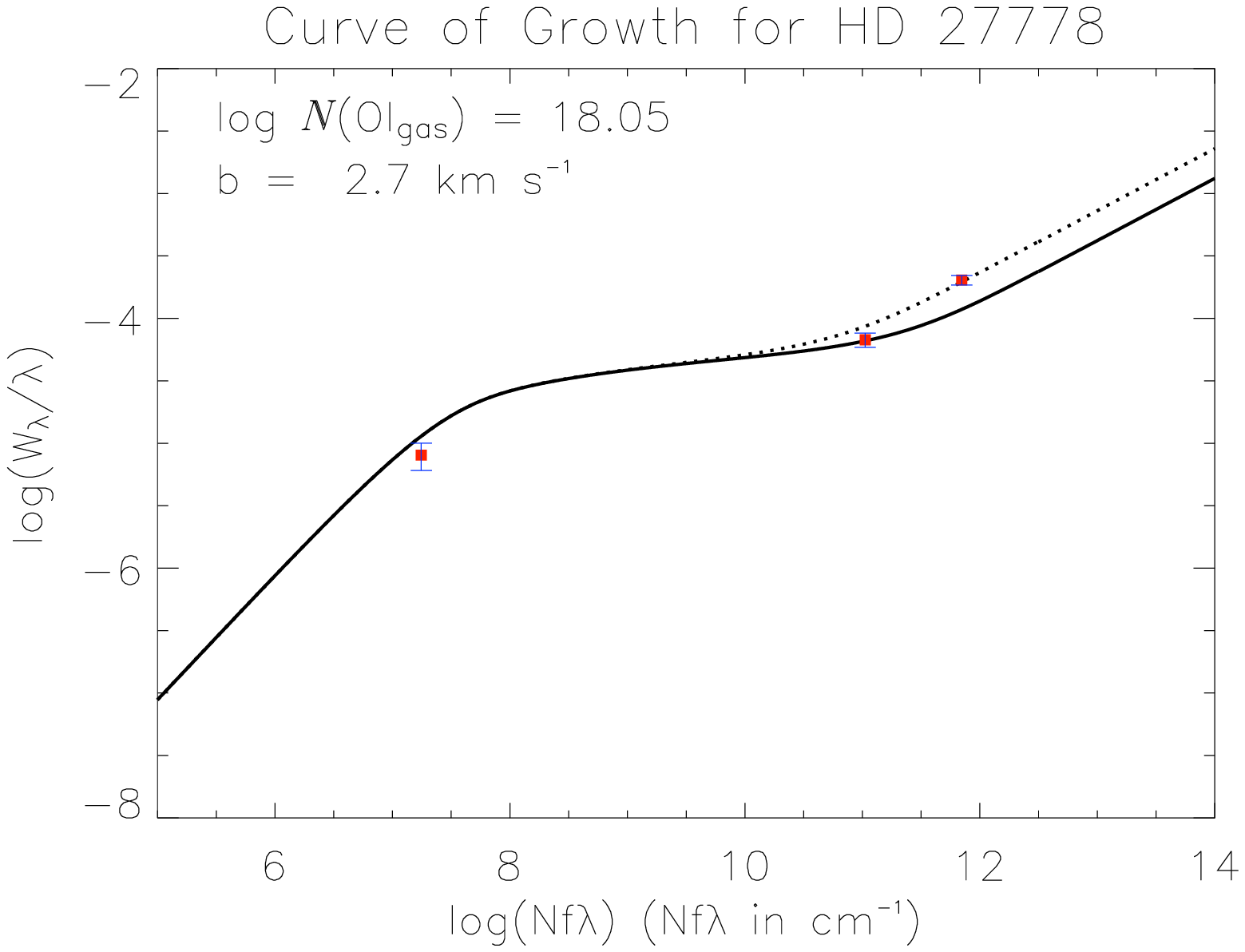}
\end{center}
\begin{center}
\epsscale{1.00}
\plottwo{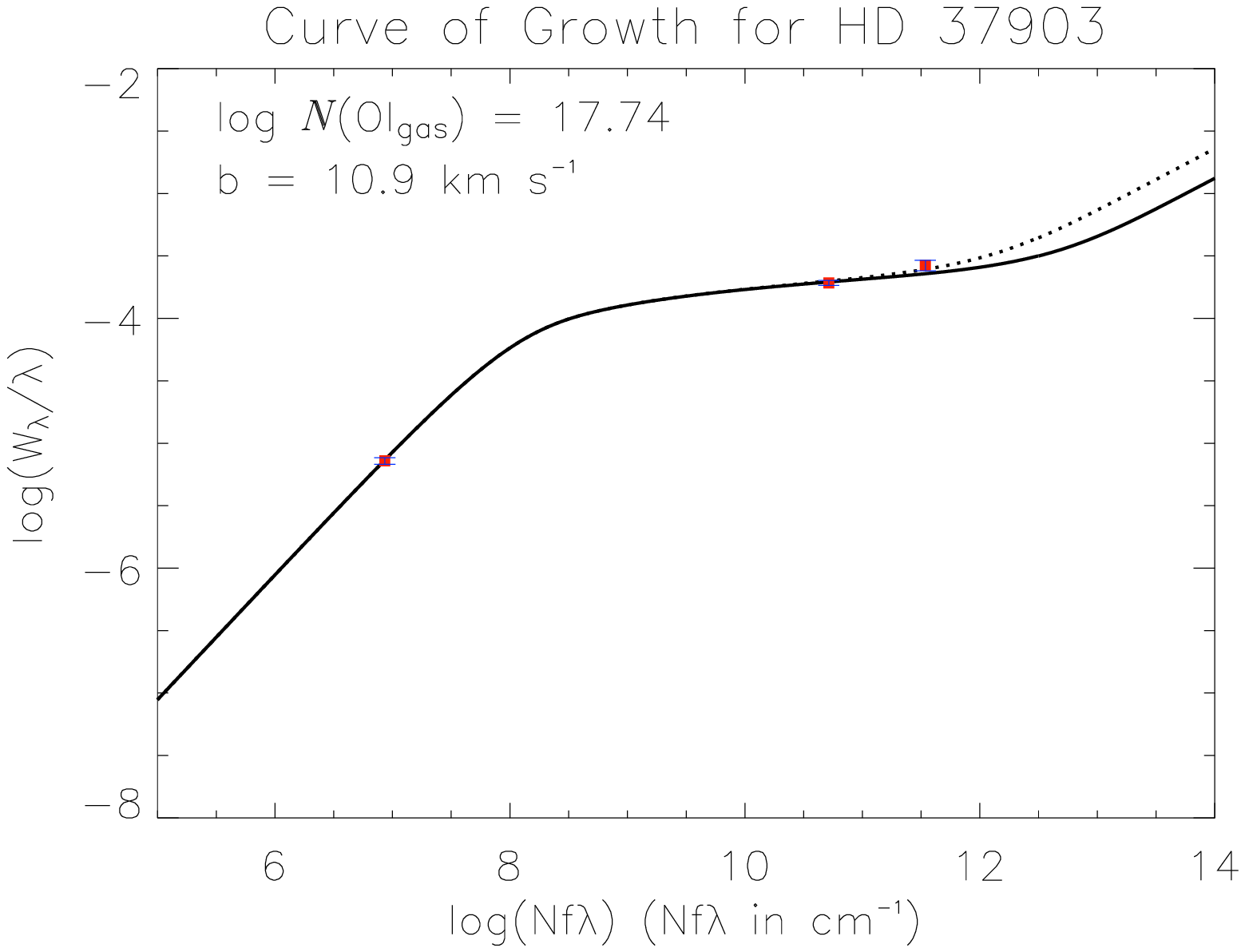}{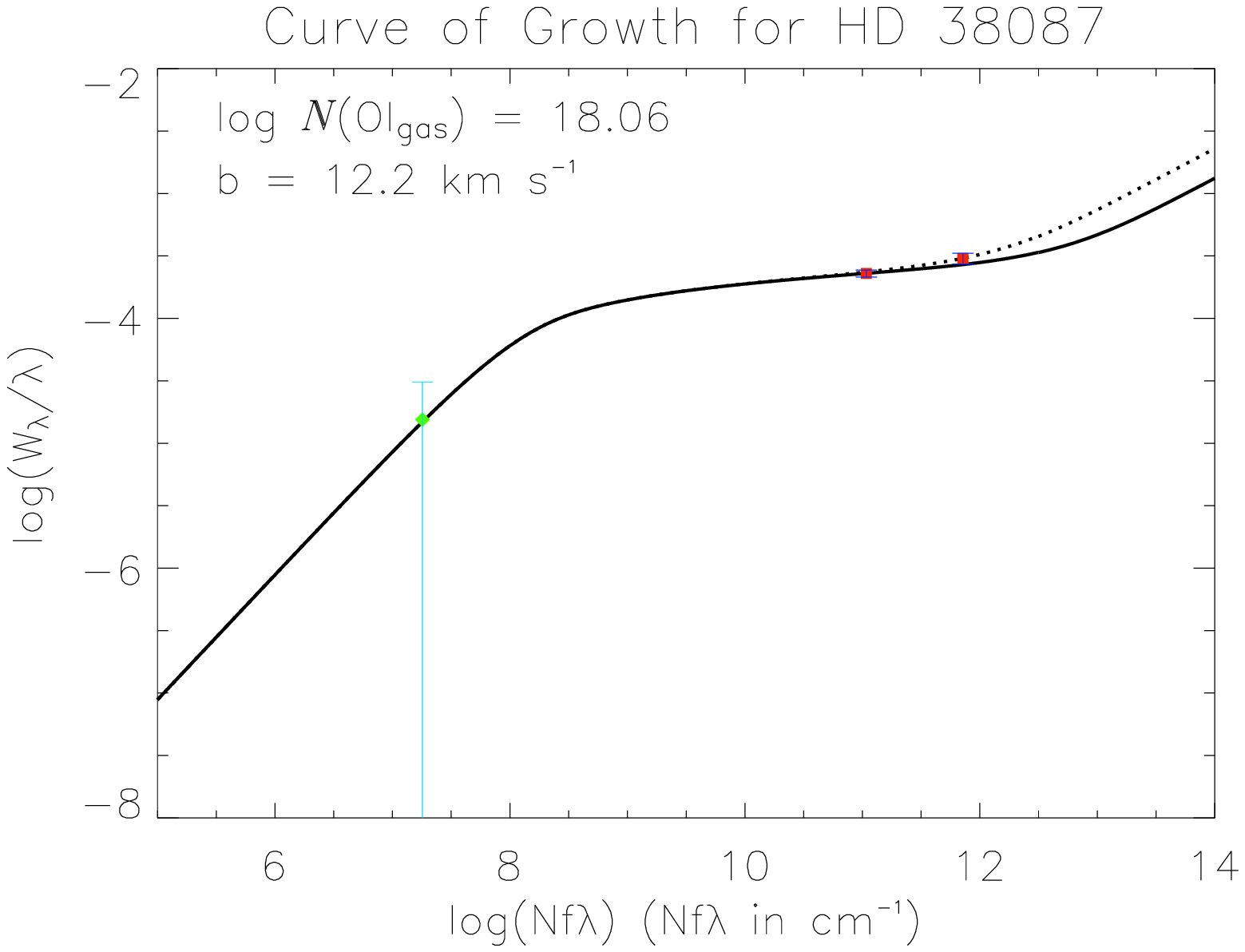}
\end{center}
\begin{center}
\epsscale{1.00}
\plottwo{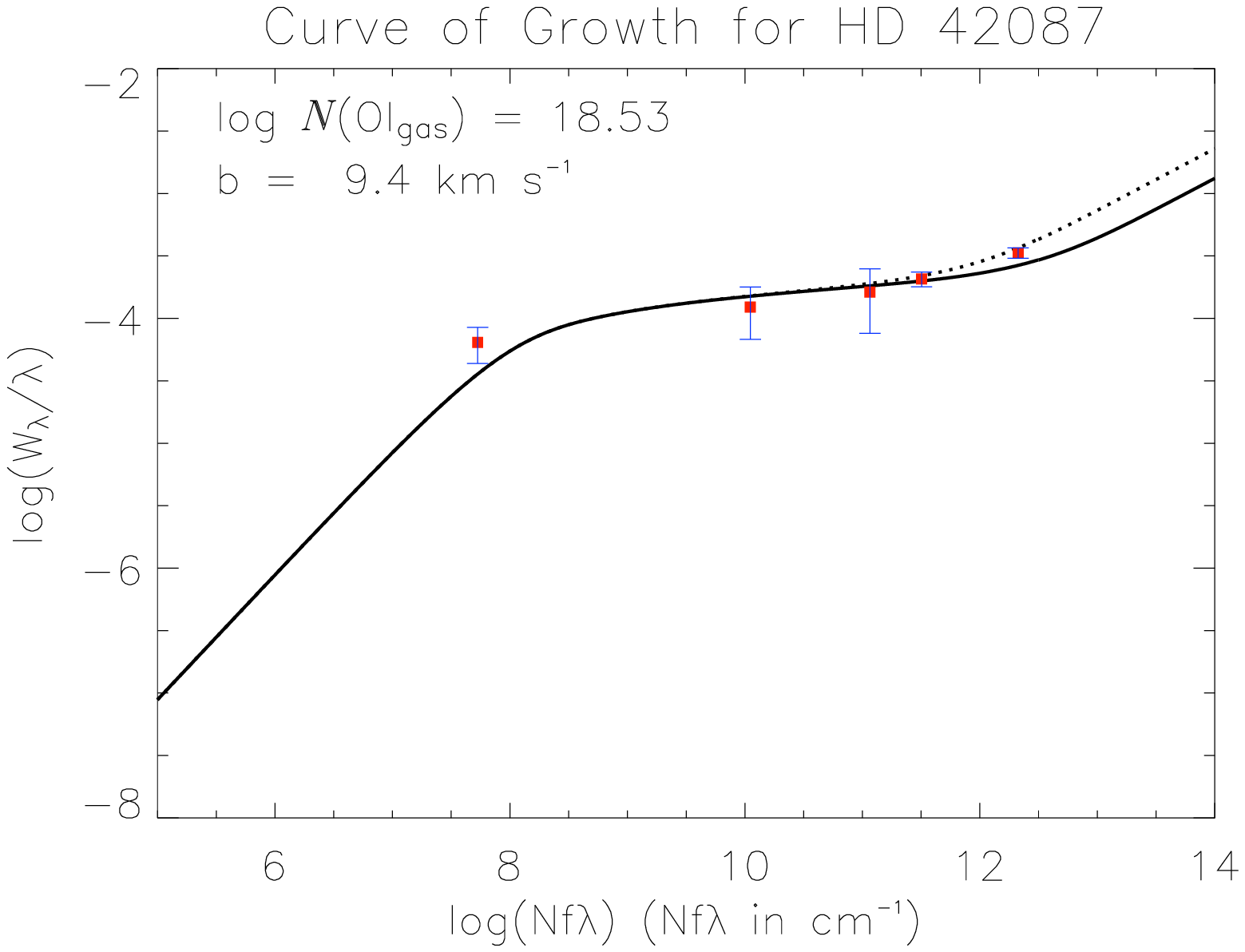}{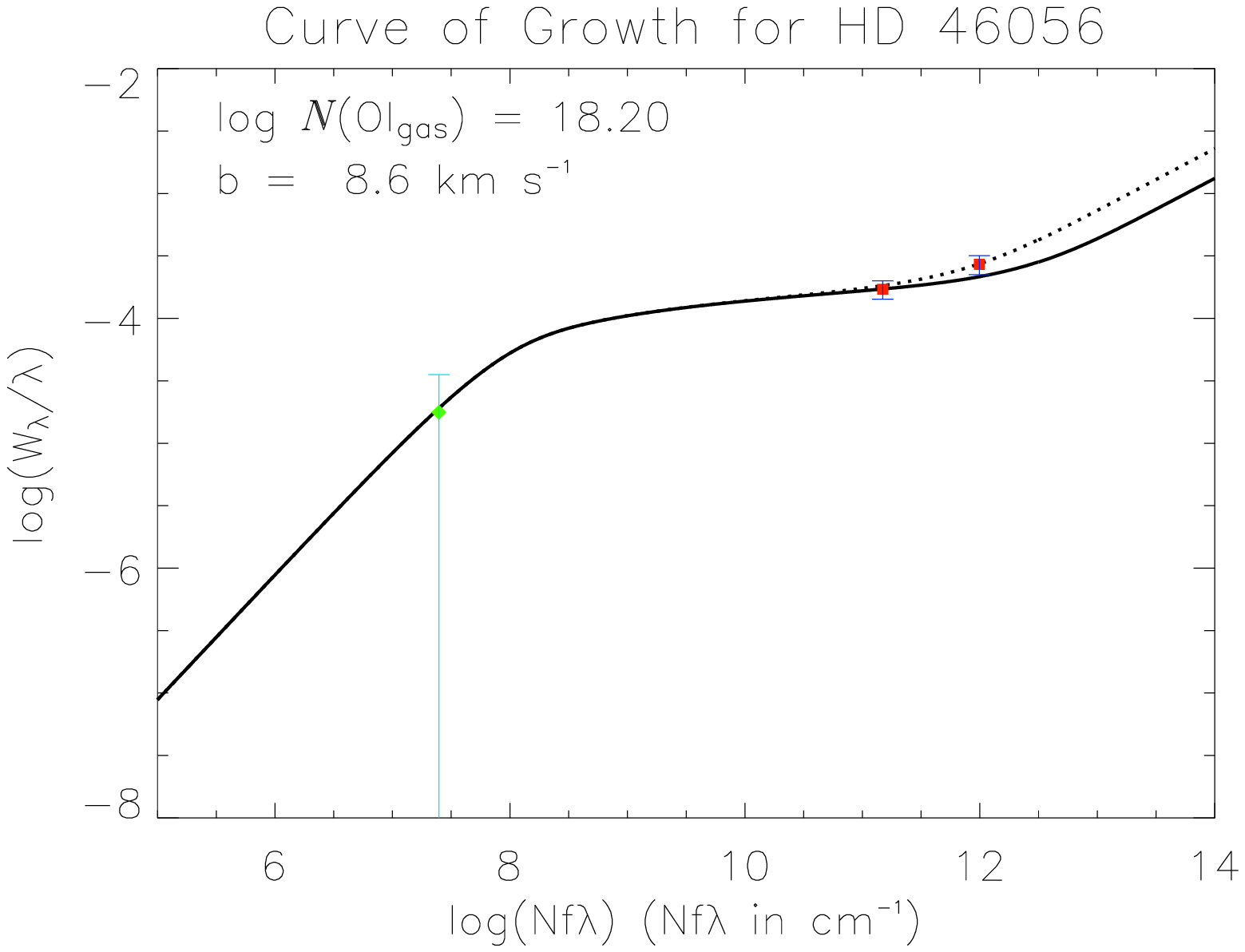}
\end{center}
\caption{Curves of growth for HD 24534, HD 27778, HD 37903, HD 38087, HD 42087, and HD 46056.  Equivalent width fits are shown as red squares with dark blue error bars; 1-$\sigma$ upper limits on the 1355 \AA{} are shown as green diamonds, with the upper limit on the light blue error bars representing a 2-$\sigma$ upper limit.  The curves of growth for HD 38087 and HD 46056 have only two equivalent widths each, and the figures represent the adopted solution from two potential solutions (see \S\ref{ss:twopointcogs} and Table \ref{twopointcogs}).}
\label{fig:cogs1-6}
\end{figure}

\begin{figure}[t!]
\begin{center}
\epsscale{1.00}
\plottwo{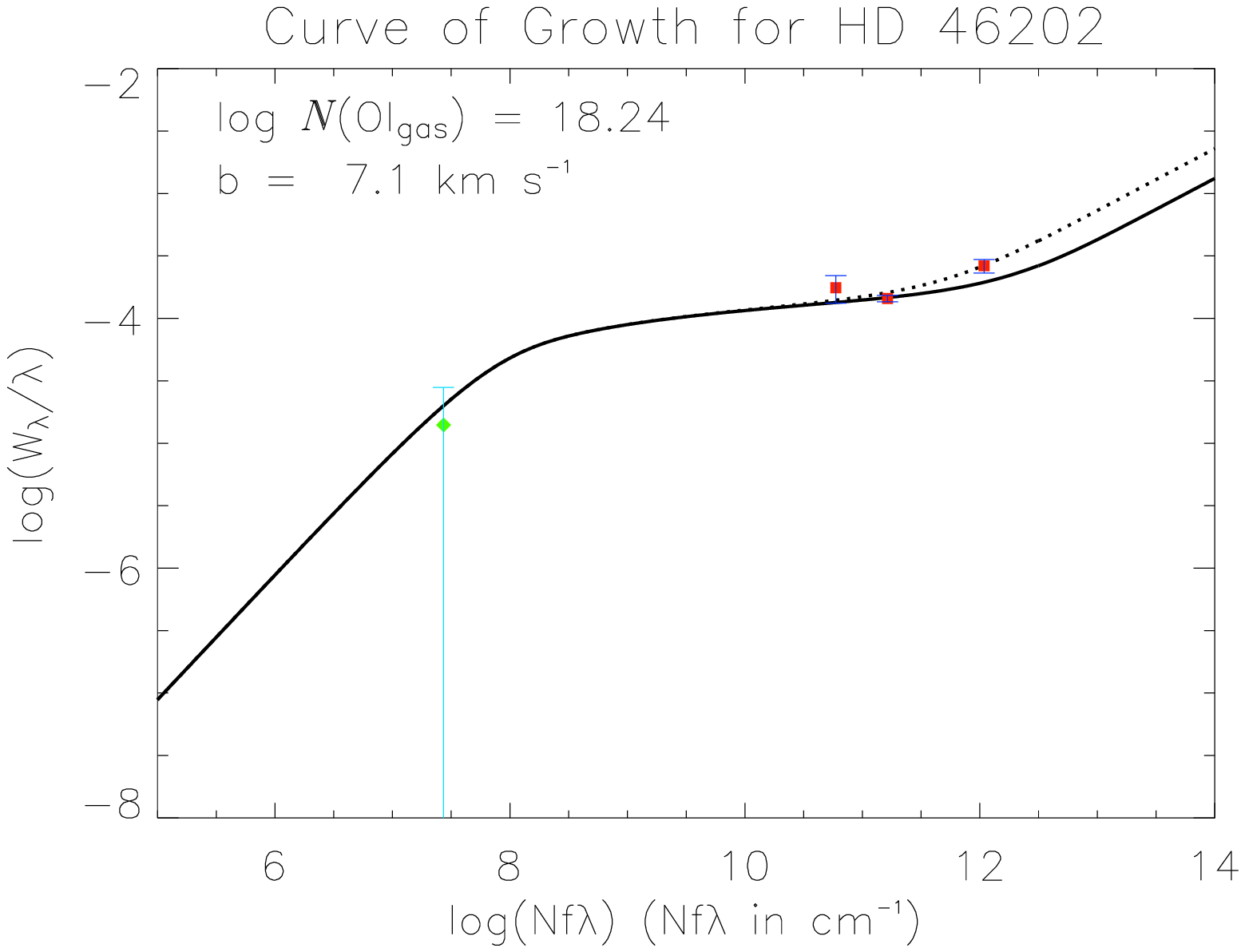}{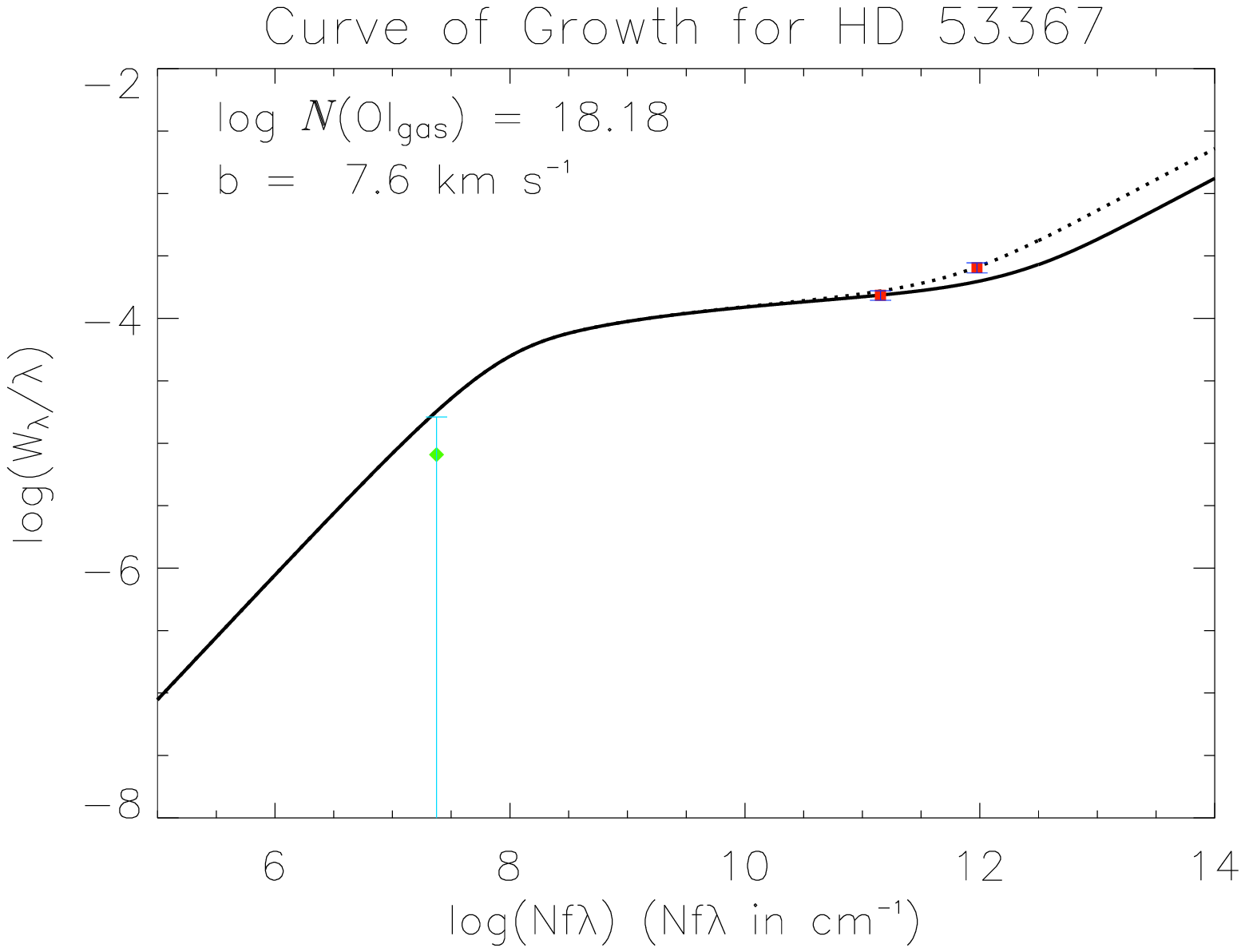}
\end{center}
\begin{center}
\epsscale{1.00}
\plottwo{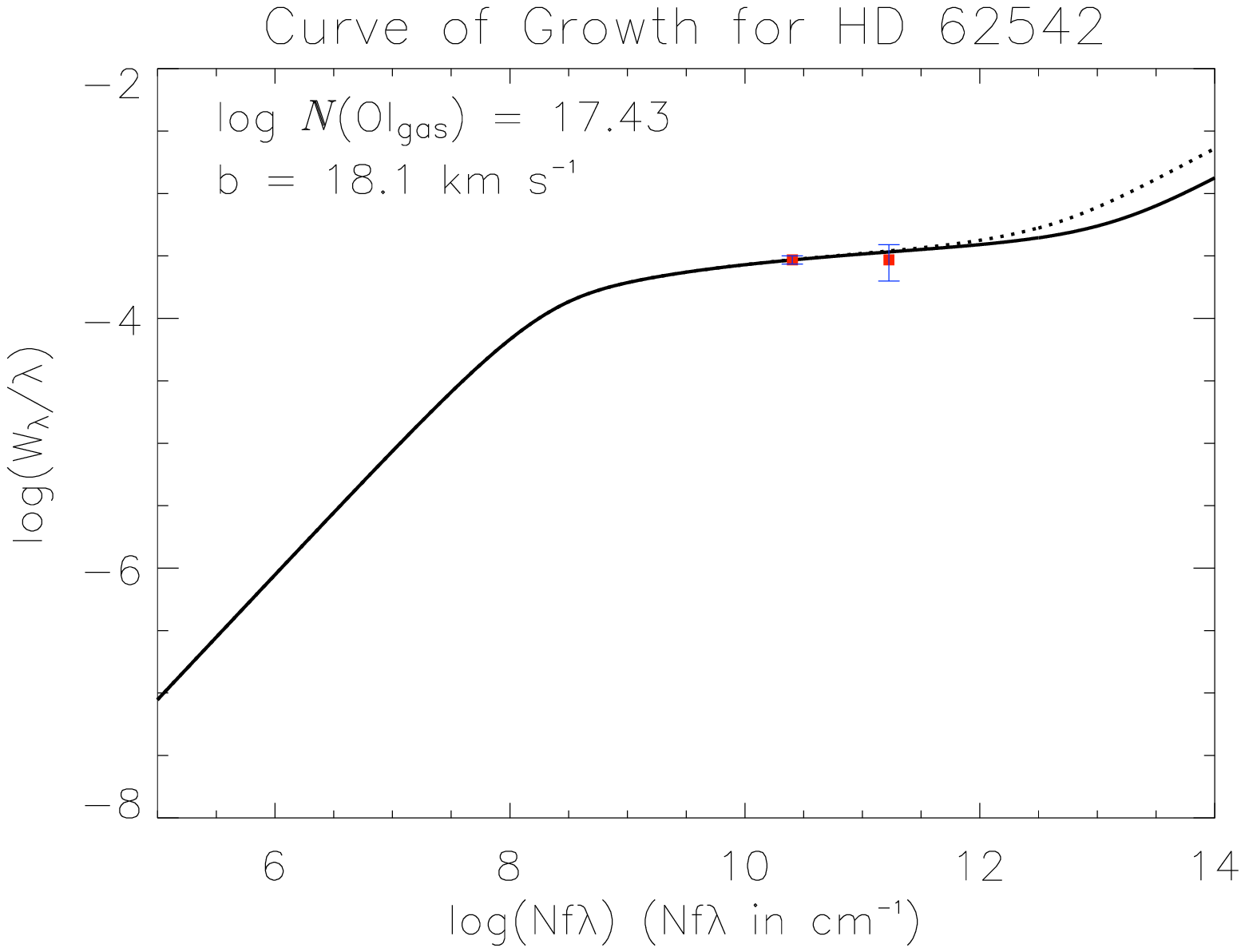}{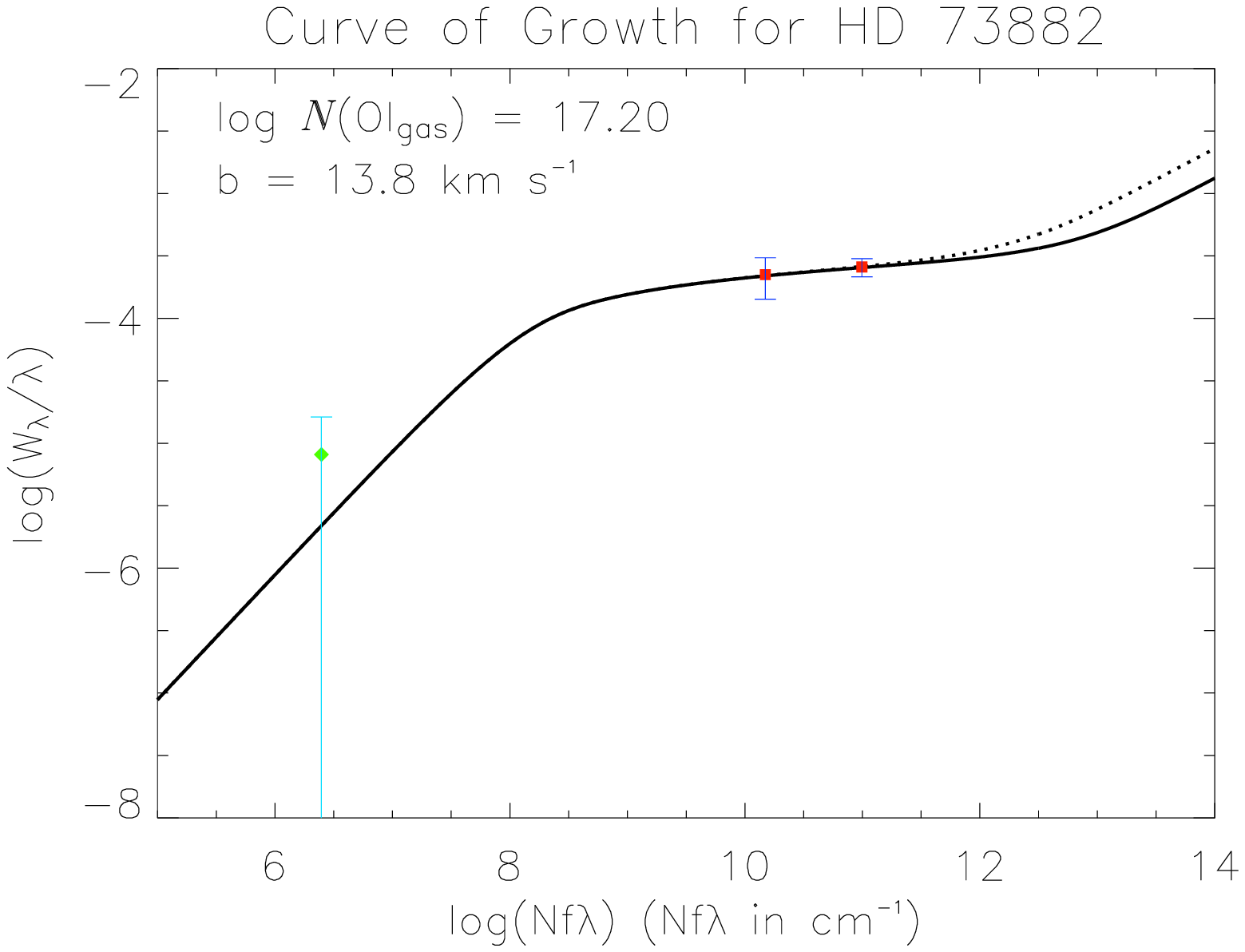}
\end{center}
\begin{center}
\epsscale{1.00}
\plottwo{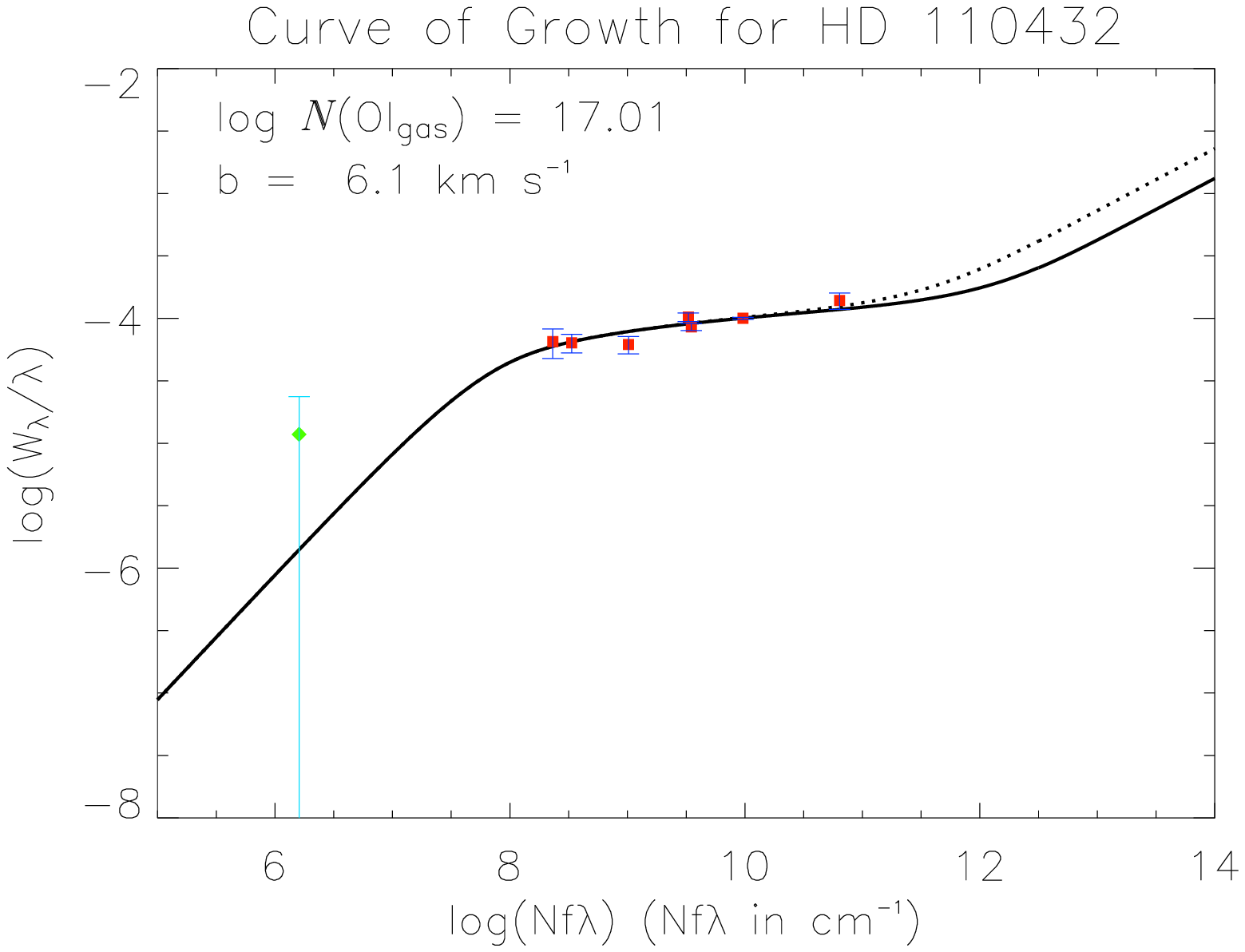}{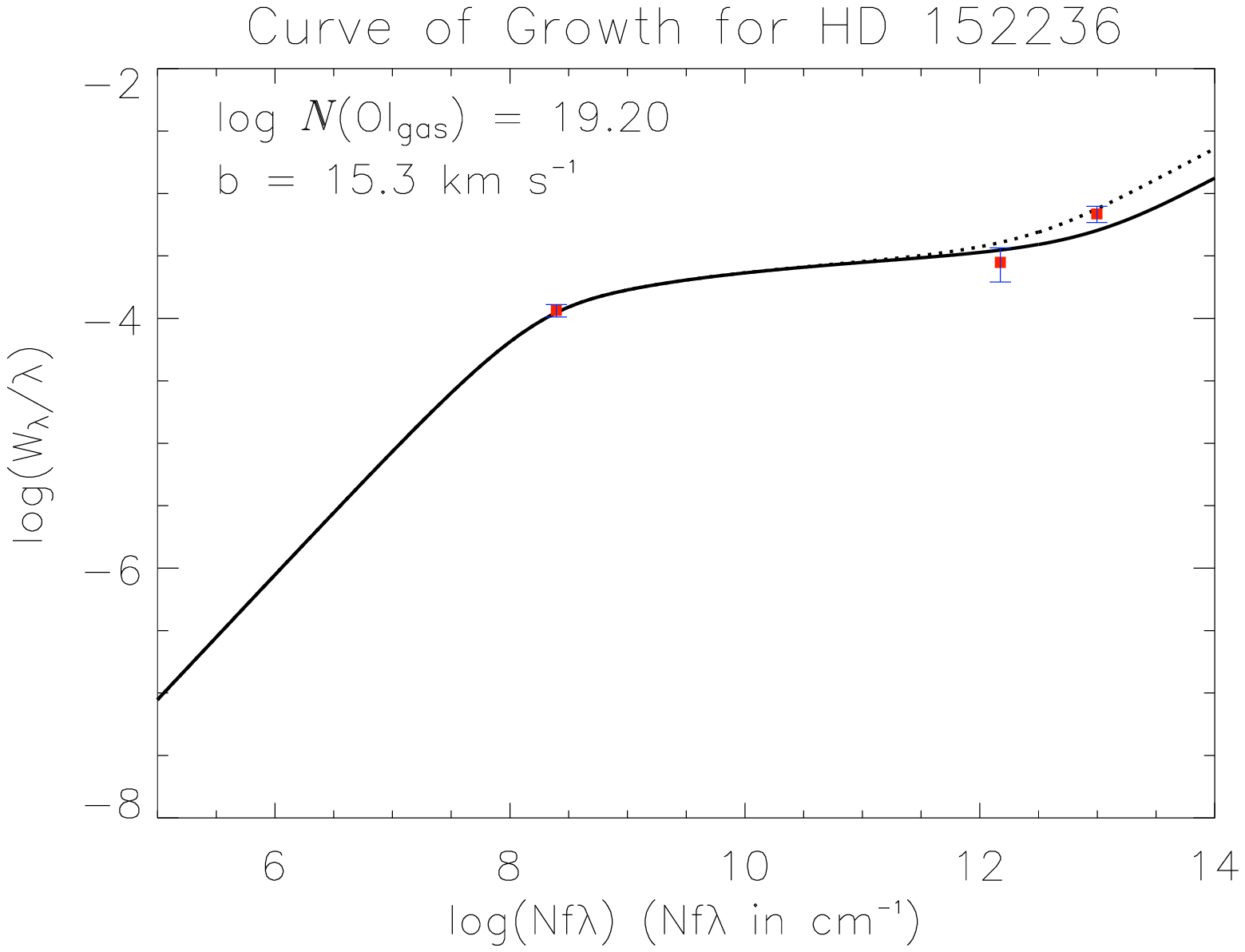}
\end{center}
\caption{Curves of growth for HD 46202, HD 53367, HD 62542, HD 73882, HD 110432, and 152236.  Equivalent width fits are shown as red squares with dark blue error bars; 1-$\sigma$ upper limits on the 1355 \AA{} line are shown as green diamonds, with the upper limit on the light blue error bars representing a 2-$\sigma$ upper limit.  The curves of growth for HD 53367, HD 62542, and HD 73882 each have only two equivalent widths.  For HD 53367 the figure represents the adopted solution from two potential solutions (see \S\ref{ss:twopointcogs} and Table \ref{twopointcogs}).}
\label{fig:cogs7-12}
\end{figure}

\begin{figure}[t!]
\begin{center}
\epsscale{1.00}
\plottwo{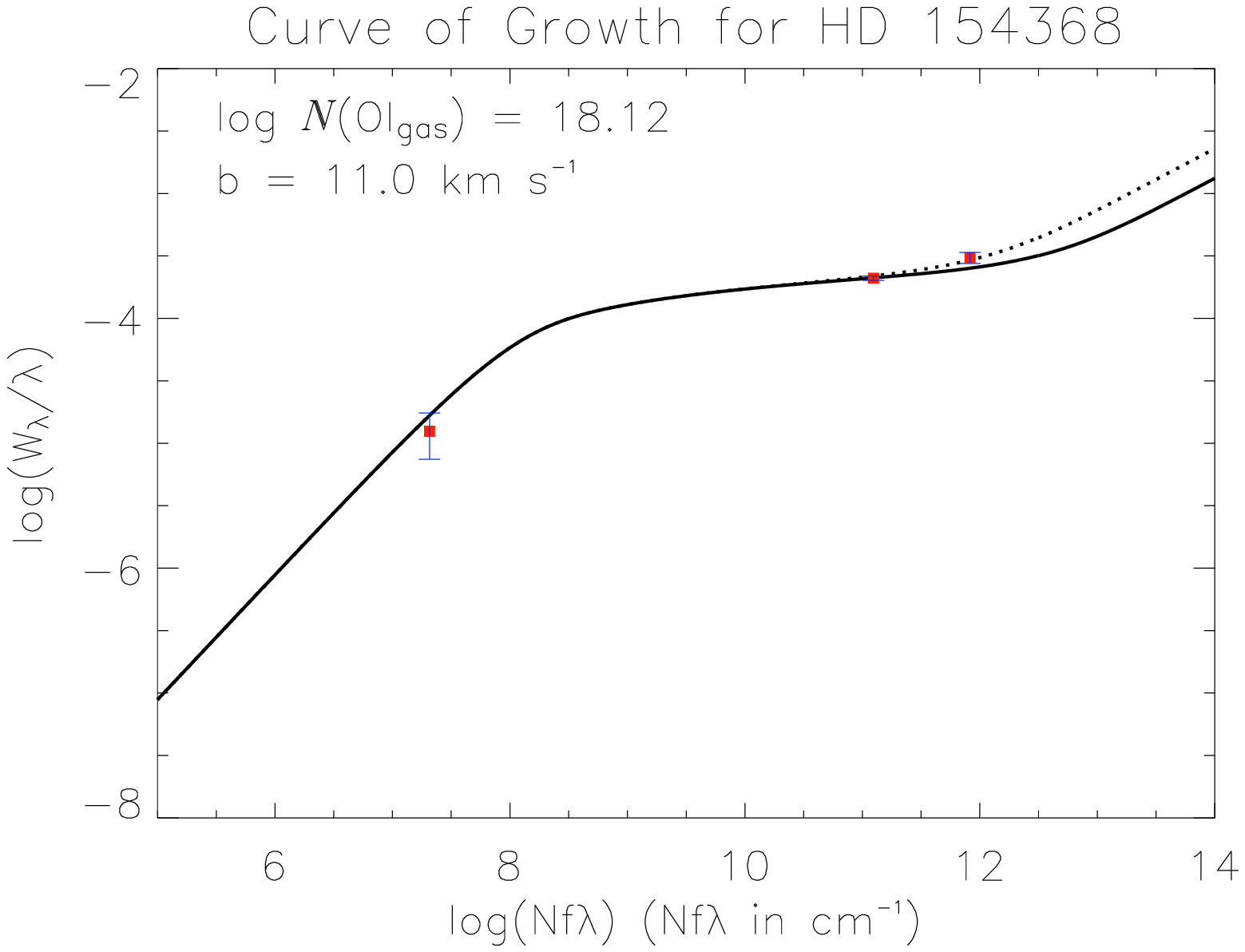}{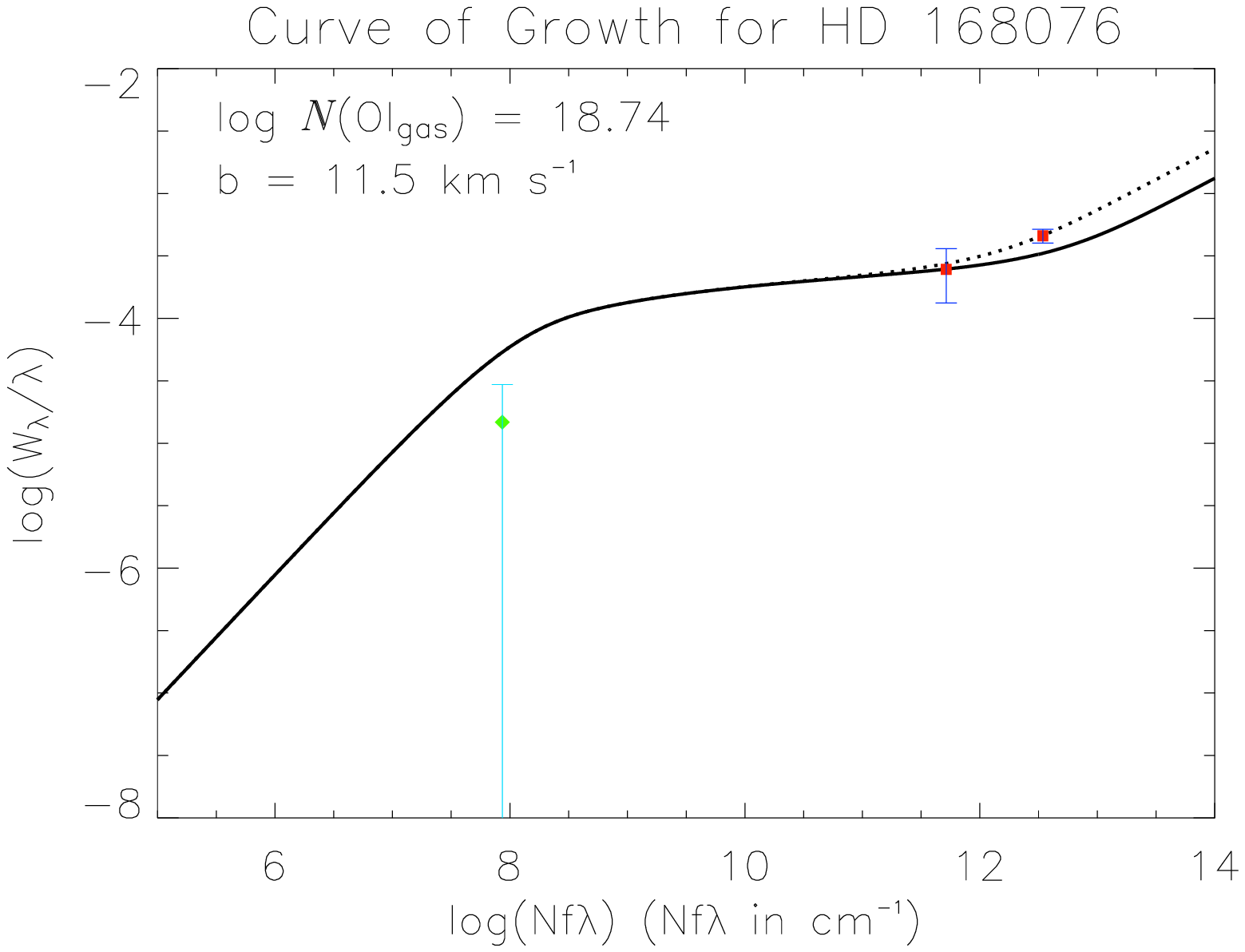}
\end{center}
\begin{center}
\epsscale{1.00}
\plottwo{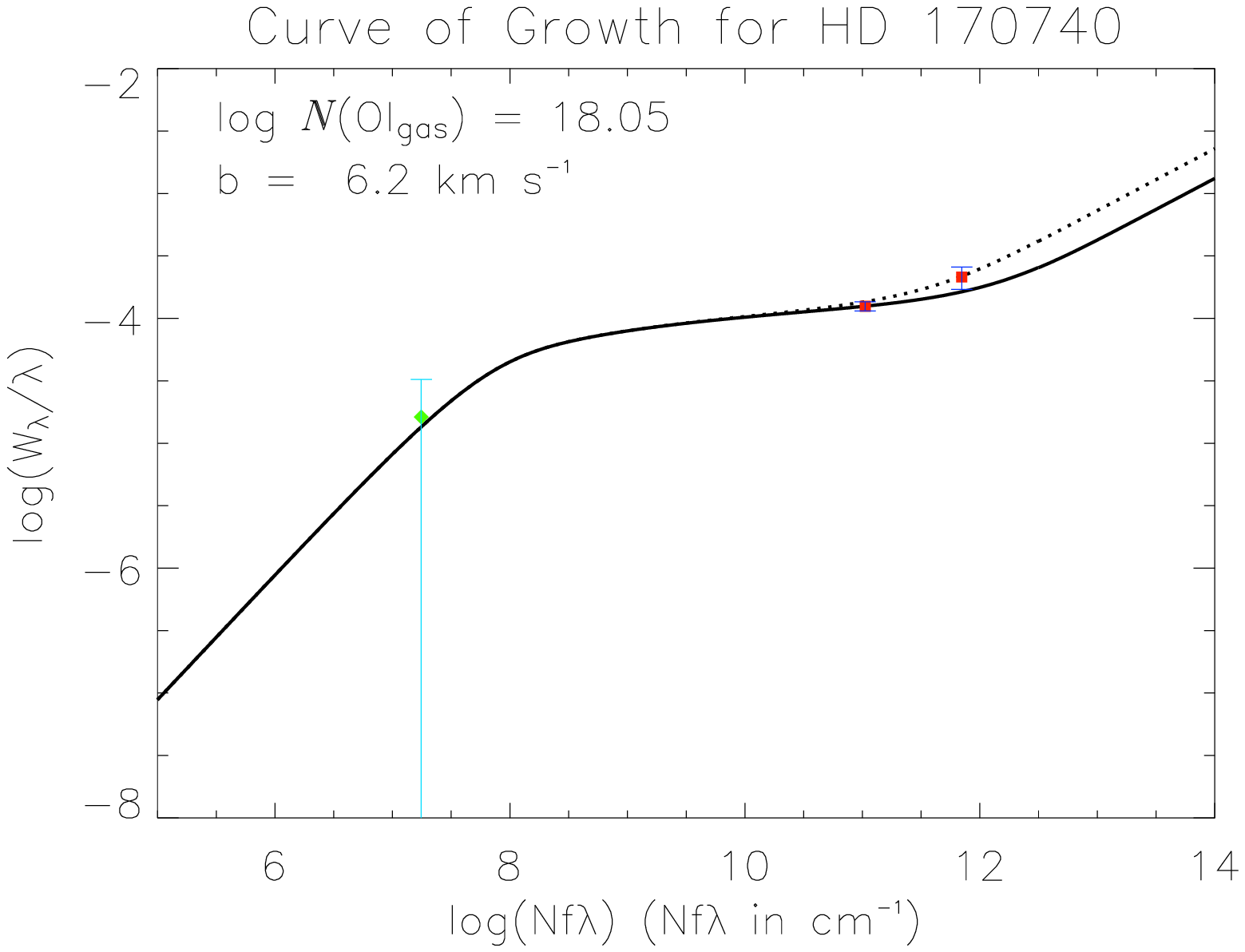}{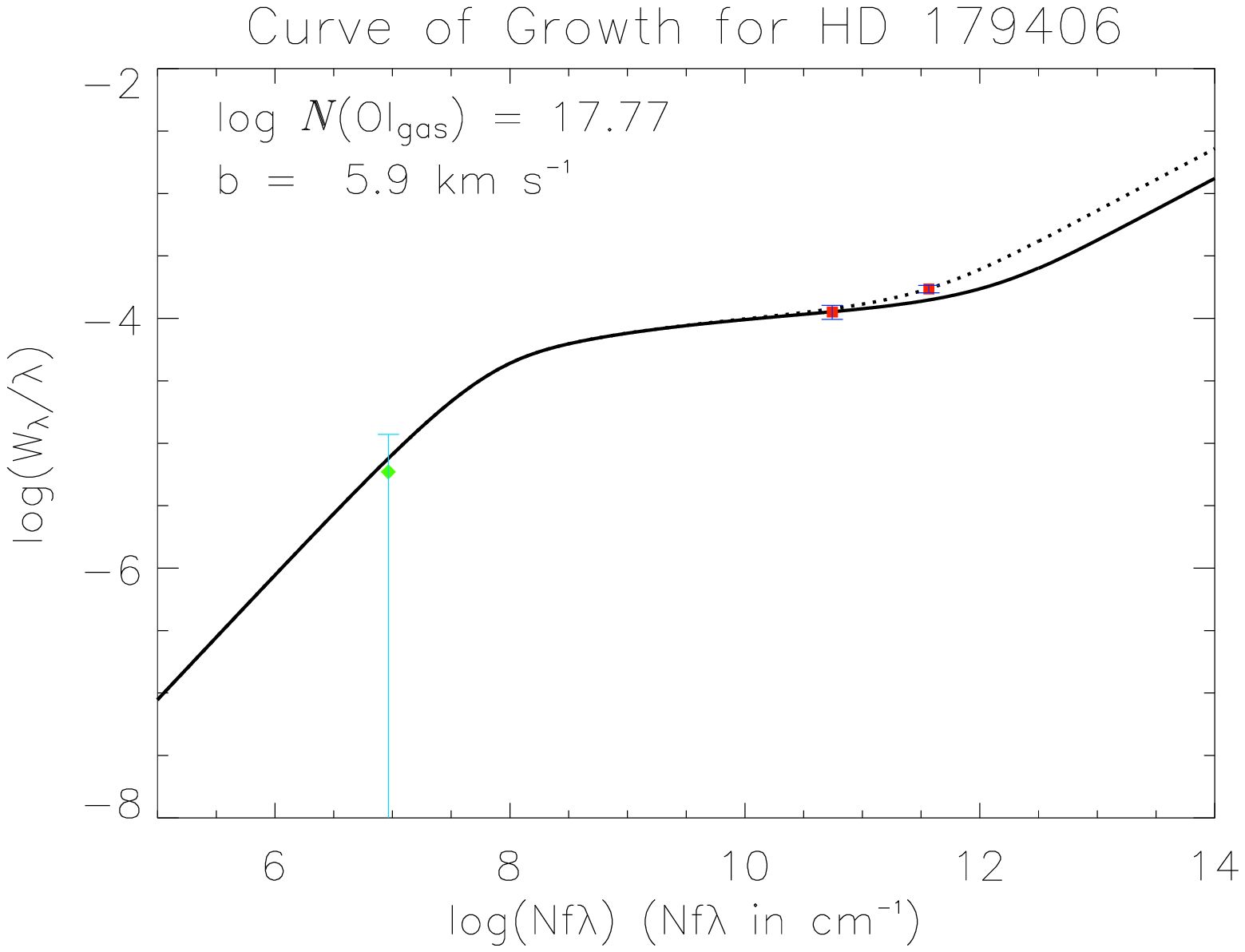}
\end{center}
\begin{center}
\epsscale{1.00}
\plottwo{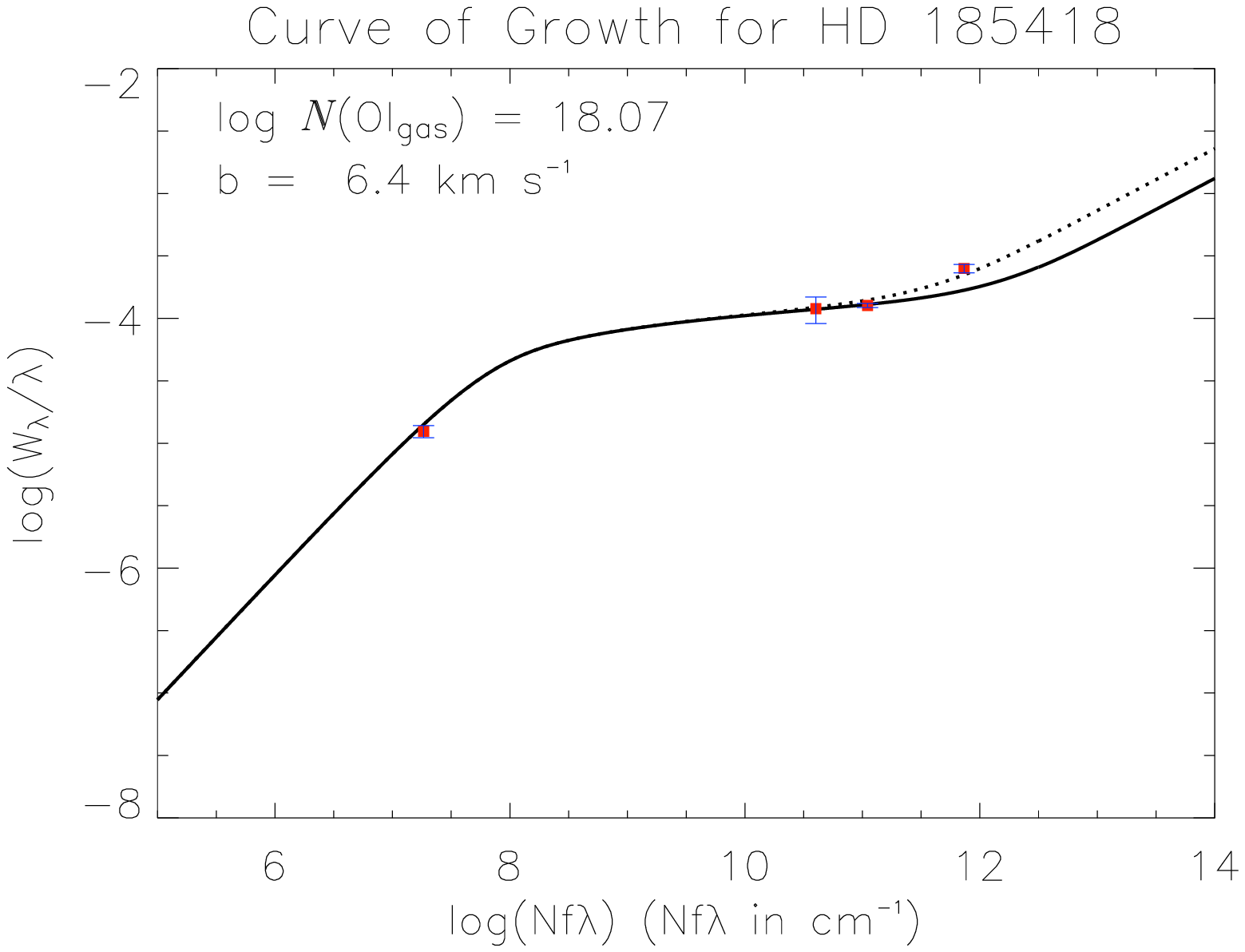}{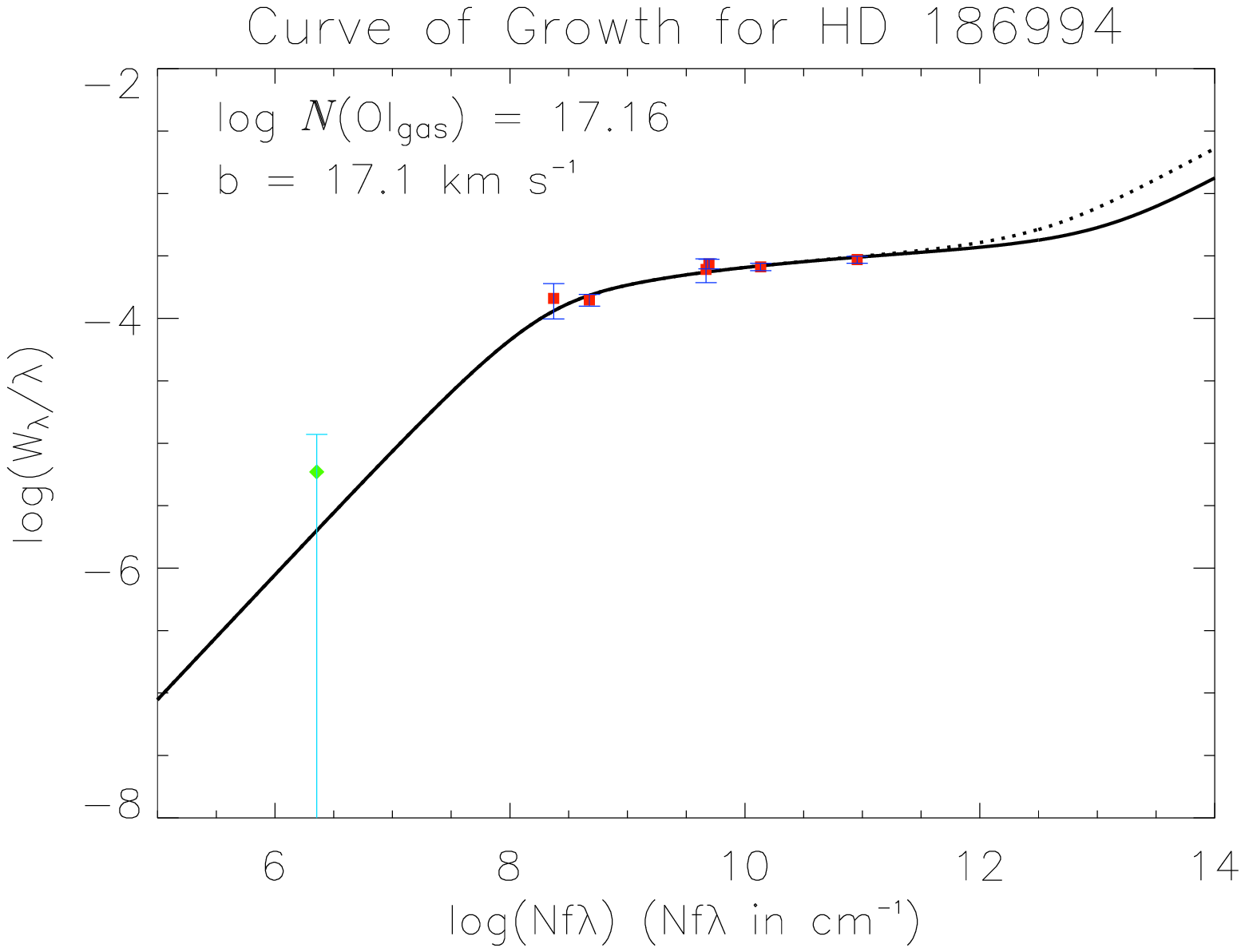}
\end{center}
\caption{Curves of growth for HD 154368, HD 168076, HD 170740, HD 179406, HD 185418, and HD 186994.  Equivalent width fits are shown as red squares with dark blue error bars; 1-$\sigma$ upper limits on the 1355 \AA{} line are shown as green diamonds, with the upper limit on the light blue error bars representing a 2-$\sigma$ upper limit.  The curves of growth for HD 168076, HD 170740, and HD 179406 each have only two equivalent widths.  For HD 170740 and HD 179406 the figures represent the adopted solution from two potential solutions (see \S\ref{ss:twopointcogs} and Table \ref{twopointcogs}).}
\label{fig:cogs13-18}
\end{figure}

\begin{figure}[t!]
\begin{center}
\epsscale{1.00}
\plottwo{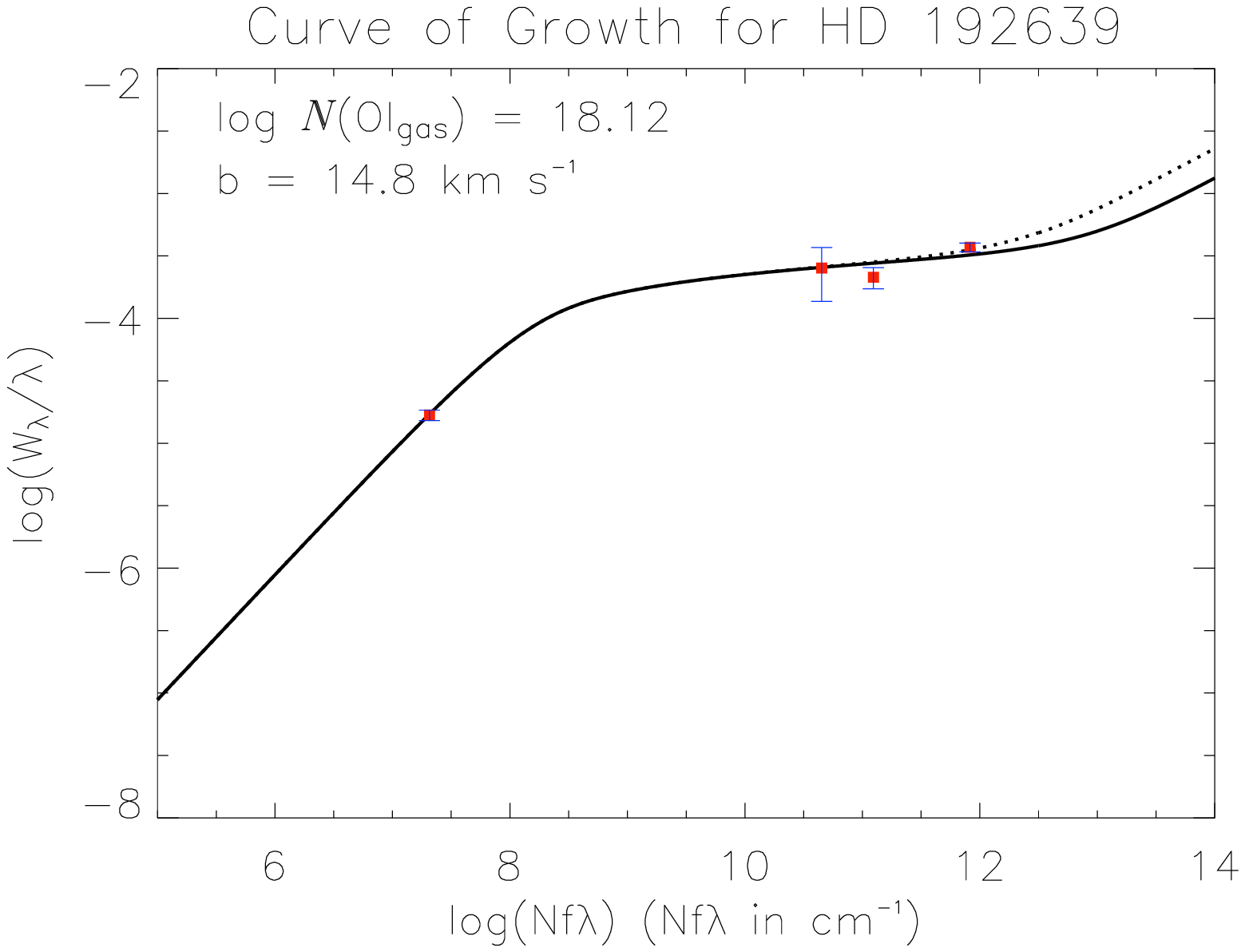}{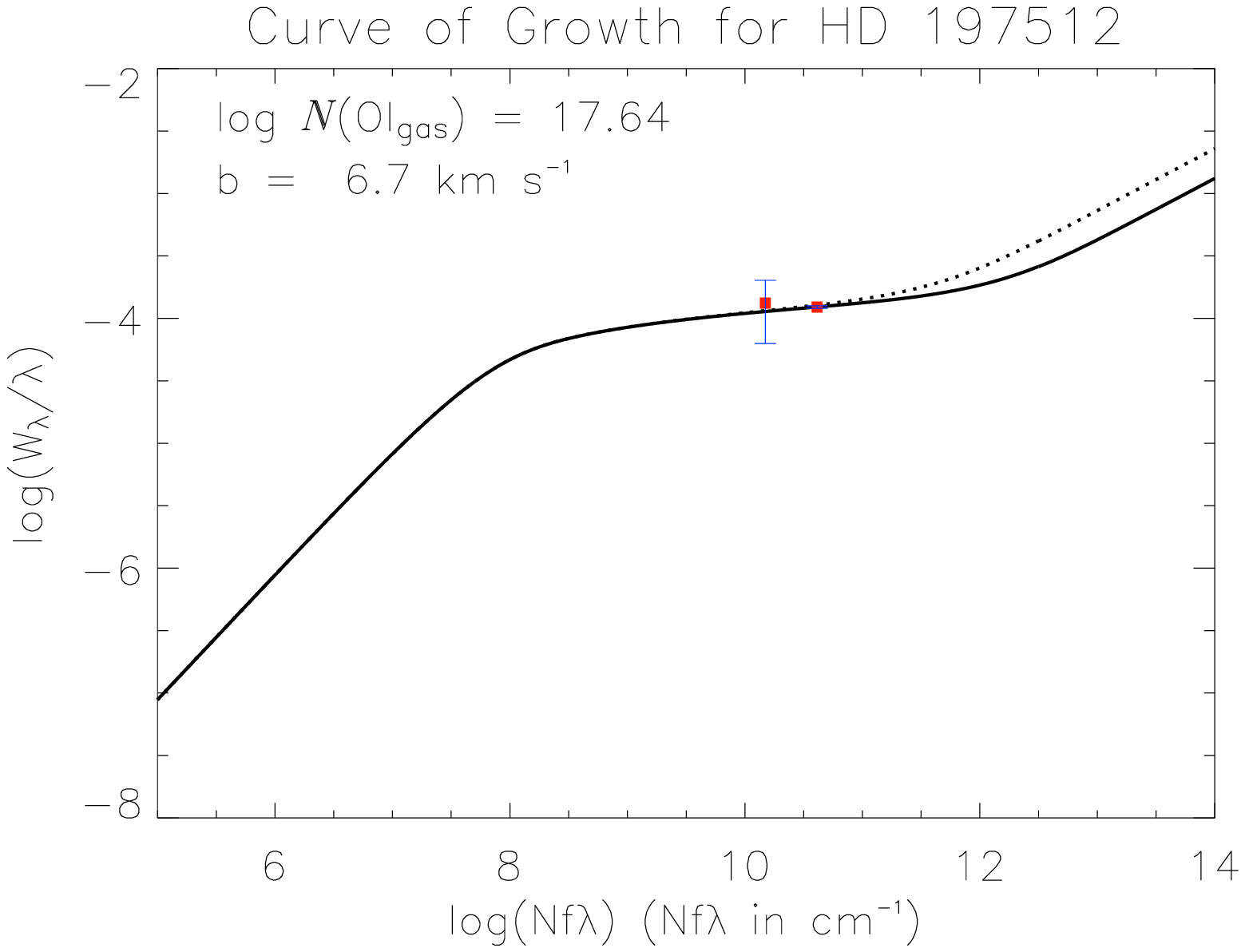}
\end{center}
\begin{center}
\epsscale{1.00}
\plottwo{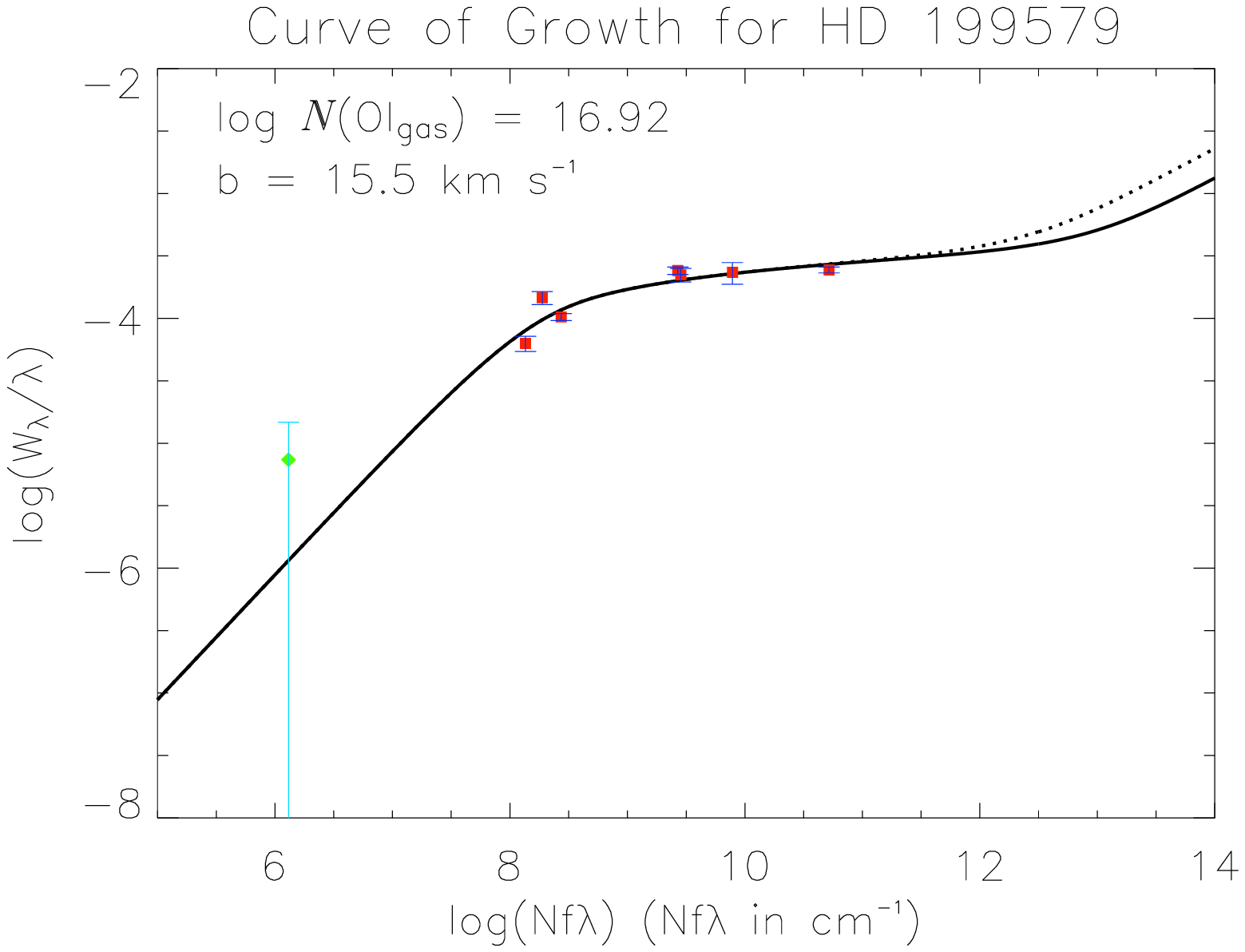}{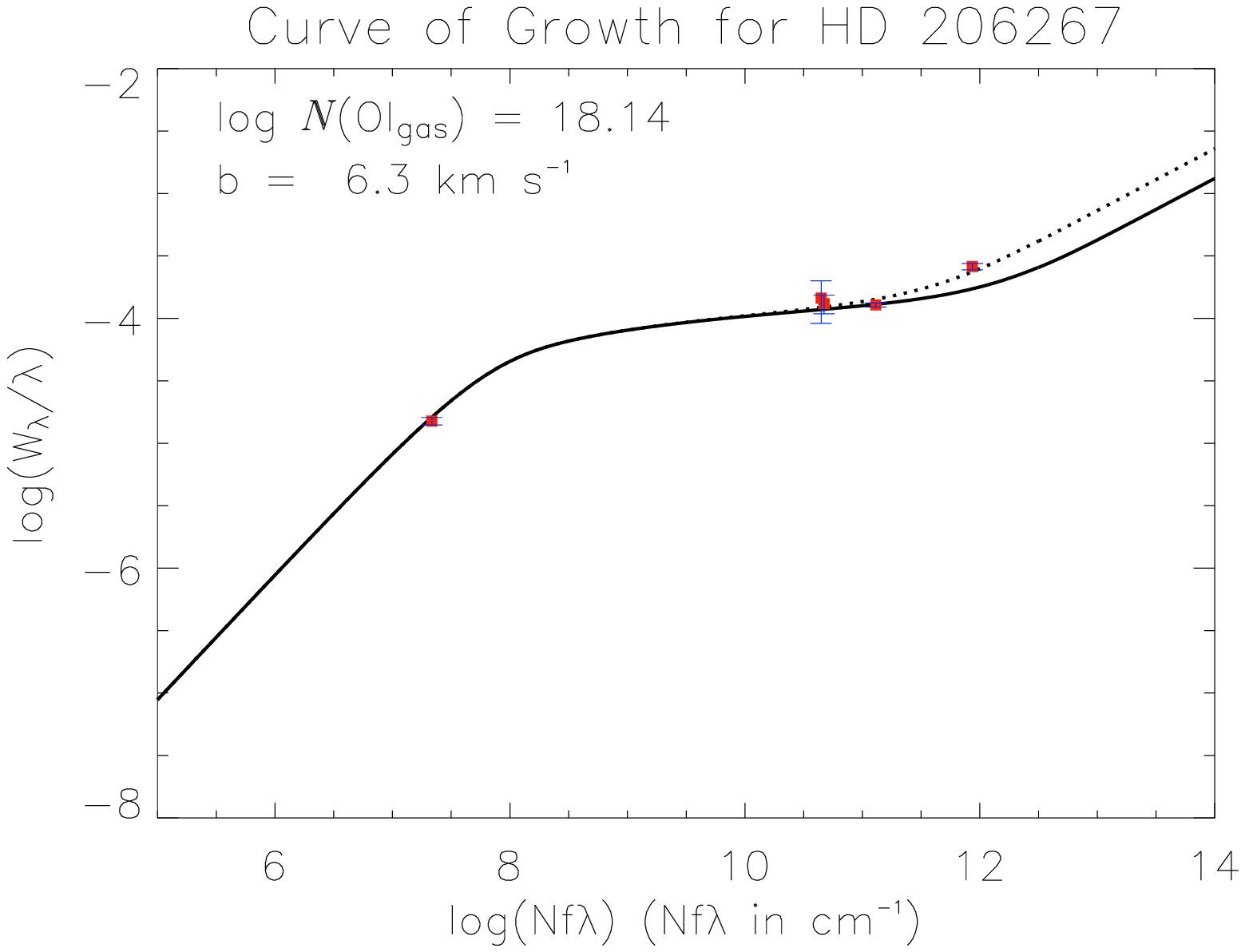}
\end{center}
\begin{center}
\epsscale{1.00}
\plottwo{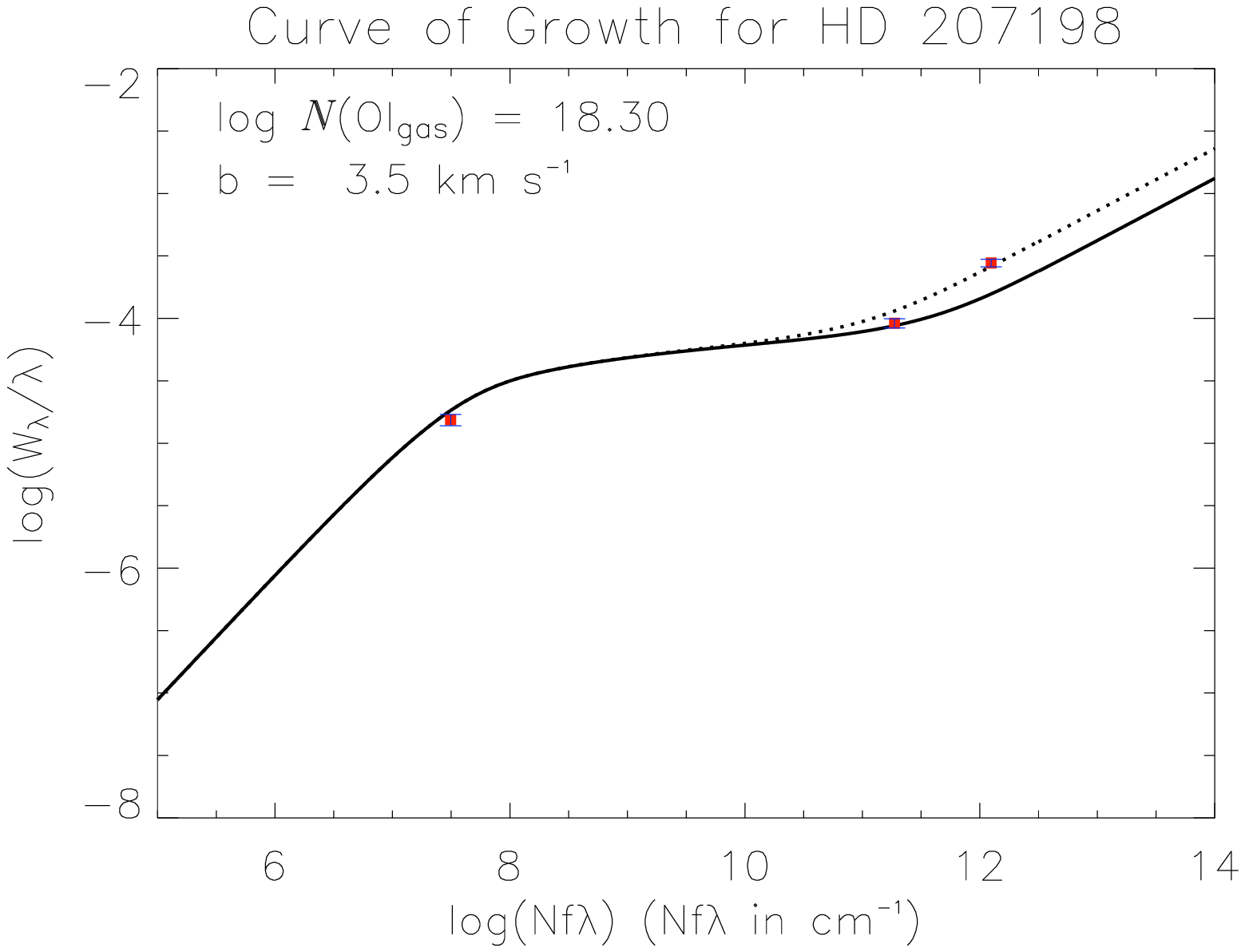}{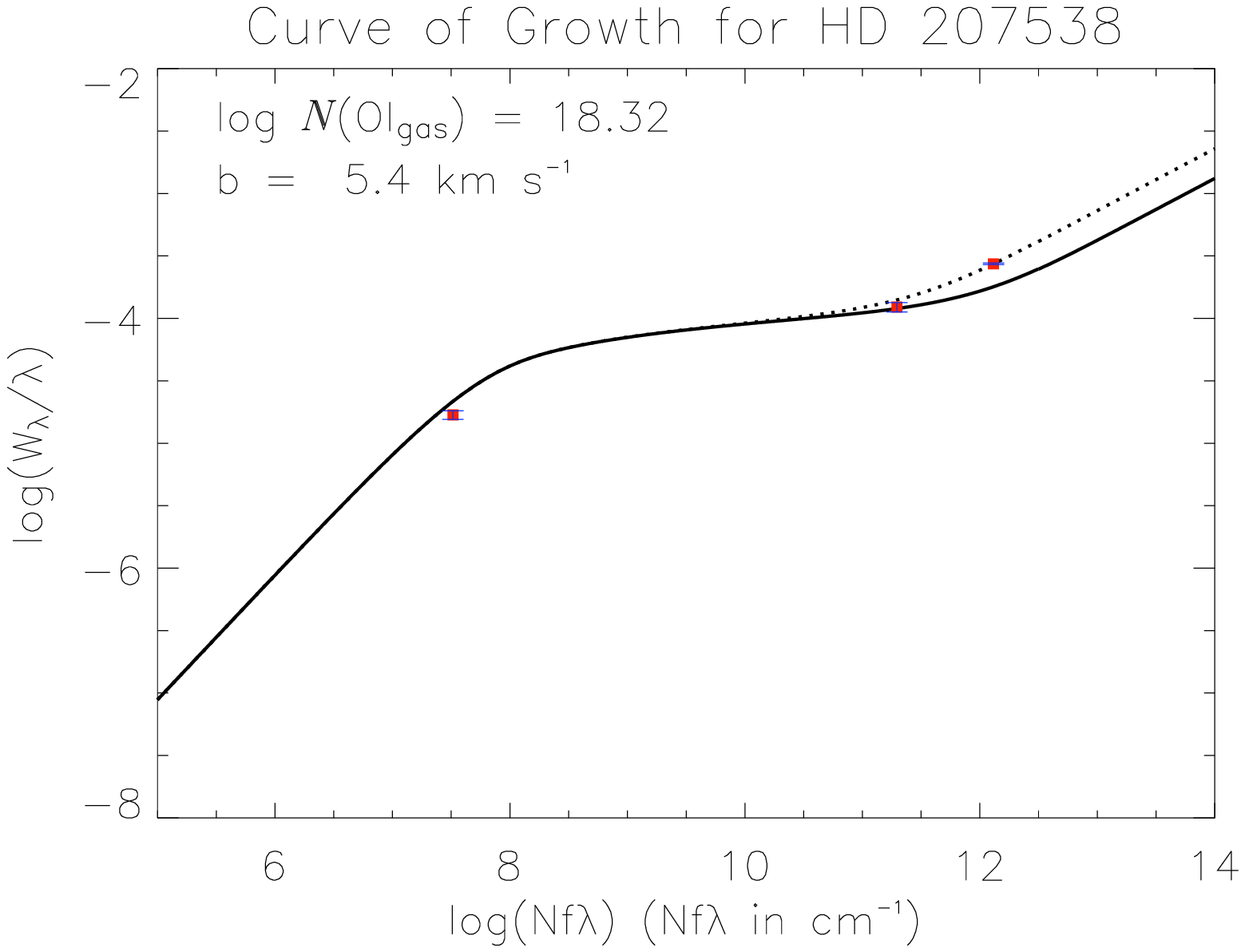}
\end{center}
\caption{Curves of growth for HD 192639, HD 197512, HD 199579, HD 206267, HD 207198, and HD 207538.  Equivalent width fits are shown as red squares with dark blue error bars; 1-$\sigma$ upper limits on the 1355 \AA{} line are shown as green diamonds, with the upper limit on the light blue error bars representing a 2-$\sigma$ upper limit.  The curve of growth for HD 197512 has only two equivalent widths.}
\label{fig:cogs19-24}
\end{figure}

\begin{figure}[t!]
\begin{center}
\epsscale{1.00}
\plottwo{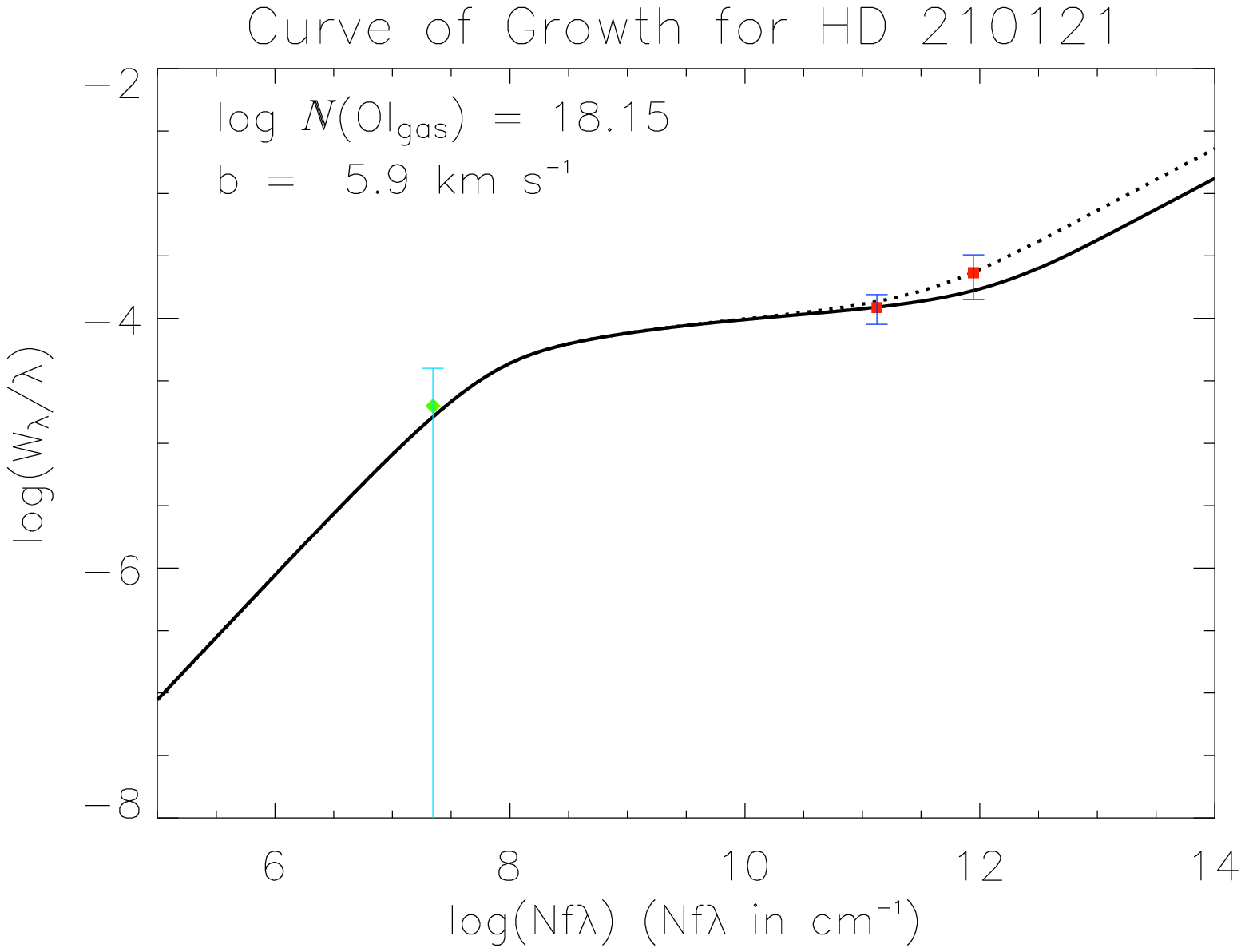}{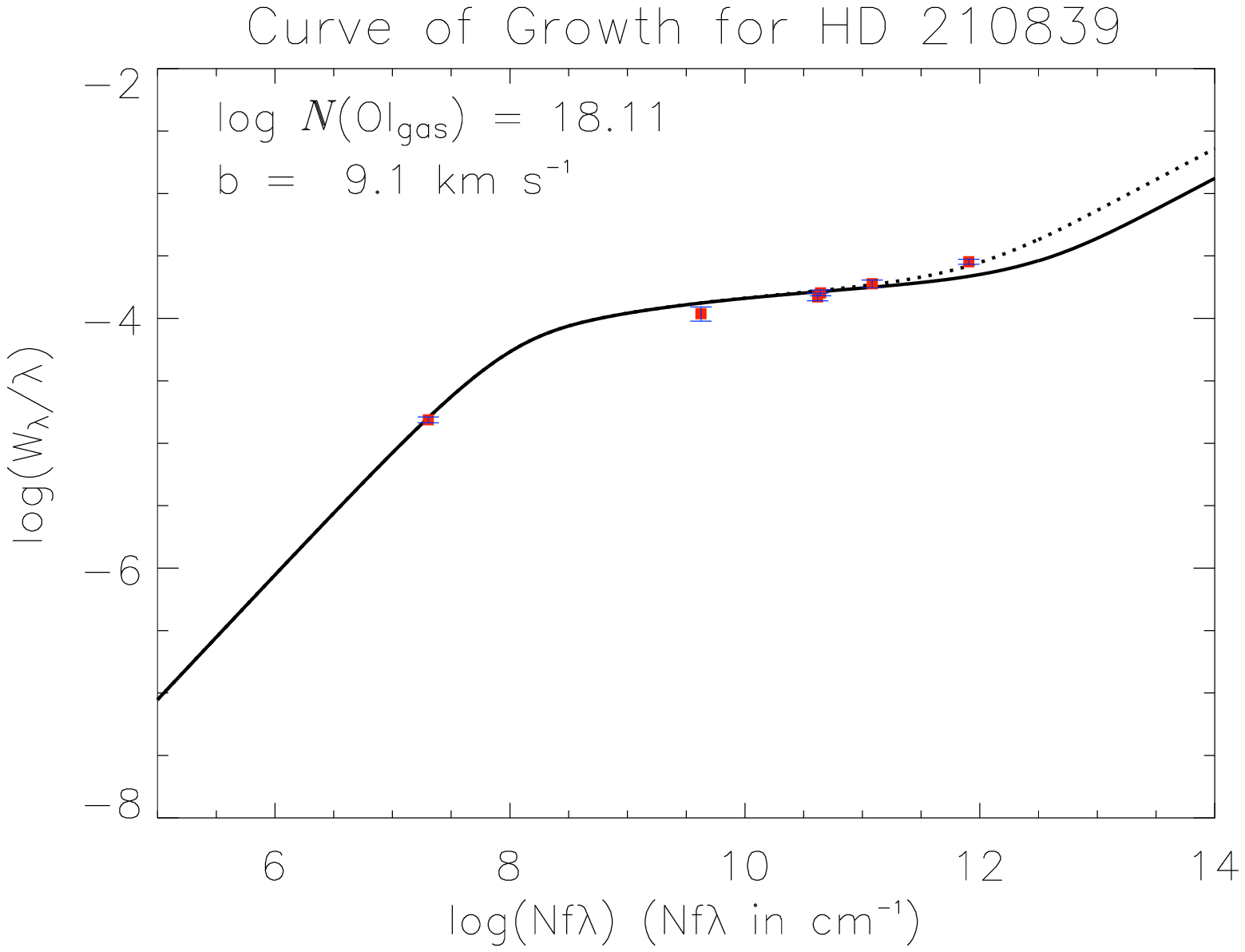}
\end{center}
\caption{Curves of growth for HD 210839 and HD 210121.  Equivalent width fits are shown as red squares with dark blue error bars; 1-$\sigma$ upper limits on the 1355 \AA{} line are shown as green diamonds, with the upper limit on the light blue error bars representing a 2-$\sigma$ upper limit.  The curve of growth for HD 210121 has only two equivalent widths, and the figure represents the adopted solution from two potential solutions (see \S\ref{ss:twopointcogs} and Table \ref{twopointcogs})}
\label{fig:cogs25-26}
\end{figure}

We calculated upper limits for many of the absorption lines seen in the {\it FUSE} data, and for the 1355 \AA{} line when it was undetected in the {\it IUE} data.  The calculated upper limits for the {\it FUSE} lines are very large.  Because these upper limits are so large, they did not provide a useful constraint on the oxygen column density for any of our sightlines, and are not included.  However, the 1355 \AA{} line provided relevant constraints for a few sightlines, and 1- and 2-$\sigma$ upper limits are included in our curve of growth plots (Figures \ref{fig:cogs1-6}-\ref{fig:cogs25-26}).  Most of the 1355 \AA{} line upper limits are consistent with our adopted results at the 2-$\sigma$ level.  Two exceptions are HD 53367 and HD 168076 (both sightlines for which we measure the equivalent widths of only two absorption lines; discussed in \S\ref{ss:twopointcogs}).  However, our results for both of these poorly-constrained sightlines agree with the upper limits at the 3-$\sigma$ level.  This discussion neglects saturation effects, which would tend to reconcile equivalent width upper limits with the column density and $b$-value of a given solution.

\section{COLUMN DENSITIES}
\label{s:columndensities}
For each sightline, the equivalent widths were fit to a curve of growth.  Each curve of growth was constructed using the damping constant of either the 1039 \AA{} line ($\gamma=1.87\times10^{8}$) or the 1302 \AA{} line ($\gamma=5.65\times10^{8}$).  For each damping constant, a family of curves was constructed with $b$-values ranging from 0.1-$25.0\kmpers$, in increments of 0.1$\kmpers$.  For each sightline, the set of equivalent widths was fit to the constructed family of curves by calculating $\chi^2$ for each $b$-value and column densities from $\logOI=15.00$ to $\logOI=25.00$, in increments of 0.01 dex.  All lines except for the 1302 \AA{} line were compared to the curves of growth with the 1039 \AA{} line damping constant, while the 1302 \AA{} was compared to a curve of growth with the same $b$-value but its own damping constant.  The damping constants of lines weaker than the 1039 \AA{} line do not need to be considered, as those lines are unlikely to be found on the portions of the curve of growth where the curve is strongly dependent on the damping constant.

These calculations result in a 250$\times$1001 array of $\chi^2$ values, one for every pair of column density and $b$-value.  The smallest $\chi^2$ is selected as the best solution.  With consideration for the number of degrees of freedom of the fit (number of equivalent widths fit minus the two degrees of freedom in the construction of the curve of growth), we determine the 1-, 2-, and 3-$\sigma$ confidence intervals.  The best fits for curves of growth with at least three measured equivalent widths are reported in Table \ref{coldensities}, with 1-$\sigma$ errors.  Curves of growth with fewer than three measured equivalent widths are discussed in \S\ref{ss:twopointcogs} and \S\ref{ss:onepointcogs}.

Table \ref{hydrogentable} summarizes the data on our program stars.  Molecular hydrogen column densities are taken from recent FUSE studies \citep{Rachford, Rachford2}.  In these papers $\NHmol$ is determined for each sightline by fitting several low-$J$ lines.  Atomic hydrogen densities are taken from several sources, listed in Table \ref{hydrogentable}.  We adopt literature values where they exist.  For four sightlines (HD 27778, HD 40893, HD 53367, and HD 62542), we estimated the total hydrogen column densities using the relationship given by \citet{BohlinSavDrake}, and then inferred total atomic hydrogen column densities based on our molecular hydrogen results.  Of these four sightlines, we measured only one equivalent width for the sightline towards HD 40893 (see \S\ref{ss:onepointcogs}) and two equivalent widths for HD 53367 and HD 62542 (see \S\ref{ss:twopointcogs}).  We consider only the result for the sightline towards HD 27778 to be a well-constrained result for the oxygen column density.  We believe that potential systematic effects from using this relationship will be largely alleviated by the conservative error assumed in HI column density ($\pm0.30{\rm \: dex}$).

In discussion of subsolar or supersolar O/H ratios, we adopt ${\rm(O/H)}_{\odot}=500\pm75\ppm$ for simplicity.  We believe this value is representative of the most recent results for the solar oxygen abundance \citep{Allende, Holweger, Asplund}, without selecting any particular value regarding a question with an ever-changing answer (see \S\ref{ss:dust}).

We report on the whole of our results in \S\ref{s:results}.  In the rest of this section we describe sightlines that require special analysis.  Adopted curves of growth are shown for all sightlines in Figures \ref{fig:cogs1-6}-\ref{fig:cogs25-26}.

\subsection{Outliers}
\label{ss:outliers}
Several sightlines produced what seem to be anomalous values for O/H.  Specifically, four sightlines with at least three data points on the curve of growth have O/H ratios with 1-$\sigma$ errors that do not overlap with the range of 50-100\% of our adopted solar O/H ratio of $500\pm75\ppm$.  Since O/H ratios that are supersolar or less than 50\% solar are very atypical of what is found in the current literature, we examine these outliers first before we draw conclusions on the sample as a whole.  We discuss alternate solutions if the least well-constrained lines are omitted.  We also make direct profile fits by constructing theoretical absorption profiles to the 1039 \AA{} line to see if we can place constraints on either $b$-value or column density.  This is particularly useful when there are multiple solutions that are not excluded to a given confidence level in the $\chi^2$ array that we calculate, and the differing solutions have different $b$-values.  We comment on these four sightlines below.

\subsubsection{HD 37903}
\label{sss:HD37903}
HD 37903 is the only sightline for which we have {\it HST} data that meets our criterion as an outlier.  We derive a column density of $\logOI=17.74$, while in previous work \citet{Cartledge} derive a column density of $\logOI=17.88$.  It is worthing noting the unusual spectrum of this sightline, filled with absorption from vibrationally excited $\Hmol$ as well as near-IR $\Hmol$ \citet{Meyer37903}, indicating that this sightline is subject to unusual physical coniditions that may affect elemental abundances and depletions.  A more complete discussion of the difficulties in analyzing this sightline, our results as compared to those of \citet{Cartledge}, and the best estimate of the column density is found in \S\ref{ss:compare} and \S\ref{ss:bvalues}.  Here we note that if we adopt the \citet{Cartledge} column density, this line no longer meets our outlier criterion, and as discussed in \S\ref{ss:compare}, our results may be in agreement with \citet{Cartledge} if complex velocity structure is considered.


\subsubsection{HD 42087}
\label{sss:HD42087}
The initial analysis of the sightline toward HD 42087 included the 925 \AA{}, 936 \AA{}, 1039 \AA{}, 1302 \AA{}, and 1355 \AA{} lines of OI.  The 1302 \AA{} and 1355 \AA{} lines were taken from {\it IUE} data.  The best fit for the measured equivalent widths is $\logOI=18.53^{+0.18}_{-0.22}$, on a curve of growth with $b=9.4\kmpers$.  This implies $\logOH=-2.97$, or $\OH=1068\ppm$.  However, there are many questions concerning the equivalent widths.  The 1355 \AA{} line is a very poorly determined point, as it is barely a 3-$\sigma$ detection.  The equivalent width is much larger than would be expected as compared to other lines (e.g., it is more than one-third the equivalent width of the much stronger 1039 \AA{} line), and may be subject to significant uncertainty in the continuum of the {\it IUE} spectra.  The 925 \AA{} and 936 \AA{} lines are even more uncertain, approximately 2-$\sigma$ detections.  If these questionable lines are removed, however, we are left with only a two point curve of growth.  This would leave us with a very poor constraint, but a slightly higher $b$-value of $10.2\kmpers$, and the best fit for the oxygen column density would be reduced somewhat to $\logOI=18.40$.  Direct profile fitting of the 1039 \AA{} line does not provide a useful constraint on either column density or $b$-value in this case.  Due to the lack of a significant constraint on the two-point curve of growth, we do not adopt an alternate solution for the column density.   Instead, we accept the initially determined column density with some reservation, and note that the 2-$\sigma$ error of the solution with all lines included overlaps with the range we specified as our criterion for when a sightline is not an outlier.

\subsubsection{HD 152236}
\label{sss:HD152236}
Toward HD 152236, the 1039 \AA{}, 1302 \AA{}, and 1355 \AA{} lines were observed.  Despite a seemingly well-fit curve of growth, the resulting column density ($\logOI=19.20$) is the highest of any of the sightlines in our sample, which also results in one of the highest O/H ratios in the sample ($2276\ppm$).  The $b$-value of $15.3\kmpers$ is also larger than we typically see in other sightlines.

The {\it IUE} spectrum of HD 152236 has an extremely uncertain continuum due to a variety of stellar features.  This uncertain continuum affects our measurement of both the 1302 \AA{} and 1355 \AA{} lines, and leads us to believe that the equivalent widths of these two lines may have much larger errors than the standard error assumed from the errors in the fit parameters and photon noise.  The large uncertainties and low resolution in the {\it IUE} spectra make profile fitting of the 1302 \AA{} and 1355 \AA{} lines useless, but the 1039 \AA{} line may provide a constraint.  Direct profile fitting of the 1039 \AA{} line shows the line is best fit by a $b$-value of $14.6\kmpers$ and a column density of $\logOI=18.57$.  This would correspond to an O/H ratio of $537\ppm$.  However, this fit is not well-constrained, and the profile is still relatively consistent with the larger column density from our curve of growth.  The $b$-values of the curve of growth and the profile fit are consistent because in both cases the greatest constraint is the 1039 \AA{} line.  Therefore, we can be confident that we have determined the $b$-value to some level of accuracy, but column density cannot be well-constrained.

\subsubsection{HD 199579}
\label{sss:HD199579}
Our initial analysis of HD 199579 results in a column density of $\logOI=16.92$ and a $b$-value of $15.5\kmpers$.  This column density corresponds to an O/H ratio of $47\ppm$.  The equivalent widths used in this analysis, however, are better determined than in most of the other sightlines with O/H ratios flagged as potentially anomalous.  Ultimately this may be a case of real, significant depletion, but only if the $b$-value of $15.5\kmpers$---which is among the largest measured in this survey, and larger than any $b$-value measured among the sightlines with {\it HST} data that we consider our best-constrained sightlines---is reasonable.  Direct profile fitting of the 1039 \AA{} line favors a solution of $\logOI\sim18.0$ and $b\sim10\kmpers$ over the larger $b$-value and smaller column density, but the results are not conclusive.  A solution with $\logOI\sim18.0$ and $b\sim10\kmpers$ would not be a case of enhanced depletion.

\subsubsection{Conclusions Regarding Outliers}
\label{sss:concoutliers}
In conclusion, we find that two of the four cases of apparent outliers (HD 42087 and HD152236) have a great deal of uncertainty in the some of the equivalent widths used in their curves of growth, particularly in the lines measured with {\it IUE}.  Both of these sightlines are supersolar in O/H ratio if our values are adopted.  The other two sightlines (HD 37903 and HD 199579) are below 50\% solar in O/H ratio if our values are adopted.  However, comparison of our results with \citet{Cartledge} may shed light on the case with HD 37903 (see \S\ref{ss:compare} and \S\ref{ss:bvalues}).  The adoption of our value would imply moderately enhanced depletion of oxygen in this sightline.  In the case of HD 199579, profile fitting of the 1039 \AA{} line suggests a slightly different solution (that is not subsolar) may be more appropriate, though the results are not conclusive.

\subsection{Two-Point Curves of Growth}
\label{ss:twopointcogs}
On many sightlines, we were not able to measure the equivalent widths of more than two absorption lines.  With only two equivalent widths, we can obtain a unique solution by finding the column density and $b$-value that minimize $\chi^2$.  However, since the number of degrees of freedom in the fit (number of fitted points minus the number of degrees of freedom in constructing the curve of growth) on a two-point curve of growth is zero, the confidence intervals are undefined and we cannot provide 1-$\sigma$ uncertainties on these fits.

Furthermore, given the nature of the free parameters in finding a best fit on the curve of growth, it is often the case with only two equivalent widths that there are two local minima of $\chi^2$, consisting of a solution with a higher $b$-value and a lower column density, and a solution with a lower $b$-value and a higher column density.  Usually these solutions place the two measured equivalent widths near either transition region of the curve of growth, i.e. where the linear portion transitions into the flat protion or where the flat portion transitions into the square-root portion.

We measure the equivalent widths of only two OI absorption lines for HD 38087, HD 46056, HD 53367, HD 62542, HD 73882, HD 168076, HD 170740, HD 179406, HD 197512, and HD 210121.  In all cases except HD 62542, HD 73882, HD 168076, and HD 197512 we find two local minima in the $\chi^2$ analysis.  These results, with solutions corresponding to both local $\chi^2$ minima, are given in Table \ref{twopointcogs}.  Errors quoted in this table are ranges for which the absolute (i.e. unreduced) $\chi^2\leq0.1$, since we cannot convert $\chi^2$ into a confidence interval with only two data points and two free parameters.  Where two solutions are present, we also comment on the adopted solution for each sightline.

The adopted solutions are selected on the basis of one or more criteria.  The first criterion is consistency with the profile of the 1039 \AA{} line, which can at times give a good constraint on the $b$-value of the sightline.  This is the case for HD 46056, HD 53367, HD 170740.  If the profile of the 1039 \AA{} line is inconclusive, we choose the solution with a column density that is more consistent with our results for other sightlines in terms of O/H ratio and $b$-value.  We do this for HD 38087, HD 179406, and HD 210121.  In all six cases with multiple solutions, we adopt the solution with the smaller $b$-value (justified by direct profile fitting in three cases).  Adopted solution $b$-values are $12.2\kmpers$ or less, while five of the six $b$-values of the alternate solutions are greater than or equal to $18.0\kmpers$, with the sixth alternate solution $b$-value (for HD 179406) equal to $13.6\kmpers$.  A $b$-value as large as $18.0\kmpers$ is larger than would be expected in typical interstellar medium conditions and inconsistent with the $b$-values of our better-constrained sightlines, so our selection of the solutions with a smaller $b$-value seems well justified.  For the case of HD 179406, where the $b$-value of both solutions is less than $18.0\kmpers$, we note that \citet{Hanson} find a column density of $\logOI=17.79$, in close agreement with our adopted solution of $\logOI=17.77$.  They also find a $b$-value of $\sim6\kmpers$, a value that is within the errors of our $b$-value of $5.9\kmpers$ (see Table \ref{comparison} and discussion in \S\ref{ss:compare}).

The remaining four cases of HD 62542, HD 73882, HD 168076, and HD 197512 do not have two local minima in the $\chi^2$ analysis.  However, all of these sightlines have very large errors in column density and $b$-value that are inclusive of what we consider to be reasonable values.

We add these solutions to our plots of O/H abundance against various physical parameters of the various sightlines in Figures \ref{fig:Htot}-\ref{fig:distance}.  These plots show that the adopted solutions for the two-point curves of growth, though not tightly constrained and showing some scatter, agree to within the errors with an O/H ratio that is no more than solar and no more than 50\% depleted from the solar value.  Thus, while we cannot give them significant weight for lack of constraint, we note that our adopted solutions for these sightlines are consistent with the rest of our sample in terms of column density and O/H ratio, and the selection of these solutions is justified by reasonable constraints on their $b$-values.

\subsection{One-Point Curves of Growth}
\label{ss:onepointcogs}
The sightlines towards HD 40893, HD 164740, HD 167971, and HD 203938 each had only one observed aborption line of oxygen each.  For HD 167971 the 1302 \AA{} line was observed with {\it IUE}, and for HD 40893, HD 164740, and HD 203938 the 1039 \AA{} line was observed with {\it FUSE}.  The equivalent widths for these lines are reported with the equivalent widths from the other sightlines in Table \ref{eqwidths}.  Also reported is an upper limit from {\it IUE} data on the 1355 \AA{} line for HD 167971.  However, since one equivalent width does not provide a unique solution for the column density without an assumed $b$-value, we cannot determine column densities for these sightlines.  It has been suggested, e.g. \citet{York}, that the $b$-value of NI can be used as a proxy for the $b$-value of OI, though our results may suggest otherwise (see \S\ref{ss:bvalues}).  Depending on the answer to this question, further work on NI could be combined with our results to determine the OI column density for a given sightline.  However, such a column density would still be poorly constrained if the equivalent widths were on the flat region of the curve of growth.  Since we do not report on column densities, the minimal results from these sightlines are not included in our final analysis of abundances and depletions.

\subsection{Comparison With Previous Work}
\label{ss:compare}
Nine of our sightlines have been examined in earlier work, including two sightlines that were examined twice previously (HD 24534 and HD 185418) and another sightline examined three times previously (HD 192639).  We compare our results to this previous work in Table \S\ref{comparison}.  Of these nine sightlines, five (HD 154368, HD 179406, HD 185418, HD 192639, and HD 210839) show very close agreement with the previous work.  These five sightlines include the two special cases of HD 154368 and HD 179406.  In our analysis of the sightline toward HD 154368, we used the \citet{Snow154368} value for the equivalent width of the 1355 \AA{} line, so the results we are comparing are not completely independent.  However, our measured equivalent widths for the 1039 \AA{} and 1302 \AA{} lines, in conjunction with the \citet{Snow154368} equivalent width for the 1355 \AA{} line, produce a consistent curve of growth.  \citet{Snow154368} also measured an equivalent width for the 1302 \AA{} line using a theoretical simulation based on their 1355 \AA{} line results, but the theoretical fit for this line, when plotted against the spectral data, appears to be too narrow.  The errors on this equivalent width are also very small.  We elect to use our {\it IUE} equivalent width, which was larger but with more conservative error bars.  HD 179406 was a sightline for which we measured equivalent widths for only two absorption lines, and thus we have multiple solutions.  However, our adopted solution for this sightline is consistent with the solution of \citet{Hanson} in both column density and $b$-value (see \S\ref{ss:twopointcogs} and Table \ref{twopointcogs}).

Our results do not agree to within errors with previous results for HD 27778, HD 37903, and HD 207198, all previously studied by \citet{Cartledge}.  Our result for HD 24534 agrees with a recent result by \citet{Knauth}, but is significantly smaller than an older result by \citet{SnowXPer}.

Our result for HD 27778 is $\logOI=18.05$, while \citet{Cartledge} finds $\logOI=17.83$.  \citet{Cartledge} use a profile fit for the 1355 \AA{} line that contains two components separated by $4\kmpers$, with $b$-values of $1.9\kmpers$ and $2.2\kmpers$.  This corresponds to an ``effective $b$-value'' (see \S\ref{ss:bvalues}) of $\sim3\kmpers$, which is what our curve of growth analysis and a single-component Gaussian approximation to the line found.  We generated simulated profiles of all three of our measured lines (1039 \AA{}, 1302 \AA{}, and 1302 \AA{}) in an attempt to discern where this inconsistency arises.  Simulated profiles to the 1039 \AA{} line do not show a detectable difference between our results and those of \citet{Cartledge}.  However, the 1302 \AA{} line provides some evidence in favor of our value.  The column density and $b$-value we derive from our curve of growth method provide a very accurate fit to the profile of the 1302 \AA{} line, while the corresponding values from \citet{Cartledge} produce a profile that is too narrow in the core and that underestimates the damping wings.  If favor of the results of \citet{Cartledge} is the fact that our column density generates a simulated profile of the 1355 \AA{} line that is too strong.  However, if we take a lower column density ($\logOI=17.96$) and higher $b$-value ($3.4\kmpers$) that are within our 1-$\sigma$ errors, we can generate a 1355 \AA{} line profile that is consistent with the data.

Using the column density of \citet{Cartledge} but a larger $b$-value (which would be appropriate if additional weak componets of OI are hidden in the noise surrounding the 1355 \AA{} line; see \S\ref{ss:bvalues}) can fit the core of the 1302 \AA{} line profile, but still underestimates the damping wings.  It is therefore likely that the profile fitting methods used by \citet{Cartledge} did not detect some amount of saturation, and a compromise between our values for the column density ($\logOI \sim17.95$) represents the best solution.

The most complicated case of disagreement is the sightline towards HD 37903.  The spectrum of HD 37903 is rife with absorption lines from vibrationally excited $\Hmol$, and therefore oxygen lines that are distinct in the spectra of other sightlines may be contaminated in the spectrum of the sightline towards HD 37903.  The 1355 \AA{} line is narrow enough to avoid contamination; however, it is difficult to estimate the contamination from the excited lines on the 1039 \AA{} and 1302 \AA{} lines because the population of each vibrationally excited state of $\Hmol$ is not clear.  Nevertheless, reasonable upper limits on the important excited lines are not enough to account for the fact that the 1039 \AA{} and 1302 \AA{} lines are much broader than would be expected if their profiles were generated by the single-component $b$-value ($1.6\kmpers$) of the 1355 \AA{} line (even if a much larger column density is chosen, the wings grow faster than the broad core).  The only explanation for the discrepancy in apparent $b$-value between the stronger lines and the and weak 1355 \AA{} line is that there are very weak components hidden in the noise surrounding the weak line.  These weak components would not contribute significantly to the column density, but would increase the effective $b$-value of the stronger lines (see \S\ref{ss:bvalues}).

With consideration for the excited $\Hmol$ lines and inconsistent OI line profiles that complicate our analysis of HD 37903, our results for the column density still differ from those of \citet{Cartledge}, who measures a column density of $\logOI=17.88$ compared to our result of $\logOI=17.74$.  The two results are mutually exlusive at the 2-$\sigma$ confidence level.  The results of \citet{Cartledge}, which are based on a single cloud component, do a better job of reproducing the data for the 1355 \AA{} line profile than the results of our curve of growth method; our column density is too small to produce the line without a large $b$-value, but a larger $b$-value is inconsistent with the line profile.  The profiles of the 1039 \AA{} and 1302 \AA{} lines suggest a $b$-value of $\sim9\kmpers$, and do not distinguish between $\logOI=17.74$ and $\logOI=17.88$.  If we artificially reduce the weight of the 1355 \AA{} line (by artificially increasing the equivalent width uncertainty), our column density result with the curve of growth method increases, as the stronger lines dominate and the 1355 \AA{} line is ``pulled off'' the linear portion of the curve of growth to a higher column density.  To produce a consistent picture without this artificial weighting would require a curve of growth that accounted for multiple cloud components.  The likely result of this method is that the most consistent picture would come from a higher column density similar to that derived by \citet{Cartledge}.

Our column density for HD 207198 ($\logOI=18.30$) is significantly larger than the column density of $\logOI=18.13$ found by \citet{Cartledge}.  The two results are mutually exclusive at the 1-$\sigma$ confidence level but are consistent within 2-$\sigma$.  As with HD 37903, the column density of \citet{Cartledge} provides a better fit to the 1355 \AA{} line than our column density and $b$-value ($3.5\kmpers$) derived from the curve of growth.  The cloud components found by \citet{Cartledge} are at velocities of $-20$, $-16$, $-12$, and $-9\kmpers$ with $b$-values of 1.2, 1.4, 1.2, and $0.7\kmpers$, respectively.  Both that set of components and our curve of growth $b$-value do not accurately reproduce the broad profiles of the 1039 \AA{} and 1302 \AA{}.  The 1039 \AA{} line profile is more sensitive to $b$-value and relatively independent of the choice of column density, while the 1302 \AA{} profile marginally favors the smaller column density.

The final sightline where our measured oxygen column density does not agree to within errors with previous work is HD 24534, which was examined by \citet{SnowXPer} and \citet{Knauth}.  Our result for the column density of this sightline is smaller than that of \citet{SnowXPer} by $0.29\dex$.  Based on \citet{Lien}, \citet{SnowXPer} assumed the $b$-value of dominant ions in the sightline towards HD 24534 is $1.0\kmpers$, and applied an appropriate saturation correction to the 1355 \AA{} line to determine column density.  However, while our equivalent width for the 1355 \AA{} line agrees with \citet{SnowXPer}, our curve of growth analysis results in a best fit with $b=6.6\pm0.3\kmpers$, placing the 1355 \AA{} line on the linear portion of the curve, away from any significant saturation.  This $b$-value is greatly constrained by observations of the 1039 \AA{} line, which were not available to \citet{SnowXPer}.  The $b$-value of oxygen along this sightline cannot be as low as $1.0\kmpers$ unless our equivalent width for either the 1039 \AA{} line or the 1355 \AA{} line is significantly in error.  Since we have high-resolution {\it HST} data for this line of sight, we can directly constrain the $b$-value of the 1355 \AA{} line from its observed width.  The $b$-value measured for the weak line is $2.0\kmpers$ (the apparent inconsistency between this number and the value from the curve of growth is discussed in \S\ref{ss:bvalues}).  While this $b$-value may still be small enough to require a small saturation correction, it is clear that the $b$-value assumed by \citet{SnowXPer} is too small and the resulting column density is too large.  Furthermore, as discussed in \S\ref{ss:bvalues}, direct profile fitting of other absorption lines is consistent with our measured column density, but not that of \citet{SnowXPer}.  The more recent result from \citet{Knauth} agrees with our value for the column density, but details of the profile and $b$-value are not given in that paper.

\section{RESULTS AND DISCUSSION}
\label{s:results}
The value we adopt to summarize our O/H results depends on the sample we include.  If we take the results for the 10 sightlines for which {\it HST} data were available, we find a mean O/H of $421^{+47}_{-33}\ppm$, with a standard deviation of 29\%.  As discussed further in \S\ref{ss:methods}, it is fair to consider these sightlines separately since the {\it HST} data provide the best-constrained equivalent widths for the 1302 and 1355 \AA{} lines, and the presence of these two lines gives the widest range in $\log{f\lambda}$, constraining all three parts of the curve of growth.  This mean value for the O/H ratio agrees very closely with \citet{Andre} and the low-density sightline mean of \citet{Cartledge2}.  If we include all 26 sightlines for which we determined O/H (including our adopted solutions for the 10 two-point curves of growth), our mean O/H is $468^{+103}_{-46}\ppm$, with a standard deviation of 98\%.  Given the conservative error bars, this value is still consistent with a solar abundance and also relatively consistent with, though larger than, the values measured by \citet{Andre} and \citet{Cartledge2}.

\begin{figure}[t!]
\begin{center}
\epsscale{1.00}
\plottwo{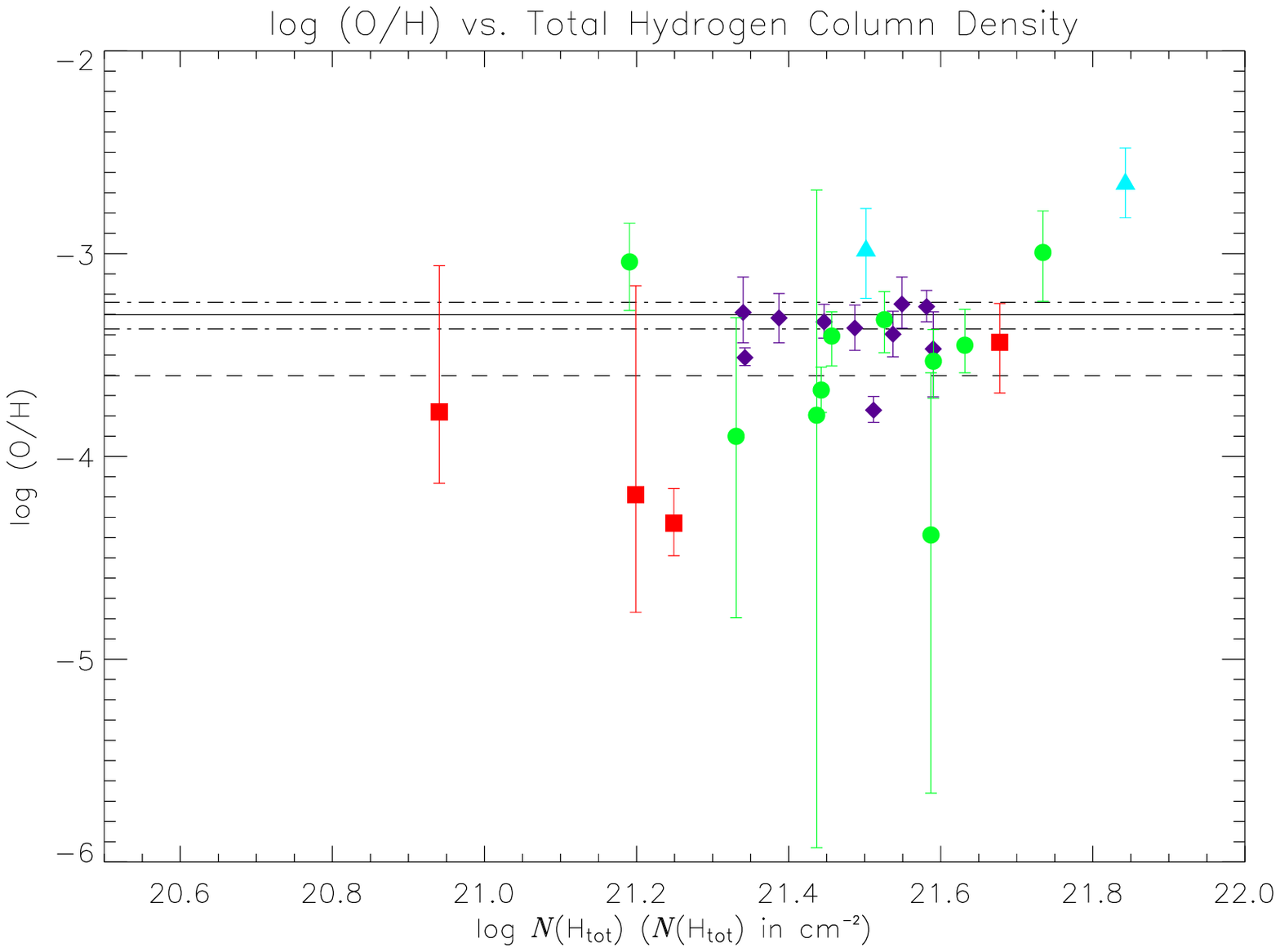}{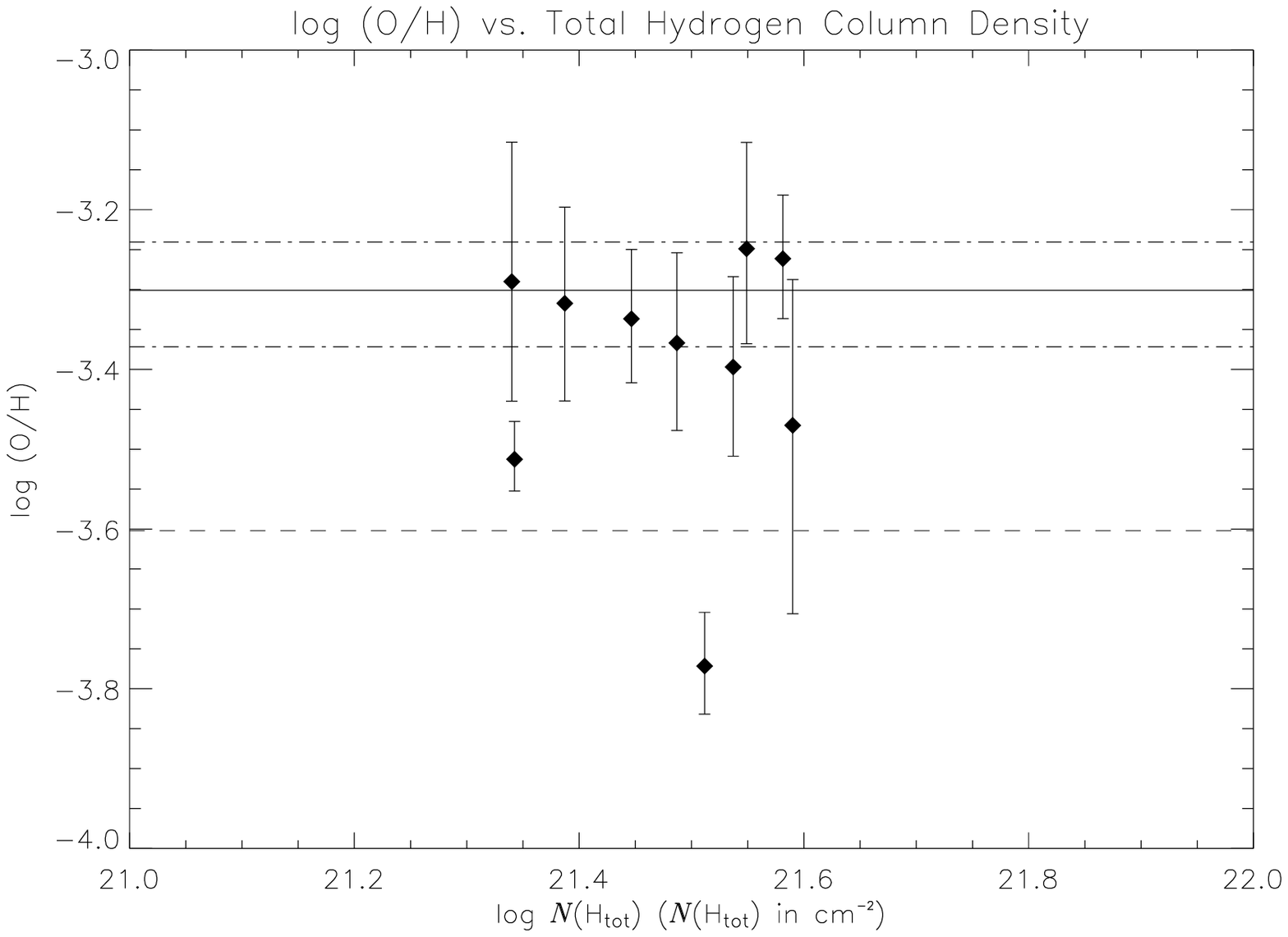}
\end{center}
\caption{O/H vs. $\Htot$.  {\bf Left}---all 26 sightlines for which we report on $N{\rm (OI)}$.  {\it Solid line}: Adopted solar O/H ratio of $500\ppm$.  {\it Dashed-dotted lines}: Errors in adopted solar O/H ratio.  {\it Dashed line}: 50\% of the adopted solar O/H ratio.  {\it Red squares}: Curves of growth with three or more equivalent widths, but no measurement of the 1355 \AA{} line.  {\it Light blue triangles}: Curves of growth with three or more equivalent widths, {\it IUE} measurement of the 1355 \AA{} line.  {\it Purple diamonds}: Curves of growth with three or more equivalent widths, {\it HST} measurement of the 1355 \AA{} line.  {\it Green circles}: Curves of growth with two equivalent widths, preferred solution if two solutions exist.  {\bf Right}---the 10 sightlines with an {\it HST} measurement of the 1355 \AA{} line.}
\label{fig:Htot}
\end{figure}

\begin{figure}[t!]
\begin{center}
\epsscale{1.00}
\plottwo{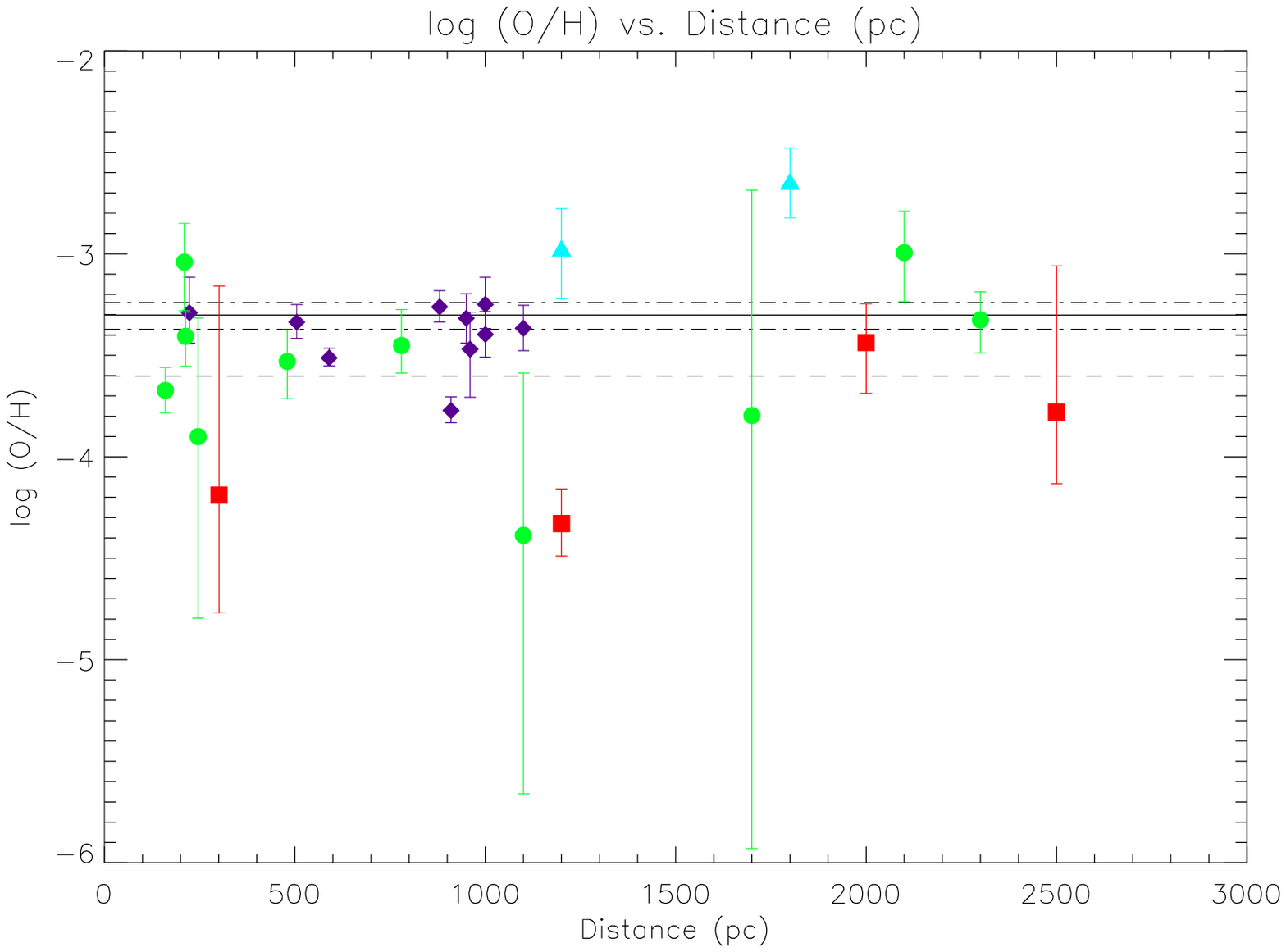}{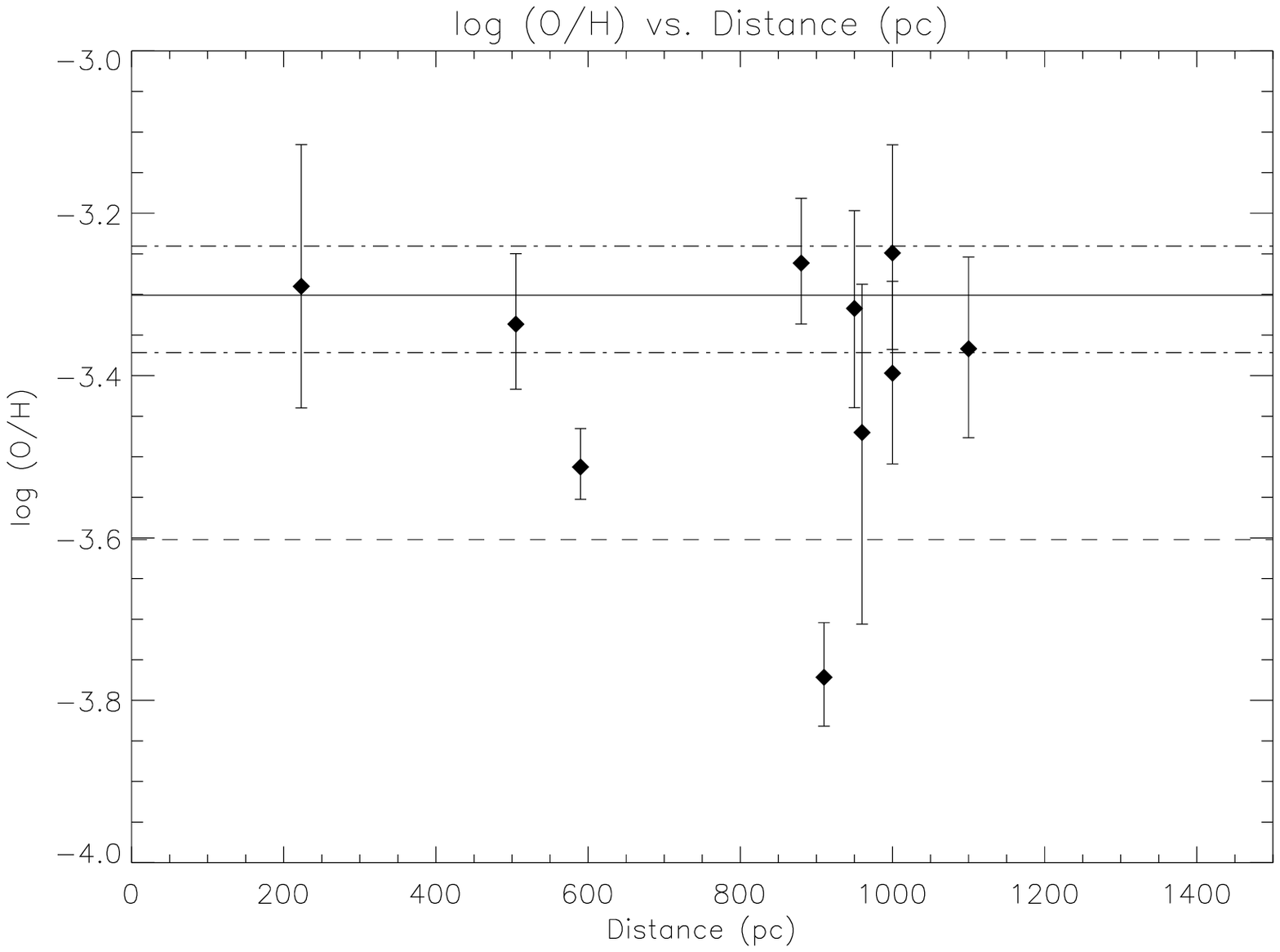}
\end{center}
\caption{O/H vs. distance.  {\bf Left}---all 26 sightlines for which we report on $N{\rm (OI)}$.  {\it Solid line}: Adopted solar O/H ratio of $500\ppm$.  {\it Dashed-dotted lines}: Errors in adopted solar O/H ratio.  {\it Dashed line}: 50\% of the adopted solar O/H ratio.  {\it Red squares}: Curves of growth with three or more equivalent widths, but no measurement of the 1355 \AA{} line.  {\it Light blue triangles}: Curves of growth with three or more equivalent widths, {\it IUE} measurement of the 1355 \AA{} line.  {\it Purple diamonds}: Curves of growth with three or more equivalent widths, {\it HST} measurement of the 1355 \AA{} line.  {\it Green circles}: Curves of growth with two equivalent widths, preferred solution if two solutions exist.  {\bf Right}---the 10 sightlines with an {\it HST} measurement of the 1355 \AA{} line.}
\label{fig:distance}
\end{figure}

We plot the logarithmic ratio of gas-phase oxygen to hydrogen against total hydrogen column density and distance in Figures \ref{fig:Htot} and \ref{fig:distance}, plotting both the general sample and the {\it HST} sample separately to show the difference in the samples.  The plots show the scatter in the O/H ratio for our less well-constrained lines of sight and no trend is apparent, except for perhaps with respect to distance to the illuminating star, i.e., the length of the sightline.  The most significant scatter in our O/H determinations occurs almost exclusively for sightlines greater than 1100 pc in length.  If, contrary to the reservation expressed in \S\ref{ss:outliers}, we accept our column densities for these sightlines, the reason for these deviations may be that over such distances we are sampling local inhomogeneities.  However, this would be in contrast to the results of \citet{Andre}, who find consistent O/H ratios for many sightlines greater than 1000 pc in length.  Additionally, if we are observing local inhomogeneities, it is difficult to know where they occur with respect to distance since our measurements of equivalent width are integrated over the entire sightline.  However, with the previous discussion of \S\ref{ss:outliers} in mind, noting the large qualitative uncertainties and potential alternate solutions for many of these sightlines, the fact that the greater scatter in our results occurs for sightlines greater than 1000 pc in length is likely coincidental.  When we consider the sample of 10 {\it HST} sightlines separately, any possible trend disappears.  The {\it HST} sightlines are at most 1100 pc in length and do not show any increased scatter with respect to distance.

\begin{figure}[t!]
\begin{center}
\epsscale{1.00}
\plotone{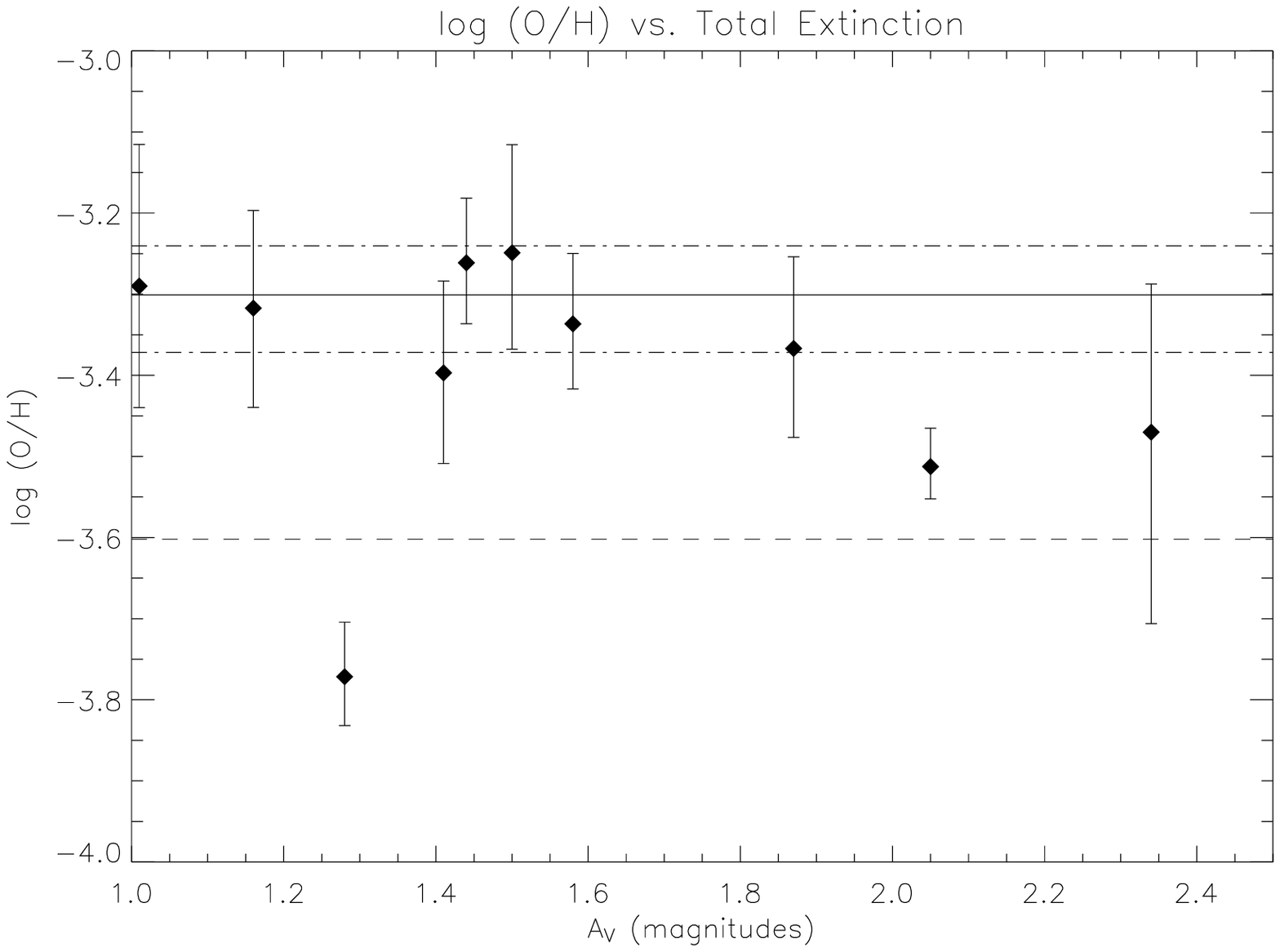}
\end{center}
\caption{O/H vs. $A_V$ for the {\it HST} sample.  {\it Solid line}: Adopted solar O/H ratio of $500\ppm$.  {\it Dashed-dotted lines}: Errors in adopted solar O/H ratio.  {\it Dashed line}: 50\% of the adopted solar O/H ratio.}
\label{fig:Av}
\end{figure}

\begin{figure}[t!]
\begin{center}
\epsscale{1.00}
\plotone{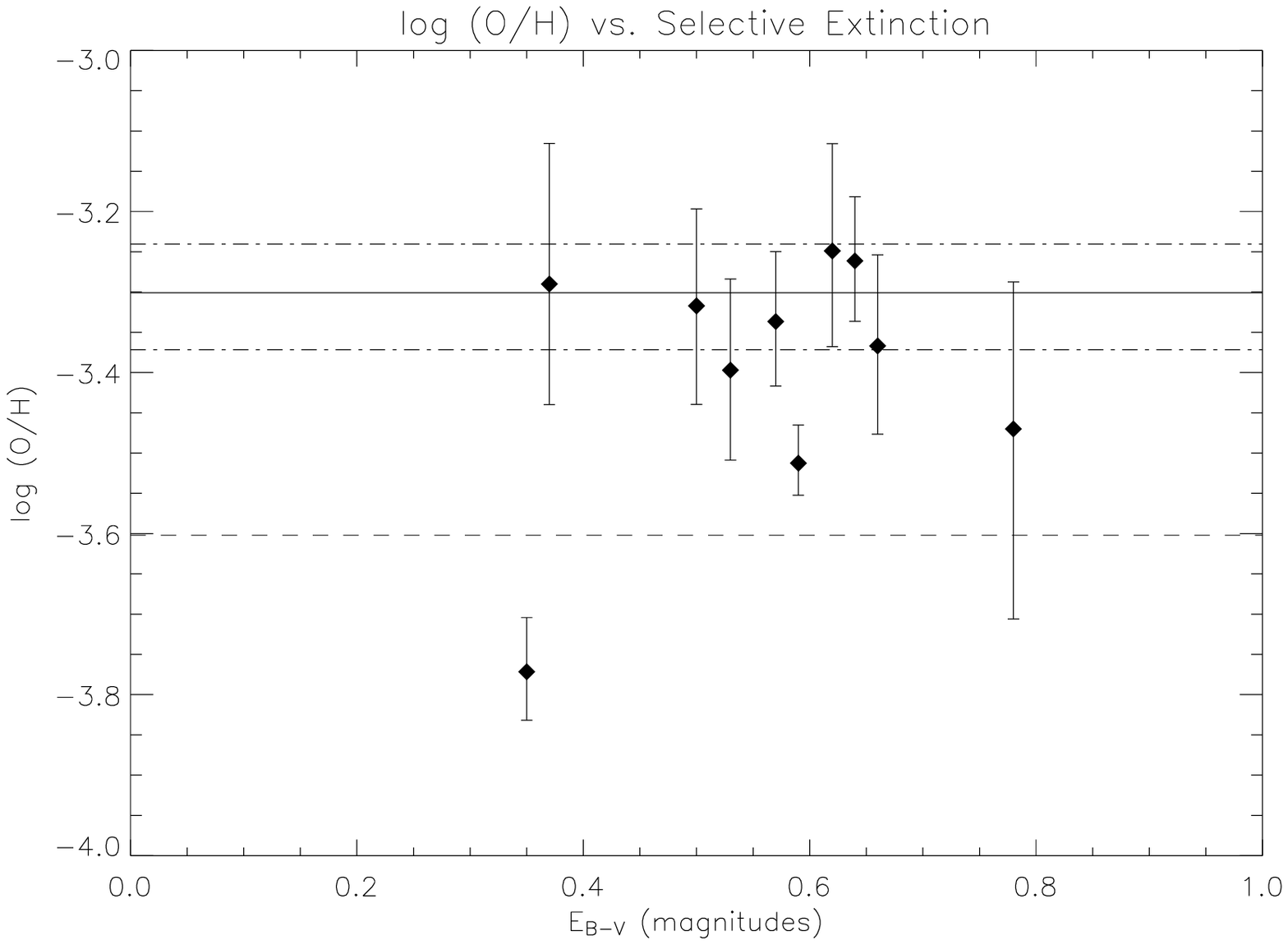}
\end{center}
\caption{O/H vs. $E_{B-V}$ for the {\it HST} sample.  {\it Solid line}: Adopted solar O/H ratio of $500\ppm$.  {\it Dashed-dotted lines}: Errors in adopted solar O/H ratio.  {\it Dashed line}: 50\% of the adopted solar O/H ratio.}
\label{fig:EB-V}
\end{figure}

\begin{figure}[t!]
\begin{center}
\epsscale{1.00}
\plotone{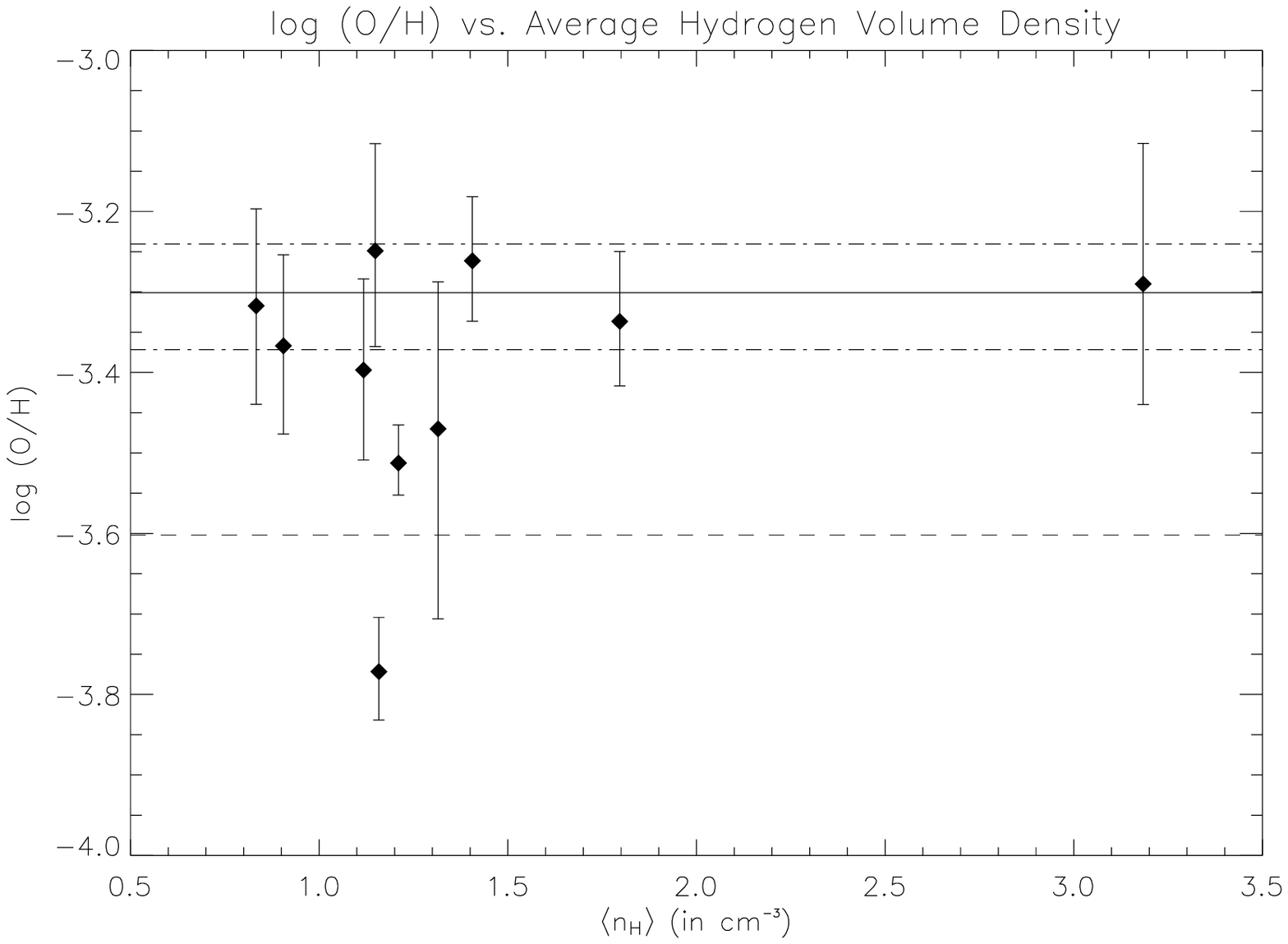}
\end{center}
\caption{O/H vs. $\nH$ for the {\it HST} sample.  {\it Solid line}: Adopted solar O/H ratio of $500\ppm$.  {\it Dashed-dotted lines}: Errors in adopted solar O/H ratio.  {\it Dashed line}: 50\% of the adopted solar O/H ratio.}
\label{fig:ntot}
\end{figure}

\begin{figure}[t!]
\begin{center}
\epsscale{1.00}
\plotone{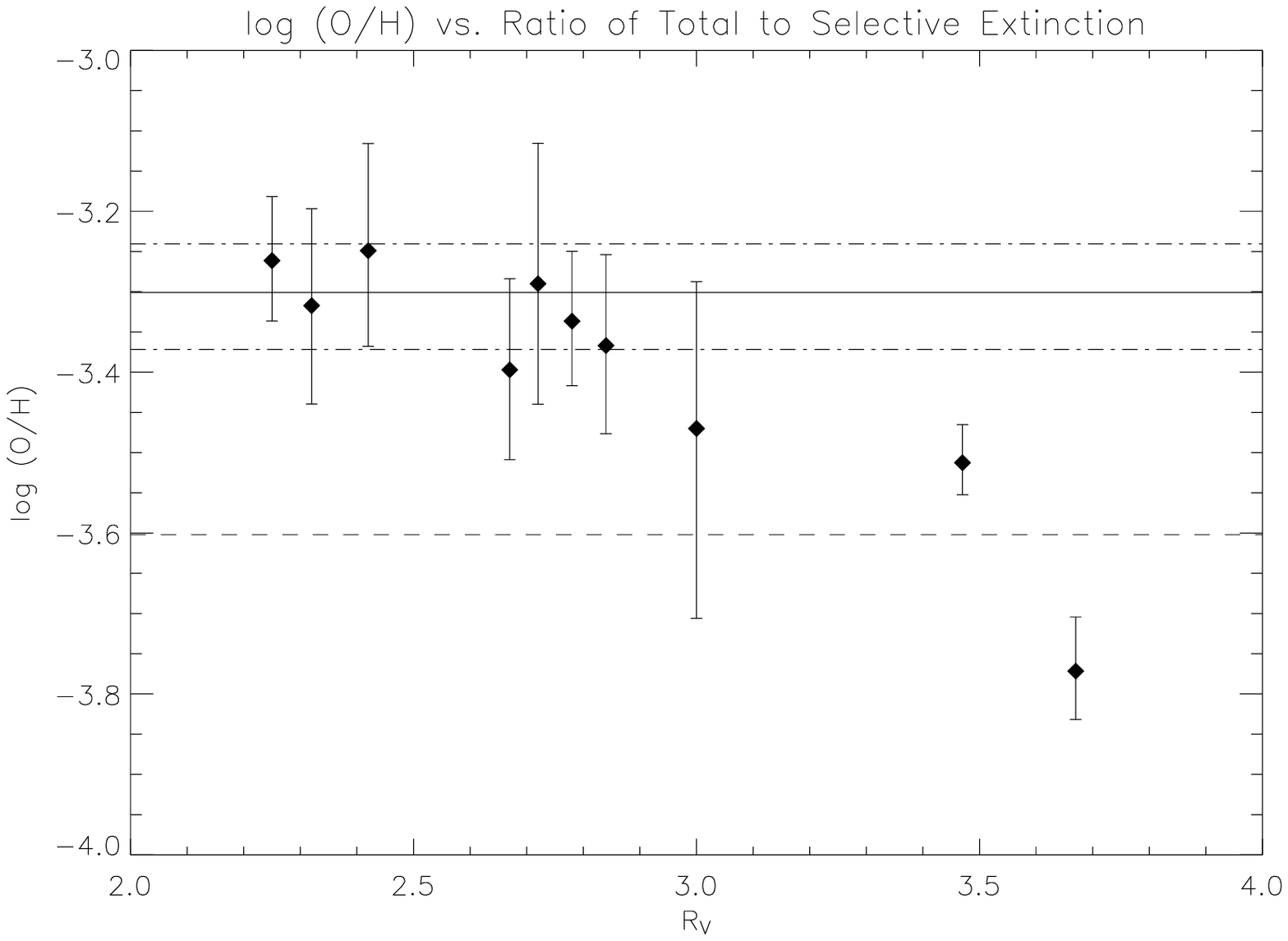}
\end{center}
\caption{O/H vs. $R_V$ for the {\it HST} sample.  {\it Solid line}: Adopted solar O/H ratio of $500\ppm$.  {\it Dashed-dotted lines}: Errors in adopted solar O/H ratio.  {\it Dashed line}: 50\% of the adopted solar O/H ratio.}
\label{fig:Rv}
\end{figure}

\begin{figure}[t!]
\begin{center}
\epsscale{1.00}
\plotone{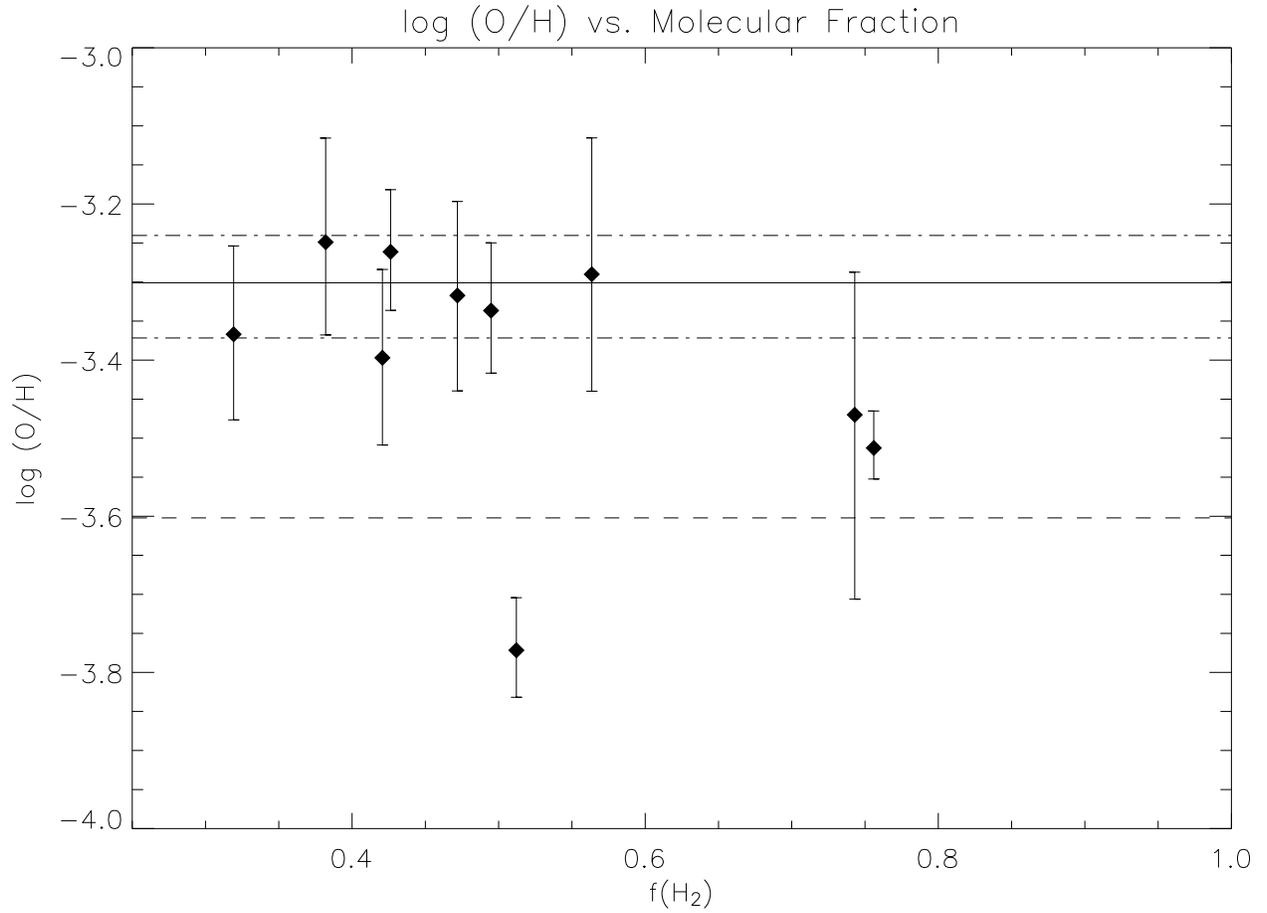}
\end{center}
\caption{O/H vs. Molecular Fraction of Hydrogen for the {\it HST} sample.  {\it Solid line}: Adopted solar O/H ratio of $500\ppm$.  {\it Dashed-dotted lines}: Errors in adopted solar O/H ratio.  {\it Dashed line}: 50\% of the adopted solar O/H ratio.}
\label{fig:molecularfraction}
\end{figure}

The {\it HST} sample is plotted against $A_V$, $E_{B-V}$, $\nH$, $R_V$, and $\fHmol$ in Figures \ref{fig:Htot}-\ref{fig:molecularfraction}.  In the {\it HST} sample we find the O/H ratios for two sightlines (HD 24534 and HD 37903) do not agree within the errors with the mean value found by \citet{Andre} or the mean value found by \citet{Cartledge2} for sightlines with a lower density than their derived critical density they of $\nH=1.5\percc$.  As discussed in \S\ref{ss:compare}, there is some discrepancy between our result and that of \citet{Cartledge} for HD 37903.  However, whether our O/H ratio or that of \citet{Cartledge} is adopted, it is the smallest O/H ratio of all 10 sightlines in the {\it HST} sample.  Our derived O/H ratio for HD 154368 is similar to that of HD 24534, but with much larger error bars that include the \citet{Andre} and low-density \citet{Cartledge2} means.

There is the possible hint of two trends in the {\it HST} sample.  HD 24534, HD37903, and HD 154368, with the lowest O/H ratios in the sample, also have the three largest values of $R_V$, the ratio of total to selective extinction, in the {\it HST} sample.  These three sightlines, along with HD 27778, are the only four sightlines in the {\it HST} sample with molecular hydrogen fractions greater than 0.5.  However, both of these trends are far from definite and we do not want to overstate the case based on only these three points in a sample of just 10 points.  Some implications of the possibility of a trend with $R_V$ are discussed in \S\ref{ss:dust}.

In addition to the lack of a trend with $A_V$ and $E_{B-V}$, we do not see a trend with respect to $\nH$, the trend that has been seen by \citet{Cartledge2}.  However, comparing our sample size of 10 sightlines to the 56 sightlines of \citet{Cartledge2} indicates the absence of evidence is not necessarily the evidence of absence.  In our {\it HST} sample, only HD 27778 and HD 210839 are more dense than the critical density determined by \citet{Cartledge2}.  Both of these sightlines have O/H ratios larger than the \citet{Andre} and low-density \citet{Cartledge2} means, though consistent within the errors.  HD 24534, HD 37903, and HD 154368 all have $\nH$ between 1.1 and $1.4\percc$.

\subsection{Evaluation of Methods}
\label{ss:methods}
The difference between our {\it HST} and general samples sheds some light on the methods that we should use.  The {\it HST} sample uses data on the 1355 \AA{} line.  Even if the line is somewhat saturated, it is weak enough that it produces a moderately strong constraint on the column density.  When combined with results from other lines, such as the 1039 \AA{} line, we can determine a $b$-value rather than merely assuming one.  Hindering this, however, is the fact that the velocity structure may be very complex, requiring multi-component curves of growth for the most accurate picture (see \S\ref{ss:bvalues}).

Our conclusion regarding the utility of the {\it FUSE} data for determining neutral oxygen column densities is mixed.  We note the case of HD 24534 discussed in \S\ref{ss:compare} where measurements of absorption lines in the {\it FUSE} wavelength region led to a significant revision in $b$-value and resulting column density compared to a previous study \citep{SnowXPer}, when used in conjunction with {\it HST} measurements of the 1355 \AA{} line, a result confirmed by another study \citep{Knauth}.  This result indicates the possible importance of using {\it FUSE} data in oxygen column density determinations.  However, {\it FUSE} lines without support from {\it HST} data can often lead to poorly constrained column densities.  Even though {\it IUE} can measure the two lines at 1302 \AA{} and 1355 \AA{}, uncertainties in the {\it IUE} spectra complicate matters, and the 1355 \AA{} line is typically not observed.  Our evaluation of the {\it FUSE} data in this context is that the far-UV lines should be viewed as a very useful tool in conjunction with good data on the 1355 \AA{} line, but the far-UV data alone have some limitations if such additional data are not available.  Compared to recent studies that use profile fitting, the curve of growth method is inferior when using the lines found in {\it FUSE} without {\it HST} measurements of the 1302 \AA{} and 1355 \AA{} lines.  However, when {\it HST} measurements of these lines are available, a curve of growth with the addition of the medium-strength {\it FUSE} lines seems to provide results similar to profile fitting in most cases.  In \S\ref{ss:compare} we discussed the fact that there are two cases of discrepancy where the evidence seems to favor previous profile fitting results (HD 37903 and HD 207198) based on consistency with the line profiles.  In both of these cases the reason for the discrepancy seems to arise from the fact that the sightline has a complex velocity structure that cannot be easily approximated by a single cloud.  However, in one case where a single cloud is a good approximation (HD 27778), the curve of growth indicates that the profile fitting methods were unable to detect some saturation, a suggestion that is reinforced by the line profiles of the stronger lines.

Our biggest points of concern with respect to systematic errors are the inhomogeneous determinations of hydrogen (particularly for sightlines where only the reddening relationship was used to determine total hydrogen) and the fact that both sightlines with an {\it IUE} measurement of the 1355 \AA{} line (HD 42087 and HD 152236) had anomalously high O/H ratios.  Future work will undoubtedly allow us to resolve the first issue and refine our results.  With respect to the second point, as discussed in \S\ref{sss:HD42087} and \S\ref{sss:HD152236}, there is a great deal of uncertainty with both of the 1355 \AA{} line measurements for these sightlines, including possible contamination from stellar lines.  It would be presumptuous to assume a systematic error in the {\it IUE} data based on only two data points that are admittedly not well-determined.  Furthermore, if our derived column densities are accurate, this would likely be the result of a selection effect:  the {\it IUE} data has the poorest S/N and resolution of all the data used, and we would only be likely to conclusively observe the 1355 \AA{} line in a sightline with a very large oxygen column density.

\subsection{$b$-values}
\label{ss:bvalues}
The mean $b$-value of the {\it HST} sample is $7.7^{+0.4}_{-0.3}\kmpers$ with a standard deviation of 49\%.  If we include all 26 sightlines, we find a mean $b$-value of $9.4^{+0.5}_{-0.3}\kmpers$ with a standard deviation of 47\%.  These $b$-values are somewhat larger than the typical $b$-values ($\sim {\rm few}\kmpers$) assumed for oxygen in many previous studies, e.g. \citet{York}.  It may also be an indication that the $b$-value for neutral oxygen is inconsistent with the $b$-values of other atomic species, which may provide insight into variability of composition in clouds along a given line of sight.  For example, these average $b$-values are on the upper end of the NI $b$-values found by \citet{York}.  Since NI has an ionization potential that is slightly smaller than oxygen, it is possible that cloud components along the line of sight subject to ionizing influences (e.g., shocks and radiation) contain a smaller than average ratio of neutral nitrogen to neutral oxygen, while at the same time being the cloud components with the largest velocity dispersions due to the energy imparted by these ionizing influences.  This would allow for the possibility that the $b$-value of oxygen is larger than the $b$-value of nitrogen.  If this scenario of differing $b$-values for different atomic species is found to be common, it may provide some insight into the dynamics and composition of various cloud components along a given sightline.

For the sightlines that have data from {\it HST}, we can attempt to directly measure the ``effective'' $b$-value of the 1302 \AA{} and 1355 \AA{} lines, albeit through different methods.  The $b$-value of the 1355 \AA{} line can be inferred through the Gaussian width of the fit.  The $b$-value of the 1302 \AA{} line is derived through the fit of a Voigt profile.  In the context of this paper, a Voigt profile is any saturated profile, one that may include damping wings.  In all cases of the 1302 \AA{} line with HST data, the damping wings were apparent.  When strong damping wings are present, the $b$-value and column can be determined simultaneously with reasonable accuracy (in our measurements, the 1302 \AA{} line was never strong enough for the profile to be completely independent of $b$-value).  The 1039 \AA{} line, which was usually well-fit by a Gaussian but occasionally required the fitting of a Voigt profile, can also be fit with the same Voigt profile program in all cases to determine $b$-value and column density (in cases of minimal saturation, the profile reduces to a Gaussian).  However, because the 1039 \AA{} line is usually on the flat portion of the curve of growth, there can be a fairly wide range of solutions for the $b$-value and column density that produce reasonable fits.  Yet if one parameter ($b$-value or column density) is fixed (for consistency with the other profiles), the other parameter can be determined.  Fitting Voigt profiles to lines below 1000 \AA{} (discussed in \S\ref{ss:1000lines}), where the spectrum had many other interfering lines and a much more uncertain continuum, was used in a few cases to determine equivalent widths, but is less useful for direct determinations of column density and $b$-value.

Through an iterative process of fitting one of the three major lines (1039 \AA{}, 1302 \AA{}, and 1355 \AA{}) for both column density and $b$-value, and then generating simulated absorption profiles for the other lines for consistency, we examined the nature of the nine sightlines where we had high-resolution {\it STIS} data.  In this process, it was initially assumed that the absorption lines were well-represented by a single-component cloud along the line of sight with a specific $b$-value.  Using a combination of results from this process and results from the literature where applicable, we also engaged in multiple-component fitting and profile generation.

Using this method, the sightlines towards HD 206267 and HD 207538 are the simplest cases to check for consistency.  The column density and $b$-value derived through the curve of growth method generated reasonable fits within the errors to the observed profiles of the three major lines.  The evidence that these sightlines are not well-represented by a single-cloud approximation is very limited from the oxygen lines.  These two sightlines are the only sightlines in our {\it HST} sample that have not previously been included in oxygen studies (see \S\ref{ss:compare}), but optical spectra of these sightlines have been published in \citet{Pan}.  While the optical spectra of elements such as potassium do reveal multiple cloud components within $\sim10\kmpers$ of each other for these sightlines, this is consistent with the $b$-values of $6.3\kmpers$ (HD 206267) and $5.4\kmpers$ (HD 207538) that were determined with the curve of growth method and verified through profile generation.

Another relatively simple case is the sightline towards HD 27778.  Using profile fitting methods, \citet{Cartledge} find two major cloud components in this sightline.  However, the weak 1355 \AA{} line does not show independent evidence of this, and can be fit well by a single Gaussian.  A consistent picture of column density and $b$-value can be achieved with the single-component approximation.

The sightline towards HD 210839 was reported on by \citet{Andre}.  Observations of the 1355 \AA{} line make it apparent that there is a complex velocity structure along this line of sight.  We used the velocity measurements of \citet{Andre} to generate profiles for the three major lines that fit the data well.  The column density used to generate these profiles was the same as the column density independently measured by \citet{Andre} and through our single-component curve of growth analysis.  The curve of growth analysis measured a $b$-value of approximately $9\kmpers$, which also generates relatively accurate profiles in the single-component approximation.  This is anecdotal evidence, along with our earlier analysis of HD 206267 and HD 207538, that sometimes even complex velocity structure can be well-approximated by a single $b$-value.

Other sightlines showed an interesting property: consistent single-component profiles could be generated for the various absorption lines with a consistent column density, but with the $b$-value increasing with the strength of the line.  This is physically possible if the lines are not best represented by a single-component cloud, but rather there are one or more small, undetected components hidden in the noise near the weak 1355 \AA{} line.  These small components would not significantly contribute to the overall column density, but would contribute to the profile of the stronger lines.  While this may be physically reasonable, simple attempts to model this phenomenon through profile generation were unsuccessful.

For the sightline towards HD 24534, appropriate profiles for all three major lines can be generated with a consistent column density of $\logOI=17.83$, but with increasing $b$-values under the single-component assumption: $2.0\kmpers$ for the 1355 \AA{} line, $6.5\kmpers$ for the 1039 \AA{} line, and $8.2\kmpers$ for the 1302 \AA{} line.  This suggests that that the line of sight is not best represented by a single cloud component, but that our column density is accurate.  The curve of growth method suggests this same column density, and a $b$-value closest to the value measured for the 1039 \AA{} line.  The curve of growth analysis caused us to reject an old literature value for the oxygen column density towards HD 24534 (see \S\ref{ss:compare}), and the profile fitting of the stronger lines confirms this.

The sightlines towards HD 185418, HD 192639, and HD 207198 present similar problems.  In all three cases, multiple components are obvious in the weak line 1355 \AA{} or have been assumed elsewhere in the literature.  Yet generating profiles for the 1039 \AA{} and 1302 \AA{} lines in a given sightline using a column density consistent with the weak line and the measured or assumed multiple components of the weak line does not produce good results, in that the generated profiles of the stronger lines are either weaker or narrower than observed in the data.  Relatively consistent results can be obtained through the single-component approximation for a consistent column density but increasing $b$-values with the strength of the absorption line.  Once again, this suggests that there are one or more hidden weak cloud components.

As discussed in \S\ref{ss:compare}, the sightline towards HD 37903 is complicated because of absorption lines from vibrationally excited $\Hmol$ in the spectrum.  However, the strongest $\Hmol$ lines that might influence the measurement of the oxygen lines have rotational excitation levels of $J=7$ or greater, and it is unlikely that the potential contamination from $\Hmol$ is solely responsible for the observed increase of $b$-value in the stronger lines with the single-component approximation.  In keeping with \citet{Cartledge}, we measure the the apparent $b$-value of the 1355 \AA{} line to be $1.6\kmpers$; however, the $b$-value of the 1302 \AA{} line is greater than $9\kmpers$ when approximated as a single component.  We may once again infer there are hidden weak components near the narrow 1355 \AA{} line.  As discussed in \S\ref{ss:compare}, a moderate range of column densities, when paired with the correct $b$-value, can produce a reasonable approximation to the 1302 \AA{} line.

If we assume that this model of very weak undetected components is correct, determining the best $b$-value and column density for this type of distribution along the line of sight would best be served by the use of a multi-component curve of growth.  While the curves of growth we have used contain two separate damping wing regions to accommodate the differing damping constants of the 1039 \AA{} and 1302 \AA{} lines, a multi-component curve of growth in this context refers to a curve of growth with the assumption of multiple cloud components, possibly each with its own $b$-value.  This type of curve of growth possesses a ``multi-tiered'' flat region.  For lack of constraining data points, we have not used multi-component curves of growth, though some of our sightlines are suggestive of the need for them.

If indeed many or most sightlines are better represented by a multi-component curve of growth, it partially mitigates the discrepancy between our study and previous studies discussed at the beginning of this section, as previous studies have measured smaller $b$-values for the weak 1355 \AA{} line than our measurements through the curve of growth method.  Multi-component curves of growth might be a good compromise between the two methods.  Whereas previous studies such as \citet{York} applied a saturation correction based on $b$-value and single-component curves of growth might have a higher $b$-value (corresponding to an approximate ``effective'' $b$-value for multiple components) and not apply the same correction, multi-component curves of growth would most effectively consider the constraints from all the observed absorption lines.  The difficulty would be in obtaining a quality constraint on each line.  Future work with higher S/N spectra and the well-constrained detection of lines below 1000 \AA{} will hopefully allow us to explore this important issue in greater depth.

\subsection{Dust Grains}
\label{ss:dust}
We have quoted a theoretical upper limit from \citet{Cardelli} on the amount of oxygen that is tied up in interstellar dust grains, ${\rm O_{dust}/H}\lesssim180\ppm$.  This limit comes from an assumption on the composition of grains, and the limited amount of other elements (Fe, Si, Mg, etc.) that are available to combine with oxygen in the form of silicates in these grains.  Using the mean of our {\it HST} sample, we find ${\rm O_{dust}/H}\approx79\ppm$, with the assumption that the total (gas + dust) oxygen content of the ISM is equal to the solar value and with the upper limit determined by the lower gas-phase O/H ratio of our best determined sample.  We do not estimate errors on this value since it is highly dependent on the solar value assumed.  Given the conservative errors on each sightline and the cumulative range in the solar O/H ratio spanned by the results of \citet{Allende}, \citet{Holweger}, and \citet{Asplund}, this upper limit on dust-phase oxygen abundance may be uncertain by a factor of two or more.

\citet{Andre} use similar reasoning (but a slightly lower ISM O/H ratio and a slightly higher solar O/H ratio) to determine ${\rm O_{dust}/H}\approx109\ppm$.  If it is first assumed that maximum amount of oxygen available for dust (${\rm O_{dust}/H}\lesssim180\ppm$) is in fact in dust and the current measurement of dust is the percentage resilient to shocks, then the \citet{Andre} result implies that 60\% of dust grains containing silicates is resilient to shocks.  The results of our {\it HST} sample are in approximate agreement with this (we find 44\% resiliency).  An alternate interpretation to this presumed resiliency is the possibility that there are more oxides than silicates in grains, allowing for the enhanced depletion of elements such as iron, magnesium, and silicon, without requiring as much depletion of oxygen.

We noted that our results hint at a trend of increasing depletion with increasing $R_V$.  Papers by \citet{CCM1, CCM2} have shown that most features of the UV extinction curve can be determined to good approximation if the single parameter of $R_V$ is known.  Other studies such as \citet{MathisWallen} and \citet{MathisWhiffen} have shown that larger values of $R_V$ require grain size distributions weighted towards larger particles.  Thus, a trend of enhanced depletion with $R_V$ would imply that the depletion of oxygen is tied to grain size.  In the coldest, most dense environments large grains can accrete materials directly from the gas phase to mantles surrounding their cores, and the previously mentioned theoretical limits on oxygen in dust (based on the assumption that all oxygen in dust is found in silicates and oxides and limited by abundances of other elements) do not necessarily apply.  This would be a means for the enhanced depletion of oxygen.  In fact, if these skewed size distributions are the result of accretion, abundant elements such as carbon, nitrogen, or oxygen would be the required source of accretion material as refractory elements such as iron and calcium can only account for a limited increase in total grain volume \citep{CCM2}.  However, grain coagulation from collisions can also account for skewed size distributions and in some cases is a better fit to the total extinction per hydrogen nucleus \citep{CCM2}.

The problem with any interpretation of our result for the amount of oxygen in dust is our underlying assumption that the solar abundance of oxygen is the correct value with which to compare the total oxygen abundance in the ISM, and therefore the value from which we derive the amount of oxygen in dust.  As discussed in \S\ref{s:intro}, the accepted solar ratio of O/H has undergone significant revision in recent years, raising questions of how well this value is known.  To this point, the most recent values \citep{Allende, Holweger, Asplund} together suggest that, within the 1-$\sigma$ errors, the true solar O/H ratio is only known to a precision of $\sim30\%$ the average value adopted in this paper.  Published values of solar O/H before \citet{Withbroe} show an even larger spread \citep{Sofia}.

\subsection{O/H of Galactic Stars}
\label{ss:stellarOH}
Even if the solar ratio of O/H was well known, would it be appropriate to use it as the standard value for the total ISM oxygen abundance?  This question has been raised concerning krypton (which should not be depleted onto grains) by \citet{CardelliMeyer}, who find the average Kr/H ratio in the ISM to be $\sim60\%$ of the solar ratio.  Subsequent studies \citep{Meyer, Andre} have reviewed potential reasons for a deficit in the local ISM of elements such as krypton and oxygen:  enrichment of the presolar nebula, migration of the Sun from a smaller Galacto-centric distance, and infall of metal-poor gas onto the ISM.  Another possibility is metallicity enhancement resulting from the late infall of exoplanets or other metal-rich material onto the parent star (in this case the Sun).  If convection has begun in the parent star, the metals would remain in the photosphere and the observed metallicity would be enhanced relative to the actual metallicity.  Our study does not place strict enough constraints to comment on the relative probability of these scenarios.  However, the fact remains that the solar value of O/H is still in some question, and even if this were not the case, our estimation of oxygen in dust is very uncertain given that we cannot assume the solar value of O/H is equivalent to the average value in the ISM.

\citet{SnowWitt} and \citet{SofiaMeyer} both performed surveys of B, F, and G stars, finding that B stars have O/H ratios of $\sim350\ppm$, while F and G stars have O/H ratios of $\sim450\ppm$.  Both of these values are smaller than the current best values for the solar O/H ratio.  Additionally the O/H ratios for B stars, expected to best represent the ISM due to their recent formation, tend to be significantly lower than the average measured gas-phase O/H ratio of the ISM.  Similarly, the O/H ratios for F and G stars are not much larger than the average measured gas-phase O/H ratio of the ISM.  Both samples show spreads that are larger than the uncertainties, a result that is in contrast with studies such as \citet{Meyer} and \citet{Andre}, which show a very consistent O/H ratio across many sightlines.

The question must be asked as to how there can be such a large variation in the O/H ratios in stars that formed from a well-mixed ISM, where O/H is constant.  The answer likely lies in aspects of stellar formation that are not yet fully understood.  For example, \citet{SnowJGR} suggests that ambipolar diffusion in star-forming clouds might be able to carry away some ions and thus deplete metals in the material that stars form from.  Furthermore, photoablation of the disks of the most massive stars may prevent the type of late metallicity ``pollution'' discussed above, thus explaining why B stars have significantly lower O/H ratios than F and G stars.  A better understanding of these issues will give further insight into whether or not the elemental abundances of the Sun or any subset of observed stars can be used as a good proxy for the abundances of the ISM.

\subsection{Future Work}
\label{ss:Future}
There is much work to be done concerning oxygen in the interstellar medium.  While this study is consistent with the picture that there is a constant depletion (relative to the solar value) of gas-phase oxygen abundance relative to hydrogen and that enhanced depletions in dense regions have not been conclusively observed, we have not increased the overall constraints already established by studies such as \citet{Meyer}, \citet{Cartledge, Cartledge2}, and \citet{Andre}.  Perhaps with continued investigation into sightlines of very high reddening, we will conclusively observe our suggested trend with $R_V$.  However, this will be very difficult to do with current instrumentation, as the highly-reddened sightlines desired produce spectra with the poorest S/N.  Further work with {\it FUSE} and the possibility of a second {\it FUSE} satellite, with wavelength coverage to observe the important 1039 \AA{} line and other lines blueward of 1000 \AA{}, would help to constrain the curve of growth in many cases.  Finally, future work by infrared space telescopes SIRTF and the {\it James Webb Space Telescope} ({\it JWST}) should help to constrain the composition and amount of interstellar dust through solid-state IR features, perhaps allowing an appropriate standard of the total (gas + dust) O/H ratio to be determined from a direct measurement of dust-phase oxygen, rather than inferring dust-phase oxygen from an assumed total O/H ratio.  Unfortunately, the futures of UV instruments that are or were to be installed on {\it HST}, specifically STIS and the Cosmic Origins Spectrograph (COS), are currently unknown.

There is also work to be done concerning other atomic species.  If oxygen does not become significantly depleted onto grains in dense environments, what elements do make up the grains?  Additionally, is the total amount of material in dust grains consistent with current models?  Is there conclusive evidence of a local elemental deficit?  Why are the krypton and sulfur abundances in the local ISM significantly subsolar?  Studies such as those by \citet{Cartledge, Cartledge2} and \citet{Andre} have reconciled the ISM value of O/H with the solar value if dust is included, but the opposite question remains:  is there enough oxygen to form the large number of grains that are required by the depletion of other elements?

\section{SUMMARY}
\label{s:summary}
In summary, we measure equivalent widths for neutral oxygen absorption lines along 30 sightlines, and determine column densities for 26 of these sightlines.  Our best-constrained column densities agree very closely with the interstellar O/H ratios derived by \citet{Andre} and \citet{Cartledge2}.  Our less well-constrained column densities show some scatter in O/H, but many agree (to within errors) with the solar value we adopt of O/H=$500\pm75\ppm$.  We find an O/H mean of $421^{+47}_{-33}\ppm$, with a standard deviation of 29\%, for our 10 best-determined points, and an O/H mean of $468^{+103}_{-46}\ppm$, with a standard deviation of 98\%, for all 26 sightlines included in our final analysis.  The two cases of an apparent supersolar abundance have questions, as discussed in \S\ref{ss:outliers}.  There are two potential cases of significantly enhanced depletion in our sample, the sightlines toward HD 37903 and HD 199579.  Our results for neither sightline are conclusive, as the O/H ratio of HD 37903 may still be consistent with the more moderately enhanced depletion found by \citet{Cartledge2}, and profile fitting of the 1039 \AA{} line in the HD 199579 spectrum suggests that a smaller $b$-value and larger column density is a better solution.

We argue for a revision of the neutral oxygen column density in the sightline towards HD 24534, studied previously by \citet{Snow154368}.  Our revision is consistent with a recent result by \citet{Knauth}.  We also suggest that the literature value for the neutral oxygen column density in the sightline towards HD 27778 \citep{Cartledge} may need revision, a suggestion supported by profile fitting of the stronger oxygen lines.

A curve of growth analysis will be accomplished best in the near future by the simultaneous use of {\it FUSE} and {\it HST} data, as {\it FUSE} data alone cannot consistently provide a strong constraint on the column density of oxygen.  When {\it HST} data are available, we argue that our curve of growth method produces results similar to profile fitting methods in most cases and can be considered an appropriate parallel technique for determining elemental column densities.

Our general sample does not show convincing evidence of enhanced depletion of oxygen that is systematic with respect to any reddening parameters along the line of sight (e.g., $A_V$, $n_H$, etc.).  However, the enhanced depletion effect with respect to $\nH$ found by \citet{Cartledge2} is small and would not be seen in our general sample.  Our better-constrained {\it HST} sample does not show the effect found by \citet{Cartledge2}, though this is based on a minimal number of points.  In the {\it HST} sample, we find some very limited evidence that depletion increases with respect to $R_V$ and $\fHmol$, but the results are far from conclusive.  Further confirmation of these possible trends and the trend with respect to $\nH$ will be important as we move closer to a definitive answer for the overall abundance of interstellar oxygen.

\acknowledgments
This research has been supported by NASA contract 2430-60020 to the University of Colorado.  We would also like to acknowledge D. C. Morton for a helpful discussion of atomic data, S. I. B. Cartledge for a discussion of our results and methods, and B. A. Keeney, J. Bally, and the anonymous referee for helpful comments on the manuscript.



\clearpage 

\clearpage \clearpage  

\clearpage \clearpage   

\clearpage \clearpage   

\clearpage \clearpage

\clearpage \clearpage

\clearpage \clearpage

\clearpage \clearpage

\clearpage \clearpage

\clearpage \clearpage

\clearpage \clearpage

\clearpage \clearpage

\clearpage \clearpage

\clearpage \clearpage

\clearpage \clearpage

\clearpage \clearpage


\begin{deluxetable}{cccccc}
\tablecolumns{6}
\tablewidth{0pc}
\tabletypesize{\footnotesize}
\tablecaption{Sightlines: Stellar Data\label{stellardata}}
\tablehead{\colhead{Star Name} & \colhead{Distance (pc)} & \colhead{Distance Reference\tablenotemark{a}} & \colhead{l} & \colhead{b} & \colhead{Spectral Type}}
\startdata
HD 24534 & 590 & 2 & 163.08 & -17.14 & O9.5pe \\
HD 27778 & 223 & 1 & 172.76 & -17.39 & B3V \\
HD 37903 & 910 & 2 & 206.85 & -16.54 & B1.5V \\
HD 38087 & 480 & 2 & 207.07 & -16.26 & B5V \\
HD 40893 & 2800 & 2 & 180.09 & +4.34 & B0IV \\
HD 42087 & 1200 & 2 & 187.75 & +1.77 & B2.5Ibe\\
HD 46056 & 2300 & 2 & 206.34 & -2.25 & O8V \\
HD 46202 & 2000 & 2 & 203.31 & -2.00 & O9V \\
HD 53367 & 780 & 2 & 223.71 & -1.90 & B0IVe \\
HD 62542 & 246 & 1 & 255.92 & -9.24 & B5V \\
HD 73882 & 1100 & 2 & 260.18 & +0.64 & O8V \\
HD 110432 & 301 & 1 & 301.96 & -0.20 & B2pe \\
HD 152236 & 1800 & 3 & 343.03 & +0.87 & B1Ia+pe \\
HD 154368 & 960 & 2 & 349.97 & +3.22 & O9Ia \\
HD 164740 & 1400 & 3 & 5.97 & -1.17 & O7.5V \\
HD 167971 & 730 & 2 & 18.25 & +1.68 & O8e \\
HD 168076 & 2100 & 2 & 16.94 & +0.84 & O5f \\
HD 170740 & 213 & 1 & 21.06 & -0.53 & B2V \\
HD 179406 & 160 & 2 & 28.23 & -8.31 & B3V \\
HD 185418 & 950 & 2 & 53.6 & -2.17 & B0.5V \\
HD 186994 & 2500 & 2 & 78.62 & +10.06 & B0III \\
HD 192639 & 1100 & 2 & 74.90 & +1.48 & O8e \\
HD 197512 & 1700 & 2 & 87.89 & +4.63 & B1V \\
HD 199579 & 1200 & 2 & 87.50 & -0.30 & O6Ve\\
HD 203938 & 700 & 2 & 90.56 & -2.23 & B0.5IV \\
HD 206267 & 1000 & 3 & 99.29 & +3.74 & O6f \\
HD 207198 & 1000 & 2 & 103.14 & +6.99 & O9IIIe \\
HD 207538 & 880 & 2 & 101.60 & +4.67 & B0V \\
HD 210121 & 210 & 1 & 56.88 & -44.46 & B3V \\
HD 210839 & 505 & 1 & 103.83 & +2.64 & O6If \\
\enddata
\tablenotetext{a}{References:---(1) Hipparcos parallax of 4-$\sigma$ precision or better.  (2) Spectroscopic distance modulus.  (3) Member of an OB association, cluster, or multiple-star system.}
\end{deluxetable}

\clearpage \clearpage

\begin{deluxetable}{cccccccc}
\tablecolumns{8}
\tablewidth{0pc}
\tabletypesize{\scriptsize}
\tablecaption{Sightlines: Hydrogen Data\label{hydrogentable}}
\tablehead{\colhead{Sightline} & \colhead{$\logHmol$} & \colhead{$\Hmol$ Reference\tablenotemark{a}} & \colhead{$\logHI$} & \colhead{HI Reference\tablenotemark{b}} & \colhead{$\logHtot$} & \colhead{$\nH$} & \colhead{$\fHmol$}}
\startdata
HD 24534 & $20.92\pm0.04$ & 1 & $20.73\pm0.06$ & 1 & $21.34\pm0.03$ & 1.21 & 0.76 \\
HD 27778 & $20.79\pm0.06$ & 1 & $20.98\pm0.30$ & 2 & $21.34^{+0.16}_{-0.11}$ & 3.18 & 0.56 \\
HD 37903 & $20.92\pm0.06$ & 2 & $21.20\pm0.10$ & 1 & $21.51^{+0.06}_{-0.05}$ & 1.16 & 0.51 \\
HD 38087 & $20.64\pm0.07$ & 2 & $21.48\pm0.15$ & 3 & $21.59^{+0.13}_{-0.12}$ & 2.63 & 2.63 \\
HD 40893 & $20.58\pm0.05$ & 2 & $21.28\pm0.30$ & 4 & $21.43^{+0.23}_{-0.20}$ & 0.31 & 0.29 \\
HD 42087 & $20.52\pm0.12$ & 2 & $21.40\pm0.11$ & 1 & $21.50^{+0.10}_{-0.09}$ & 0.86 & 0.21 \\
HD 46056 & $20.68\pm0.06$ & 2 & $21.38\pm0.14$ & 1 & $21.53\pm0.11$ & 0.47 & 0.29 \\
HD 46202 & $20.68\pm0.07$ & 2 & $21.58\pm0.15$ & 1 & $21.68^{+0.13}_{-0.12}$ & 0.77 & 0.20 \\
HD 53367 & $21.04\pm0.05$ & 2 & $21.32\pm0.30$ & 4 & $21.63^{+0.18}_{-0.13}$ & 1.78 & 0.51 \\
HD 62542 & $20.81\pm0.21$ & 1 & $20.93\pm0.30$ & 2 & $21.33^{+0.17}_{-0.13}$ & 2.83 & 0.60 \\
HD 73882 & $21.11\pm0.08$ & 1 & $21.11\pm0.15$ & 3 & $21.59\pm0.07$ & 1.14 & 0.67\\
HD 110432 & $20.64\pm0.04$ & 1 & $20.85\pm0.15$ & 5 & $21.20^{+0.08}_{-0.07}$ & 1.70 & 0.55 \\
HD 152236 & $20.73\pm0.12$ & 2 & $21.77\pm0.13$ & 1 & $21.84^{+0.12}_{-0.11}$ & 1.26 & 0.15 \\
HD 154368 & $21.16\pm0.07$ & 1 & $21.00\pm0.05$ & 6 & $21.59^{+0.05}_{-0.04}$ & 1.32 & 0.74 \\
HD 164740 & $20.19\pm0.10$ & 2 & $21.95\pm0.15$ & 3 & $21.96^{+0.15}_{-0.14}$ & 1.16 & 0.03 \\
HD 167971 & $20.85\pm0.12$ & 1 & $21.60\pm0.30$ & 7 & $21.73^{+0.24}_{-0.02}$ & 2.40 & 0.26 \\
HD 168076 & $20.68\pm0.08$ & 1 & $21.65\pm0.23$ & 1 & $21.73^{+0.21}_{-0.18}$ & 0.84 & 0.18 \\
HD 170740 & $20.86\pm0.08$ & 1 & $21.15\pm0.15$ & 1 & $21.46^{+0.09}_{-0.08}$ & 4.36 & 0.51 \\
HD 179406 & $20.73\pm0.07$ & 2 & $21.23\pm0.15$ & 8 & $21.44^{+0.11}_{-0.11}$ & 5.62 & 0.39 \\
HD 185418 & $20.76\pm0.05$ & 1 & $21.11\pm0.15$ & 3 & $21.39^{+0.09}_{-0.08}$ & 0.83 & 0.47 \\
HD 186994 & $19.59\pm0.04$ & 2 & $20.90\pm0.15$ & 9 & $20.94\pm0.14$ & 0.11 & 0.09 \\
HD 192639 & $20.69\pm0.05$ & 1 & $21.32\pm0.12$ & 1 & $21.49\pm0.09$ & 0.91 & 0.32 \\
HD 197512 & $20.66\pm0.05$ & 1 & $21.26\pm0.15$ & 3 & $21.44^{+0.11}_{-0.10}$ & 0.52 & 0.33 \\
HD 199579 & $20.53\pm0.04$ & 1 & $21.04\pm0.11$ & 1 & $21.25\pm0.07$ & 0.48 & 0.38 \\
HD 203938 & $21.00\pm0.06$ & 1 & $21.48\pm0.15$ & 3 & $21.70^{+0.10}_{-0.09}$ & 2.32 & 0.40 \\
HD 206267 & $20.86\pm0.04$ & 1 & $21.30\pm0.15$ & 7 & $21.54^{+0.10}_{-0.09}$ & 1.12 & 0.42 \\
HD 207198 & $20.83\pm0.04$ & 1 & $21.34\pm0.17$ & 1 & $21.55^{+0.12}_{-0.10}$ & 1.15 & 0.38 \\
HD 207538 & $20.91\pm0.06$ & 1 & $21.34\pm0.12$ & 1 & $21.58^{+0.08}_{-0.07}$ & 1.41 & 0.43 \\
HD 210121 & $20.75\pm0.12$ & 1 & $20.63\pm0.15$ & 10 & $21.19^{+0.08}_{-0.07}$ & 2.40 & 0.73 \\
HD 210839 & $20.84\pm0.04$ & 1 & $21.15\pm0.10$ & 1 & $21.45\pm0.06$ & 1.80 & 0.49 \\
\enddata
\tablenotetext{a}{References for $\Hmol$:---(1) \citet{Rachford}.  (2) \citet{Rachford2}.}
\tablenotetext{b}{References for HI:---(1) \citet{Diplas}.  (2) \citet{Rachford}, $\NHI=5.8\times10^{21}E_{B-V}-2\NHmol$; errors in $\NHI$ assumed to be $\pm0.30$ dex.  (3) \citet{FitzMassa}; errors in $\NHI$ assumed to be $\pm0.15$ dex.  (4) \citet{Rachford2}, $\NHI=5.8\times10^{21}E_{B-V}-2\NHmol$; errors in $\NHI$ assumed to be $\pm0.30$ dex.  (5) \citet{Rachford110432}.  (6) \citet{Snow154368}.  (7) \citet{Rachford}, Ly$\alpha$ profile fitting.  (8) \citet{Hanson}.  (9) \citet{Savage}.  (10) \citet{WeltyFowler}.}
\end{deluxetable}

\clearpage \clearpage

\begin{deluxetable}{cccc}
\tablecolumns{4}
\tablewidth{0pc}
\tabletypesize{\footnotesize}
\tablecaption{Sightlines: Reddening Data\label{reddeningtable}}
\tablehead{\colhead{Sightline} & \colhead{$E_{B-V}$ (magnitudes)} & \colhead{$A_V$ (magnitudes)} & \colhead{$R_V$}}
\startdata
HD 24534 & 0.59 & 2.05 & 3.47 \\
HD 27778 & 0.37 & 1.01 & 2.72 \\
HD 37903 & 0.35 & 1.28 & 3.67 \\
HD 38087 & 0.29 & 1.61 & 5.57 \\
HD 40893 & 0.46 & 1.13 & 2.46 \\
HD 42087 & 0.36 & 1.10 & 3.06 \\
HD 46056 & 0.50 & 1.30 & 2.60 \\
HD 46202 & 0.49 & 1.39 & 2.83 \\
HD 53367 & 0.74 & 1.76 & 2.38 \\
HD 62542 & 0.35 & 0.99 & 2.83 \\
HD 73882 & 0.70 & 2.36 & 3.37 \\
HD 110432 & 0.51 & 2.02 & 3.95 \\
HD 152236 & 0.68 & 2.24 & 3.29 \\
HD 154368 & 0.78 & 2.34 & 3.00 \\
HD 164740 & 0.87 & 4.66 & 5.36 \\
HD 167971 & 1.08 & 3.42 & 3.17 \\
HD 168076 & 0.78 & 2.77 & 3.55 \\
HD 170740 & 0.48 & 1.30 & 2.71 \\
HD 179406 & 0.33 & 0.94 & 2.86 \\
HD 185418 & 0.50 & 1.16 & 2.32 \\
HD 186994 & 0.17 & 0.53\tablenotemark{a} & 3.10\tablenotemark{a} \\
HD 192639 & 0.66 & 1.87 & 2.84 \\
HD 197512 & 0.32 & 0.75 & 2.35 \\
HD 199579 & 0.37 & 1.09 & 2.95 \\
HD 203938 & 0.74 & 2.15 & 2.91 \\
HD 206267 & 0.53 & 1.41 & 2.67 \\
HD 207198 & 0.62 & 1.50 & 2.42 \\
HD 207538 & 0.64 & 1.44 & 2.25 \\
HD 210121 & 0.40 & 0.83 & 2.08 \\
HD 210839 & 0.57 & 1.58 & 2.78 \\
\enddata
\tablenotetext{a}{Independent measurements of $R_V$ and $A_V$ not available; $R_V$ assumed to be 3.10; $A_V$=3.10$\times E_{B-V}$}
\end{deluxetable}

\clearpage \clearpage

\begin{deluxetable}{cccc}
\tablecolumns{4}
\tablewidth{0pc}
\tablecaption{Absorption Lines Observed\label{linetable}}
\tablehead{\colhead{Wavelength (\AA{})} & \colhead{Oscillator Strength $f$} & \colhead{Damping Constant $\gamma$} & \colhead{Data Used}}
\startdata
919.917 & $1.77\times10^{-4}$ & \nodata\tablenotemark{a} & {\it FUSE} \\
921.857 & $1.08\times10^{-3}$ & \nodata\tablenotemark{a} & {\it FUSE} \\
922.200 & $2.45\times10^{-4}$ & \nodata\tablenotemark{a} & {\it FUSE} \\
925.4461 & $3.54\times10^{-4}$ & \nodata\tablenotemark{a} & {\it FUSE} \\
936.6295 & $3.65\times10^{-3}$ & \nodata\tablenotemark{a} & {\it FUSE} \\
976.4481 & $3.31\times10^{-3}$ & \nodata\tablenotemark{a} & {\it FUSE} \\
1039.2304 & $9.07\times10^{-3}$ & $1.87\times10^{8}$ & {\it FUSE} \\
1302.1685 & $4.80\times10^{-2}$ & $5.65\times10^{8}$ & {\it HST}, {\it IUE} \\
1355.5977 & $1.16\times10^{-6}$ & $5.56\times10^{3}$ & {\it HST}, {\it IUE} \\
\enddata
\tablenotetext{a}{Value of $\gamma$ not reported in \citet{Morton03} for this absorption line.}
\end{deluxetable}

\clearpage \clearpage

\begin{deluxetable}{cccc}
\tablecolumns{4}
\tablewidth{0pc}
\tabletypesize{\scriptsize}
\tablecaption{Absorption Lines in the {\it FUSE} Wavelength Region Not Used in This Study\label{nolinetable}}
\tablehead{\colhead{Wavelength (\AA{})} & \colhead{Oscillator Strength $f$} & \colhead{Damping Constant $\gamma$} & \colhead{Reason Unused\tablenotemark{a}}}
\startdata
912.159 & $1.40\times10^{-5}$ & \nodata\tablenotemark{b} & Ly-$\alpha$ \\
912.321 & $7.69\times10^{-5}$ & \nodata\tablenotemark{b} & Ly-$\alpha$ \\
912.321 & $1.60\times10^{-5}$ & \nodata\tablenotemark{b} & Ly-$\alpha$ \\
912.498 & $8.84\times10^{-5}$ & \nodata\tablenotemark{b} & Ly-$\alpha$ \\
912.500 & $1.84\times10^{-5}$ & \nodata\tablenotemark{b} & Ly-$\alpha$ \\
912.723 & $2.13\times10^{-5}$ & \nodata\tablenotemark{b} & Ly-$\alpha$ \\
912.729 & $1.02\times10^{-4}$ & \nodata\tablenotemark{b} & Ly-$\alpha$ \\
912.964 & $2.49\times10^{-5}$ & \nodata\tablenotemark{b} & Ly-$\alpha$ \\
912.965 & $1.19\times10^{-5}$ & \nodata\tablenotemark{b} & Ly-$\alpha$ \\
913.250 & $2.93\times10^{-5}$ & \nodata\tablenotemark{b} & Ly-$\alpha$ \\
913.259 & $1.14\times10^{-4}$ & \nodata\tablenotemark{b} & Ly-$\alpha$ \\
913.596 & $1.67\times10^{-4}$ & \nodata\tablenotemark{b} & Ly-$\alpha$ \\
913.620 & $3.48\times10^{-5}$ & \nodata\tablenotemark{b} & Ly-$\alpha$ \\
913.913 & $2.00\times10^{-4}$ & \nodata\tablenotemark{b} & Ly-$\alpha$ \\
914.057 & $4.18\times10^{-5}$ & \nodata\tablenotemark{b} & Ly-$\alpha$ \\
914.504 & $2.43\times10^{-4}$ & \nodata\tablenotemark{b} & Ly-$\alpha$ \\
914.570 & $5.07\times10^{-5}$ & \nodata\tablenotemark{b} & Ly-$\alpha$ \\
915.057 & $2.98\times10^{-4}$ & \nodata\tablenotemark{b} & Ly-$\alpha$ \\
915.057 & $6.25\times10^{-5}$ & \nodata\tablenotemark{b} & Ly-$\alpha$ \\
915.057 & $3.74\times10^{-4}$ & \nodata\tablenotemark{b} & Ly-$\alpha$ \\
915.057 & $7.86\times10^{-5}$ & \nodata\tablenotemark{b} & Ly-$\alpha$ \\
916.815 & $4.74\times10^{-4}$ & \nodata\tablenotemark{b} & Ly-$\alpha$ \\
916.960 & $1.01\times10^{-4}$ & \nodata\tablenotemark{b} & Ly-$\alpha$ \\
918.044 & $6.14\times10^{-4}$ & \nodata\tablenotemark{b} & $\Hmol$ \\
918.222 & $1.32\times10^{-4}$ & \nodata\tablenotemark{b} & $\Hmol$ \\
919.658 & $7.92\times10^{-4}$ & \nodata\tablenotemark{b} & $\Hmol$ \\
924.950 & $1.54\times10^{-3}$ & \nodata\tablenotemark{b} & $\Hmol$ \\
925.0173 & \nodata\tablenotemark{c} & \nodata\tablenotemark{b} & $\Hmol$ \\
925.0173 & \nodata\tablenotemark{c} & \nodata\tablenotemark{b} & $\Hmol$ \\
925.0174 & \nodata\tablenotemark{c} & \nodata\tablenotemark{b} & $\Hmol$ \\
929.5168 & $2.29\times10^{-3}$ & \nodata\tablenotemark{b} & $\Hmol$ \\
929.5969 & \nodata\tablenotemark{c} & \nodata\tablenotemark{b} & $\Hmol$ \\
929.5970 & \nodata\tablenotemark{c} & \nodata\tablenotemark{b} & $\Hmol$ \\
929.5971 & \nodata\tablenotemark{c} & \nodata\tablenotemark{b} & $\Hmol$ \\
930.2566 & $5.37\times10^{-4}$ & \nodata\tablenotemark{b} & $\Hmol$ \\
930.6998 & \nodata\tablenotemark{c} & \nodata\tablenotemark{b} & $\Hmol$ \\
936.4358 & \nodata\tablenotemark{c} & \nodata\tablenotemark{b} & $\Hmol$ \\
936.7551 & \nodata\tablenotemark{c} & \nodata\tablenotemark{b} & Too weak\tablenotemark{d} \\
936.7552 & \nodata\tablenotemark{c} & \nodata\tablenotemark{b} & Too weak\tablenotemark{d} \\
936.7554 & \nodata\tablenotemark{c} & \nodata\tablenotemark{b} & Too weak\tablenotemark{d} \\
937.8405 & $8.77\times10^{-4}$ & \nodata\tablenotemark{b} & $\Hmol$\tablenotemark{e} \\
938.5674 & \nodata\tablenotemark{c} & \nodata\tablenotemark{b} & $\Hmol$ \\
948.6855 & $6.31\times10^{-3}$ & \nodata\tablenotemark{b} & $\Hmol$ \\
948.8977 & \nodata\tablenotemark{c} & \nodata\tablenotemark{b} & $\Hmol$ \\
948.8980 & \nodata\tablenotemark{c} & \nodata\tablenotemark{b} & $\Hmol$ \\
948.8983 & \nodata\tablenotemark{c} & \nodata\tablenotemark{b} & $\Hmol$ \\
950.8846 & $1.58\times10^{-3}$ & \nodata\tablenotemark{b} & $\Hmol$ \\
952.2059 & \nodata\tablenotemark{c} & \nodata\tablenotemark{b} & $\Hmol$ \\
971.7371 & \nodata\tablenotemark{c} & \nodata\tablenotemark{b} & $\Hmol$ \\
971.7376 & \nodata\tablenotemark{c} & \nodata\tablenotemark{b} & $\Hmol$ \\
971.7382 & \nodata\tablenotemark{c} & \nodata\tablenotemark{b} & $\Hmol$ \\
974.0700 & $1.56\times10^{-5}$ & \nodata\tablenotemark{b} & $\Hmol$, too weak \\
979.2718 & \nodata\tablenotemark{c} & \nodata\tablenotemark{b} & Too weak\tablenotemark{d} \\
988.5778 & $5.53\times10^{-4}$ & \nodata\tablenotemark{b} & Multiplet self-blending \\
988.6549 & $8.30\times10^{-3}$ & \nodata\tablenotemark{b} & Multiplet self-blending \\
988.7734 & $4.65\times10^{-2}$ & \nodata\tablenotemark{b} & Multiplet self-blending \\
1025.7616 & $1.63\times10^{-2}$ & $1.02\times10^{8}$ & Ly-$\beta$ \\
1025.7626 & $2.91\times10^{-3}$ & $1.02\times10^{8}$ & Ly-$\beta$ \\
1025.7633 & $1.94\times10^{-4}$ & $1.02\times10^{8}$ & Ly-$\beta$ \\
1026.4733 & $1.43\times10^{-9}$ & $4.59\times10^{7}$ & $\Hmol$, too weak \\
1026.4744 & $7.30\times10^{-9}$ & $4.59\times10^{7}$ & $\Hmol$, too weak \\
1026.4757 & $8.71\times10^{-9}$ & $4.58\times10^{7}$ & $\Hmol$, too weak \\
1047.3756 & $7.08\times10^{-8}$ & $2.93\times10^{7}$ & Too weak \\
\enddata
\tablenotetext{a}{Describes the reason the line was not observed or conclusively measured, e.g. an atomic or $\Hmol$ line blends with the OI line or makes the continuum too uncertain to measure the OI line, or the OI line is too weak to observe in typical {\it FUSE} spectra.}
\tablenotetext{b}{Value of $\gamma$ not reported in \citet{Morton03} for this absorption line.}
\tablenotetext{c}{Value of the oscillator strength not reported in \citet{Morton03} for this absorption line.}
\tablenotetext{d}{Strength of line unknown, but not observed in the highest S/N spectra.}
\tablenotetext{e}{May also be contaminated by FeII line.}
\end{deluxetable}

\clearpage \clearpage

\begin{deluxetable}{cccccccccc}
\tablecolumns{10}
\tablewidth{0pc}
\tabletypesize{\scriptsize}
\tablecaption{Measured Equivalent Widths\label{eqwidths}}
\tablehead{\colhead{Sightline} & \colhead{$W_{919}$} & \colhead{$W_{921}$} & \colhead{$W_{922}$} & \colhead{$W_{925}$} & \colhead{$W_{936}$} & \colhead{$W_{976}$} & \colhead{$W_{1039}$} & \colhead{$W_{1302}$} & \colhead{$W_{1355}$} \\ \colhead{} & \colhead{(m\AA{})} & \colhead{(m\AA{})} & \colhead{(m\AA{})} & \colhead{(m\AA{})} & \colhead{(m\AA{})} & \colhead{(m\AA{})} & \colhead{(m\AA{})} & \colhead{(m\AA{})} & \colhead{(m\AA{})}}
\startdata
HD 24534 & \nodata & \nodata & \nodata & $73\pm36$ & $94\pm9$ & $101\pm9$ & $128\pm2$ & $286\pm7$ & $11.1\pm0.3$ \\
HD 27778 & \nodata & \nodata & \nodata & \nodata & \nodata & \nodata & $70\pm9$ & $264\pm23$ & $10.9\pm2.7$ \\
HD 37903 & \nodata & \nodata & \nodata & \nodata & \nodata & \nodata & $200\pm9$ & $347\pm34$ & $9.8\pm0.6$ \\
HD 38087 & \nodata & \nodata & \nodata & \nodata & \nodata & \nodata & $238\pm15$ & $393\pm39$ & $\leq21$ \\
HD 40893 & \nodata & \nodata & \nodata & \nodata & \nodata & \nodata & $275\pm34$ & \nodata & \nodata \\
HD 42087 & \nodata & \nodata & \nodata & $114\pm51$ & $152\pm81$ & \nodata & $215\pm29$ & $436\pm41$ & $87\pm28$ \\
HD 46056 & \nodata & \nodata & \nodata & \nodata & \nodata & \nodata & $178\pm30$ & $353\pm61$ & $\leq24$ \\
HD 46202 & \nodata & \nodata & \nodata & \nodata & $165\pm41$ & \nodata & $150\pm9$ & $343\pm43$ & $\leq19$ \\
HD 53367 & \nodata & \nodata & \nodata & \nodata & \nodata & \nodata & $159\pm14$ & $332\pm31$ & $\leq11$ \\
HD 62542 & \nodata & \nodata & \nodata & \nodata & \nodata & \nodata & $306\pm23$ & $383\pm124$ & \nodata \\
HD 73882 & \nodata & \nodata & \nodata & \nodata & \nodata & \nodata & $233\pm85$ & $336\pm56$ & $\leq11$ \\
HD 110432 & \nodata & $57\pm9$ & $60\pm16$ & $59\pm10$ & $80\pm5$ & $100\pm8$ & $104\pm2$ & $181\pm27$ & $\leq16$ \\
HD 152236 & \nodata & \nodata & \nodata & \nodata & \nodata & \nodata & $292\pm89$ & $894\pm132$ & $157\pm18$ \\
HD 154368 & \nodata & \nodata & \nodata & \nodata & \nodata & \nodata & $218\pm9$ & $399\pm40$ & $16.9\pm6.8$\tablenotemark{a} \\
HD 164740 & \nodata & \nodata & \nodata & \nodata & \nodata & \nodata & $190\pm15$ & \nodata & \nodata \\
HD 167971 & \nodata & \nodata & \nodata & \nodata & \nodata & \nodata & \nodata & $667\pm267$ & $\leq28$ \\
HD 168076 & \nodata & \nodata & \nodata & \nodata & \nodata & \nodata & $257\pm119$ & $597\pm75$ & $\leq20$ \\
HD 170740 & \nodata & \nodata & \nodata & \nodata & \nodata & \nodata & $130\pm11$ & $279\pm57$ & $\leq22$ \\
HD 179406 & \nodata & \nodata & \nodata & \nodata & \nodata & \nodata & $117\pm15$ & $224\pm15$ & $\leq8$ \\
HD 185418 & \nodata & \nodata & \nodata & \nodata & $112\pm27$ & \nodata & $132\pm5$ & $327\pm25$ & $16.9\pm1.9$ \\
HD 186994 & $133\pm42$ & \nodata & \nodata & $130\pm14$ & $256\pm23$ & $241\pm52$ & $269\pm18$ & $385\pm25$ & $\leq8$ \\
HD 192639 & \nodata & \nodata & \nodata & \nodata & $237\pm109$ & \nodata & $222\pm43$ & $483\pm39$ & $22.8\pm2.2$ \\
HD 197512 & \nodata & \nodata & \nodata & \nodata & $124\pm65$ & \nodata & $128\pm3$ & \nodata & \nodata \\
HD 199579 & $58\pm8$ & \nodata & $135\pm16$ & $95\pm6$ & $209\pm26$ & $236\pm16$ & $243\pm48$ & $318\pm17$ & $\leq10$ \\
HD 203938 & \nodata & \nodata & \nodata & \nodata & \nodata & \nodata & $71\pm11$ & \nodata & \nodata \\
HD 206267 & \nodata & \nodata & \nodata & \nodata & $123\pm21$ & $142\pm53$ & $133\pm5$ & $339\pm20$ & $20.4\pm1.4$ \\
HD 207198 & \nodata & \nodata & \nodata & \nodata & \nodata & \nodata & $95\pm8$ & $362\pm25$ & $20.9\pm2.2$ \\
HD 207538 & \nodata & \nodata & \nodata & \nodata & \nodata & \nodata & $128\pm11$ & $356\pm5$ & $22.9\pm1.8$ \\
HD 210121 & \nodata & \nodata & \nodata & \nodata & \nodata & \nodata & $127\pm34$ & $302\pm118$ & $\leq27$ \\
HD 210839 & \nodata & \nodata & \nodata & $101\pm13$ & $150\pm8$ & $145\pm10$ & $197\pm14$ & $370\pm17$ & $20.9\pm1.1$ \\
\enddata
\tablenotetext{a}{Equivalent width taken from \citet{Snow154368}}
\end{deluxetable}

\clearpage \clearpage

\begin{deluxetable}{ccccc}
\tablecolumns{5}
\tablewidth{0pc}
\tablecaption{Column Density Results:  Sightlines with Three or More Measured Equivalent Widths\label{coldensities}}
\tablehead{\colhead{Sightline} & \colhead{$\logOI$} & \colhead{$b$-value} & \colhead{$\logOH$\tablenotemark{a}} & \colhead{O/H (ppm)\tablenotemark{a}}}
\startdata
HD 24534 & $17.83^{+0.04}_{-0.03}$ & $6.6\pm0.3$ & -3.51 & $307^{+36}_{-27}$ \\
HD 27778 & $18.05\pm0.09$ & $2.7^{+0.7}_{-0.8}$ & -3.29 & $513^{+254}_{-150}$ \\
HD 37903 & $17.74^{+0.04}_{-0.03}$ & $10.9\pm0.6$ & -3.77 & $169^{+29}_{-22}$ \\
HD 42087 & $18.53^{+0.18}_{-0.22}$ & $9.4^{+3.0}_{-2.5}$ & -2.97 & $1068^{+602}_{-467}$ \\
HD 46202 & $18.24^{+0.16}_{-0.20}$ & $7.1\pm0.7$ & -3.44 & $365^{+203}_{-160}$ \\
HD 110432 & $17.01^{+1.03}_{-0.56}$ & $6.1^{+1.1}_{-1.4}$ & -4.19 & $65^{+629}_{-48}$ \\
HD 152236 & $19.20\pm0.13$ & $15.3^{+5.5}_{-3.3}$ & -2.64 & $2276^{+1044}_{-774}$ \\
HD 154368 & $18.12^{+0.18}_{-0.23}$ & $11.0^{+0.9}_{-0.7}$ & -3.47 & $339^{+177}_{-143}$ \\
HD 185418 & $18.07\pm0.09$ & $6.4^{+0.5}_{-0.6}$ & -3.32 & $482^{+154}_{-119}$ \\
HD 186994 & $17.16^{+0.72}_{-0.29}$ & $17.1^{+2.7}_{-3.4}$ & -3.78 & $166^{+707}_{-93}$ \\
HD 192639 & $18.12^{+0.08}_{-0.07}$ & $14.8^{+2.6}_{-2.5}$ & -3.37 & $430^{+128}_{-97}$ \\
HD 199579 & $16.92^{+0.16}_{-0.14}$ & $15.5^{+2.3}_{-1.9}$ & -4.33 & $47^{+23}_{-15}$ \\
HD 206267 & $18.14\pm0.07$ & $6.3^{+0.7}_{-0.6}$ & -3.40 & $401^{+120}_{-92}$ \\
HD 207198 & $18.30^{+0.08}_{-0.06}$ & $3.5^{+0.6}_{-0.7}$ & -3.25 & $564^{+203}_{-136}$ \\
HD 207538 & $18.32\pm0.03$ & $5.4^{+1.0}_{-0.9}$ & -3.26 & $548^{+111}_{-88}$ \\
HD 210839 & $18.11^{+0.07}_{-0.06}$ & $9.1^{+0.8}_{-0.9}$ & -3.34 & $461^{+102}_{-78}$ \\
\enddata
\tablenotetext{a}{We adopt $\log \OHsolar=-3.30^{+0.06}_{-0.07}$, i.e. $\OHsolar=500\pm75 \ppm$}
\end{deluxetable}

\clearpage \clearpage

\begin{deluxetable}{cccccc}
\tablecolumns{6}
\tablewidth{0pc}
\tabletypesize{\footnotesize}
\tablecaption{Column Density Results: Sightlines with Two Measured Equivalent Widths\label{twopointcogs}\tablenotemark{a}}
\tablehead{\colhead{\bf{Sightline}} & \multicolumn{2}{c}{\bf{Solution \#1}} & \multicolumn{2}{c}{\bf{Solution \#2}} & \colhead{\bf{Adopted}} \\ \colhead{} & \colhead{$\logOI$} & \colhead{$b$-value}& \colhead{$\logOI$} & \colhead{$b$-value} & \colhead{\bf{Solution}}}
\startdata
HD 38087 & $16.14^{+0.21}_{-0.13}$ & $21.1^{+1.6}_{-1.9}$ & $18.06^{+0.11}_{-0.13}$ & $12.2\pm0.5$ & \#2 \\
HD 46056 & $15.66^{+0.13}_{-0.10}$ & $21.9^{+2.4}_{-2.2}$ & $18.20^{+0.10}_{-0.12}$ & $8.6\pm0.7$ & \#2 \\
HD 53367 & $15.56^{+0.05}_{-0.04}$ & $21.3\pm1.1$ & $18.18\pm0.05$ & $7.6^{+0.3}_{-0.4}$ & \#2 \\
HD 62542 & $17.43^{+0.58}_{-0.78}$ & $18.1^{+4.4}_{-2.1}$ & \nodata & \nodata & N/A \\
HD 73882 & $17.20^{+0.80}_{-1.21}$ & $13.8^{+5.3}_{-3.5}$ & \nodata & \nodata & N/A \\
HD 168076 & $18.74^{+0.07}_{-0.13}$ & $11.5^{+3.3}_{-2.6}$ & \nodata & \nodata & N/A \\
HD 170740 & $15.46^{+0.07}_{-0.06}$ & $18.0^{+2.3}_{-1.9}$ & $18.05^{+0.09}_{-0.12}$ & $6.2^{+0.4}_{-0.3}$ & \#2 \\
HD 179406 & $15.52\pm0.08$ & $13.6^{+0.7}_{-0.6}$ & $17.77\pm0.06$ & $5.9\pm0.4$ & \#2 \\
HD 197512 & $17.64^{+1.11}_{-1.57}$ & $6.7^{+3.7}_{-2.5}$ & \nodata & \nodata & N/A \\
HD 210121 & $15.39^{+0.15}_{-0.11}$ & $20.6^{+\infty}_{-4.5}$\tablenotemark{b} & $18.15^{+0.18}_{-0.22}$ & $5.9^{+0.9}_{-0.8}$ & \#2 \\
\cline{1-6} \\
\multicolumn{6}{c}{\bf{Values for Adopted Solutions}} \\
\cline{1-6} \\
\multicolumn{2}{c}{Sightline} & \multicolumn{2}{c}{$\logOH$} & \multicolumn{2}{c}{O/H} \\
\cline{1-6} \\
\multicolumn{2}{c}{HD 38087} & \multicolumn{2}{c}{-3.53} & \multicolumn{2}{c}{$295^{+128}_{-102}$} \\
\multicolumn{2}{c}{HD 46056} & \multicolumn{2}{c}{-3.33} & \multicolumn{2}{c}{$472^{+179}_{-148}$} \\
\multicolumn{2}{c}{HD 53367} & \multicolumn{2}{c}{-3.41} & \multicolumn{2}{c}{$353^{+179}_{-95}$} \\
\multicolumn{2}{c}{HD 62542} & \multicolumn{2}{c}{-3.90} & \multicolumn{2}{c}{$126^{+357}_{-111}$} \\
\multicolumn{2}{c}{HD 73882} & \multicolumn{2}{c}{-4.39} & \multicolumn{2}{c}{$41^{+218}_{-39}$} \\
\multicolumn{2}{c}{HD 168076} & \multicolumn{2}{c}{-2.99} & \multicolumn{2}{c}{$1013^{+610}_{-433}$} \\
\multicolumn{2}{c}{HD 170740} & \multicolumn{2}{c}{-3.41} & \multicolumn{2}{c}{$392^{+124}_{-113}$} \\
\multicolumn{2}{c}{HD 179406} & \multicolumn{2}{c}{-3.67} & \multicolumn{2}{c}{$214^{+64}_{-48}$} \\
\multicolumn{2}{c}{HD 197512} & \multicolumn{2}{c}{-3.71} & \multicolumn{2}{c}{$194^{+64}_{-47}$} \\
\multicolumn{2}{c}{HD 210121} & \multicolumn{2}{c}{-3.04} & \multicolumn{2}{c}{$911^{+501}_{-387}$} \\
\enddata
\tablenotetext{a}{Solutions are points of local minima in $\chi^2$ analysis; quoted errors are where the absolute $\chi^2\leq0.1$}
\tablenotetext{b}{$b$-values with an upper bound of $+\infty$ do not reflect an actual infinite upper limit, but a limit in our analysis that only considers solutions up to $b=25.0\kmpers$.}
\end{deluxetable}

\clearpage \clearpage

\begin{deluxetable}{cccc}
\tablecolumns{4}
\tablewidth{0pc}
\tablecaption{Comparison With Previous Work\label{comparison}}
\tablehead{\colhead{Sightline} & \colhead{$\logOI$} & \colhead{$\logOI$} & \colhead{Source} \\ \colhead{} & \colhead{(this work)} & \colhead{(previous work)} & \colhead{}}
\startdata
HD 24534 & $17.83^{+0.04}_{-0.03}$ & $18.12\pm0.06$\tablenotemark{a} & \citet{SnowXPer} \\
HD 24534 & $17.83^{+0.04}_{-0.03}$ & $17.87\pm0.02$ & \citet{Knauth} \\
HD 27778 & $18.05\pm0.09$ & $17.83\pm0.04$ & \citet{Cartledge} \\
HD 37903 & $17.74^{+0.04}_{-0.03}$ & $17.88\pm0.02$ & \citet{Cartledge} \\
HD 154368 & $18.12^{+0.18}_{-0.23}$\tablenotemark{b} & $18.11^{+0.10}_{-0.13}$\tablenotemark{a} & \citet{Snow154368} \\
HD 179406 & $17.77\pm0.06$\tablenotemark{c} & $17.8\pm0.2$ & \citet{Hanson} \\
HD 185418 & $18.07\pm0.09$ & $18.06\pm0.05$ & \citet{Cartledge}\\
HD 185418 & $18.07\pm0.09$ & $18.05\pm0.02$ & \citet{Andre}\\
HD 192639 & $18.12^{+0.08}_{-0.07}$ & $18.16\pm0.11$ & \citet{Sonnentrucker} \\
HD 192639 & $18.12^{+0.08}_{-0.07}$ & $18.13\pm0.02$ & \citet{Andre} \\
HD 192639 & $18.12^{+0.08}_{-0.07}$ & $18.14\pm0.04$ & \citet{Cartledge2} \\
HD 207198 & $18.30^{+0.08}_{-0.06}$ & $18.13\pm0.04$ & \citet{Cartledge} \\
HD 210839 & $18.11^{+0.07}_{-0.06}$ & $18.11^{+0.03}_{-0.02}$ & \citet{Andre} \\
\enddata
\tablenotetext{a}{Revised for \citet{Welty} $f$-value of the 1355 \AA{} line}
\tablenotetext{b}{This work utilizes the \citet{Snow154368} value for the equivalent width of the 1355 \AA{} line; therefore the results are not independent}
\tablenotetext{c}{Preferred solution from a two-point curve of growth; see Table \ref{twopointcogs} and discussion in \S\ref{ss:twopointcogs}}
\end{deluxetable}

\end{document}